\documentclass[acmtog]{acmart}

\usepackage{subcaption}
\usepackage{graphicx}
\usepackage{algpseudocode}
\usepackage{booktabs} % For formal tables
\usepackage{etoolbox}
\usepackage{subcaption}
\usepackage{multirow}
\usepackage{tikz}
\usepackage{tabularx}

\newcommand{\mydraft}{false}
\usepackage[ruled]{algorithm2e} % For algorithms

\newcommand{\FF}{\mathbf{F}}
\newcommand{\PP}{\mathbf{P}}
\newcommand{\eps}{\boldsymbol\epsilon}
\renewcommand{\SS}{\mathbf{S}}

\newcommand{\ff}{\mathbf{f}}
\newcommand{\fm}{\boldsymbol{\phi}}
\newcommand{\hP}{\hat{\mathbf{P}}}
\newcommand{\trial}{\boldsymbol{\Upsilon}}
\newcommand{\II}{\mathbf{I}}
\newcommand{\MM}{\mathbf{M}}
\newcommand{\KK}{\mathbf{K}}

\newcommand{\CC}{\mathbf{C}}

\newcommand{\xx}{\mathbf{x}}
\newcommand{\yy}{\mathbf{y}}
\newcommand{\nn}{\mathbf{n}}

\renewcommand{\AA}{\mathbf{A}}
\renewcommand{\gg}{\mathbf{g}}
\newcommand{\hh}{\mathbf{h}}

\renewcommand{\vv}{\mathbf{v}}

\newcommand{\XX}{\mathbf{X}}

\newcommand{\RR}{\mathbf{R}}
\newcommand{\TT}{\mathbf{T}}
\newcommand{\NN}{\mathbf{N}}
\newcommand{\lam}{{\lambda}}
\newcommand{\PE}{\textrm{PE}}
\newcommand{\dt}{\Delta t}
\SetAlFnt{\small}
\SetAlCapFnt{\small}
\SetAlCapNameFnt{\small}
\SetAlCapHSkip{0pt}

\algnewcommand\algorithmicforeach{\textbf{for }}
\algdef{S}[FOR]{For}[1]{\algorithmicforeach\ #1\ \algorithmicdo}

\setcopyright{rightsretained}
\begin{document}
% Title portion
\title{Position-Based Nonlinear Gauss-Seidel for Quasistatic Hyperelasticity}

\author{Yizhou Chen}
\affiliation{%
  \institution{UCLA}
  \country{USA}}
 \affiliation{%
  \institution{Epic Games, Inc}
  \country{USA}}
  
  \author{Yushan Han}
\affiliation{%
  \institution{UCLA}
  \country{USA}}
 \affiliation{%
  \institution{Epic Games, Inc}
  \country{USA}}
  
  \author{Jingyu Chen}
\affiliation{%
  \institution{UCLA}
  \country{USA}}

\author{Joseph Teran}
\affiliation{%
  \institution{UC Davis}
  \country{USA}}
 \affiliation{%
  \institution{Epic Games, Inc}
  \country{USA}}

\begin{teaserfigure}
  \centering
  \includegraphics[draft=\mydraft,width=\textwidth,trim={0px 0px 0px 0px},clip]{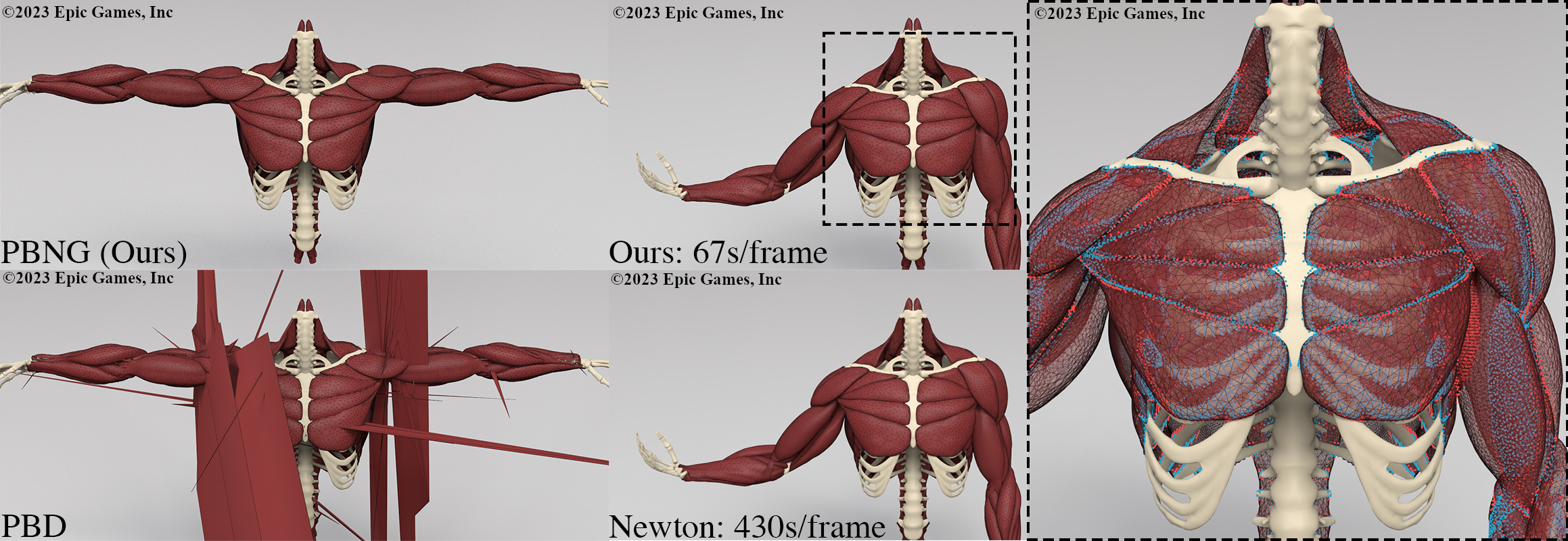}
  \caption{\textbf{Quasistatic Muscle Simulation with Collisions}. 
  Our method (PBNG) produces high-quality results visually comparable to Newton's method but with a 6x speedup.
  In this hyperelastic simulation of muscles, we use weak constraints to bind muscles together and resolve collisions.
  The rightmost image visualizes these constraints. 
  \textit{Red} indicates a vertex involved in a contact constraint. 
  \textit{Blue} indicates a vertex is bound with connective tissues.
  PBD (lower left) becomes unstable with this quasistatic example after a few iterations.}
  \label{fig:teaser}
\end{teaserfigure}

% DO NOT ENTER AUTHOR INFORMATION FOR ANONYMOUS TECHNICAL PAPER SUBMISSIONS TO SIGGRAPH 2019!
%\author{Gang Zhou}
%\orcid{1234-5678-9012-3456}
%\affiliation{%
%  \institution{College of William and Mary}
%  \streetaddress{104 Jamestown Rd}
%  \city{Williamsburg}
%  \state{VA}
%  \postcode{23185}
%\email{gang_zhou@wm.edu}
%\affiliation{%
%  \institution{Inria Paris-Rocquencourt}
%\email{beranger@inria.fr}
%\affiliation{%
% \institution{Rajiv Gandhi University}
%\email{aprna_patel@rguhs.ac.in}
%\affiliation{%
%  \institution{Tsinghua University}
%  \streetaddress{30 Shuangqing Rd}
%  \country{China}
%}
%\email{chan0345@tsinghua.edu.cn}
%\author{Ting Yan}
%\affiliation{%
%  \institution{Eaton Innovation Center}
%  \city{Prague}
%  \country{Czech Republic}}
%\email{yanting02@gmail.com}
%\author{Tian He}
%\affiliation{%
%  \institution{University of Virginia}
%  \department{School of Engineering}
%  \city{Charlottesville}
%  \state{VA}
%  \postcode{22903}
%  \country{USA}
%}
%\affiliation{%
%  \institution{University of Minnesota}
%\email{tinghe@uva.edu}
%\author{Chengdu Huang}
%\author{John A. Stankovic}
%\author{Tarek F. Abdelzaher}
%\affiliation{%
%  \institution{University of Virginia}
%  \department{School of Engineering}
%  \city{Charlottesville}
%  \state{VA}
%  \postcode{22903}
%  \country{USA}
%}

%\renewcommand\shortauthors{Zhou, G. et al}

\begin{abstract}
Position based dynamics \cite{muller:2007:pbd} is a powerful technique for simulating a variety of materials.
Its primary strength is its robustness when run with limited computational budget.
We develop a novel approach to address problems with PBD for quasistatic hyperelastic materials.
Even though PBD is based on the projection of static constraints, PBD is best suited for dynamic simulations.
This is particularly relevant since the efficient creation of large data sets of plausible, but not necessarily accurate elastic equilibria is of increasing importance with the emergence of quasistatic neural networks \cite{jin:2022:mls,luo:2020:ml,bailey:2018:fdd}.
Furthermore, PBD projects one constraint at a time.
We show that ignoring the effects of neighboring constraints limits its convergence and stability properties.
Recent works \cite{macklin:2016:xpbd} have shown that PBD can be related to the Gauss-Seidel approximation of a Lagrange multiplier formulation of backward Euler time stepping, where each constraint is solved/projected independently of the others in an iterative fashion.
We show that a position-based, rather than constraint-based nonlinear Gauss-Seidel approach solves these problems.
Our approach retains the essential PBD feature of stable behavior with constrained computational budgets, but also allows for convergent behavior with expanded budgets.
We demonstrate the efficacy of our method on a variety of representative hyperelastic problems and show that both successive over relaxation (SOR) and Chebyshev acceleration can be easily applied.
\end{abstract}


%\ccsdesc[500]{Computing methodologies}
%\ccsdesc[300]{Realtime Simulation}
%\ccsdesc[100]{Interactive Simulation}



\maketitle

\section{Introduction}
%Ever since the pioneering work of Terzopoulos et al. \shortcite{terzopoulos:1987:edm}, numerical simulation of large deformation in elastic objects has been an effective means of producing squash and stretch animation of deformable components of virtual environments.Many elasticity models and methods for their discretization have been proposed over the past decades.
In the present work, we consider large strain hyperelastic solids \cite{bonet:2008:continuum} whose governing equations are discretized in space with the finite element method (FEM) \cite{sifakis:2012:course}.
Hyperelastic solid models define continuum stresses from a notion of elastic potential energy.
In graphics applications, these models are commonly used for simulation-based enhancement of character flesh and musculature animation  \cite{mcadams:2011:mge,wang:2020:acs,smith:2018:stable,fan:2014:avm,modi:2021:emu,teran:2005:muscle}.
Wang et al. \shortcite{wang:2020:acs} provide a thorough discussion of the state-of-art.
Our primary focus is quasistatic problems where the inertia terms are negligible.
This is the most difficult case for implicit time stepping with elasticity problems since the non-convex terms in the force balance are dominant \cite{teran:2005:robust,bonet:2008:continuum,zhu:2018:qn,rabinovich:2017:injective,sorkine:2007:as-rigid-as-possible}.\\
\begin{figure*}[!h]
	\includegraphics[draft=\mydraft,width=\textwidth,trim={0 0 0 0},clip]{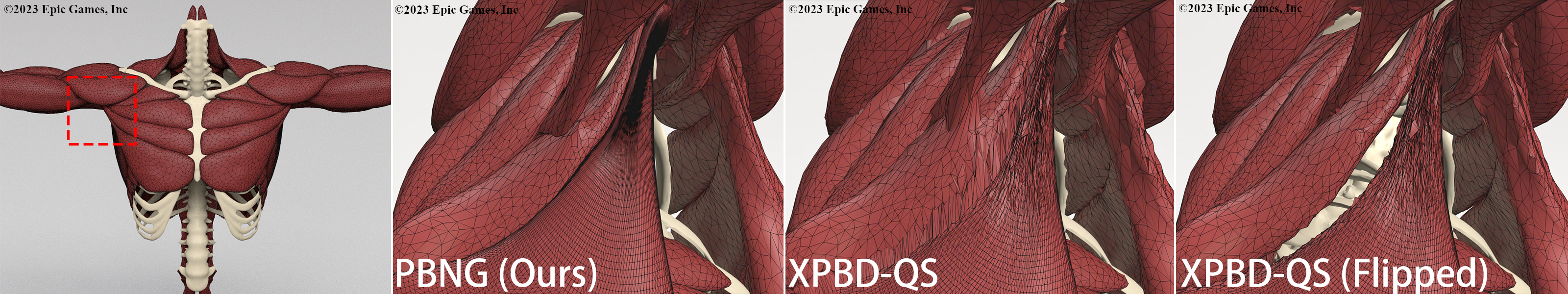}
	\caption{\textbf{PBNG vs XPBD}. Muscle simulation demonstrates iteration-order-dependent behavior with XPBD and quasistatics. A zoom-in view under the right armpit region is provided. Each method is run 130 iterations. PBNG converges to the desired solution, binding the muscles closely together. XPBD-QS and XPBD-QS (Flipped) fail to converge, leaving either artifacts or gaps between the muscles.}
	\label{fig:pbgn_vs_xpbd_muscle}
\end{figure*}\\
Quasistatic solvers are becoming increasingly important due to their use in generating training data for neural networks (or QNNS: quasistatic neural networks).
For example, various authors have shown that QNNs can be effectively trained for elastic materials in cloth and skinning applications \cite{jin:2022:mls,bertiche:2021:pbns,geng:2020:ml,jin:2020:ml,luo:2020:ml,bailey:2018:fdd}.
These networks engender real-time performance at resolutions orders of magnitude above what is achievable with any existing simulation techniques on modern hardware.
However, QNNs require tens of thousands of high-resolution equilibria for training data sets.
While the creation of these data sets is an ``off-line" process, it is still desirable to create the data set with a minimal amount of user interaction and computation time.
For example, techniques that require parameter tuning to prevent unstable/visually implausible behavior are undesirable since
the data set creation process will take longer when the user must monitor simulation data closely and frequently manually intervene.
Furthermore, extremely accurate solutions to the governing equations are not necessary since the network need only approximate visually convincing behaviors.
Therefore, simulation techniques that both generate visually plausible behavior in a minimal amount of computation with minimal user interaction/parameter tuning are ideal.\\
\\
While many methods exist for solving the FEM-discretized implicit equations of motion for hyperelastic solids (see Zhu et al. \shortcite{zhu:2018:qn} and Li et al. \shortcite{li:2019:dot} for thorough summaries), the Position Based Dynamics (PBD) approach of M{\"u}ller et al. \shortcite{muller:2007:pbd} is a natural candidate for generating training data for QNNs.
It has remarkable robustness and stability properties and can produce visually plausible results with minimal computational budgets.
However, constitutive control over PBD behavior is challenging as effective material stiffnesses etc. vary with iteration count and time step size.
The Extended Position Based Dynamics (XPBD) approach of Macklin et al. \shortcite{macklin:2016:xpbd} addressed these issues by reformulating the original PBD approach in terms of a Gauss-Seidel technique for discretizing a total Lagrange multiplier formulation of the backward Euler system for implicit time stepping.
This formulation has similarities to PBD, but with the elastic terms handled properly where PBD can be seen as the extreme case of infinite elastic modulus.
\\
\\
Despite its many strengths, PBD/XPBD has a few limitations that hinder its use in quasistatic applications.
First, XPBD is designed for backward Euler and omitting the inertial terms for quasistatics is not possible (it would require dividing by zero).
However, we show that PBD when viewed as the limit of infinite stiffness in XPBD (as detailed in Macklin et al. \shortcite{macklin:2016:xpbd}) is an approximation to the quasistatic equations.
Unfortunately, this limit incorrectly and irrevocably removes the external forcing terms. 
Second, PBD/XPBD can only discretize hyperelastic models that are quadratic in some notion of strain constraint \cite{macklin:2016:xpbd,Macklin:2021:neohookean_xpbd}.
This prevents the adoption of many models from the computational mechanics literature.
Lastly, the constraint-centric iteration in PBD/XPBD causes artifacts near vertices that appear in different types of constraints.
This effect is particularly severe in quasistatic problems.
PBD is designed to project/solve the positions involved in a single constraint at a time, ignoring the effects of adjacent constraints.
While Macklin et al. \shortcite{macklin:2016:xpbd} show this omission is justifiable via their Gauss-Seidel/Lagrange multiplier formulation of the backward Euler equations, 
we show that projecting positions based on a single constraint at a time significantly degrades convergence in both visual and numerical terms.%Indeed, the XPBD position update cannot resolve the relative magnitude of constraint stiffnesses in the infinite/PBD limit.
\\
\\
We present a position-based (rather than constraint-based) nonlinear Gauss-Seidel method that resolves the key issues with PBD/XPBD and hyperelastic quasistatic time stepping.
In our approach, we iteratively adjust the position of each simulation node to minimize the potential energy (with all other coupled nodes fixed) in a Gauss-Seidel fashion.
This makes each position update aware of all constraints that a node participates in and removes the artifacts of PBD/XPBD that arise from processing constraints separately.
Our approach maintains the essential efficiency and robustness features of PBD and has an accuracy that rivals Newton's method for the first few orders of magnitude in residual reduction.
Furthermore, unlike Newton's method, our approach is stable when the computational budget is extremely limited.
\\
\\
The minimization involved in the position update of each node amounts to a nonlinear system of equations (3 equations in 3D and 2 in 2D).
We approximate the solution with Newton's method.
The linearization of hyperelastic terms can have symmetric indefinite matrices.
We develop an inexpensive yet effective technique for projecting any isotropic potential energy density Hessian to a symmetric positive definite counterpart as in \cite{teran:2005:robust}.
However, unlike the definiteness projections in \cite{teran:2005:robust} and \cite{smith:2019:analytic}, it does not require the singular value decomposition of the deformation gradient.
Furthermore, unlike the definiteness projection in \cite{teran:2005:robust}, it does not require the solution of $3\times3$ or $2\times2$ symmetric eigensystems.
As with PBD and other Gauss-Seidel approaches, a degree of freedom coloring technique is needed for efficient parallel performance.
We provide a simple approach for this coloring and show that the position-based view tends to have far fewer colors than the constraint-based view in PBD and that this improves scalability and performance.\\
\\
Although our technique is designed for quasistatics, it is easily applicable to backward Euler discretizations of problems with inertia if we minimize the incremental potential \cite{martin:2011:ebem,gast:2015:tvcg,liu:2013:fsm,stern:2006:int,bouaziz:2014:pdf,narain:2016:admm} rather than the potential energy.
In this way, it can be compared directly with XPBD for backward Euler.
However, while our method converges to the exact solution, XPBD does not in practice since it simplifies the Lagrange multiplier system by omitting the residual of the position equations.
This omission is perfectly accurate in the first iteration, but as Macklin et al. \shortcite{macklin:2016:xpbd} point out, less so in latter iterations when constraint gradients vary significantly.
We observe that this rapid variation occurs for many hyperelastic formulations and that its omission degrades residual reduction.
However, the inclusion of this term introduces instabilities into XPBD.
Our approach does not have these limitations and is convergent even with the addition of the inertia term.\\
\\
We summarize our contributions as:
\begin{itemize}
\item A position-based, rather than constraint-based, nonlinear Gauss-Seidel technique for hyperelastic implicit time stepping.
\item A hyperelastic energy density Hessian projection to efficiently guarantee definiteness of linearized equations that does not require a singular value decomposition or symmetric eigen solves.
\item A node coloring technique that allows for efficient parallel performance of our Gauss-Seidel updates.
\end{itemize}

\begin{figure}[h]
	\includegraphics[draft=\mydraft,width=0.49\columnwidth,trim={120px 20px 240px 0px},clip]{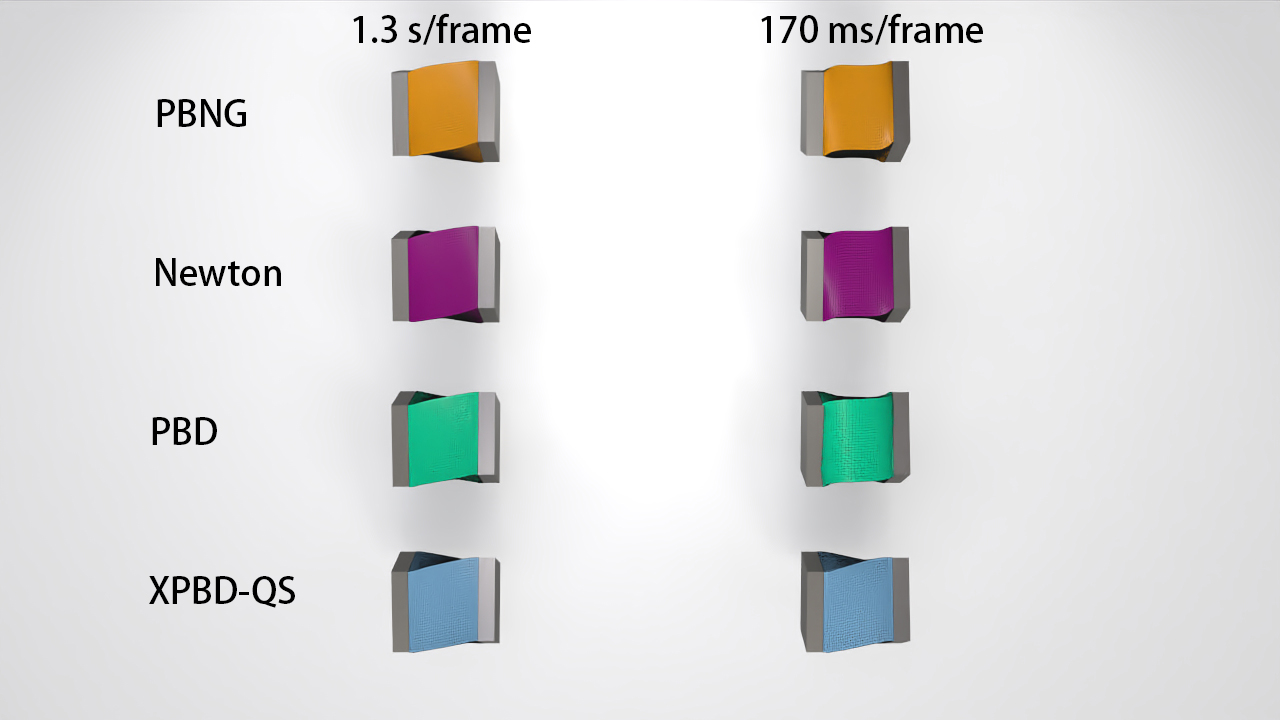}
	\includegraphics[draft=\mydraft,width=0.49\columnwidth,trim={120px 20px 240px 0px},clip]{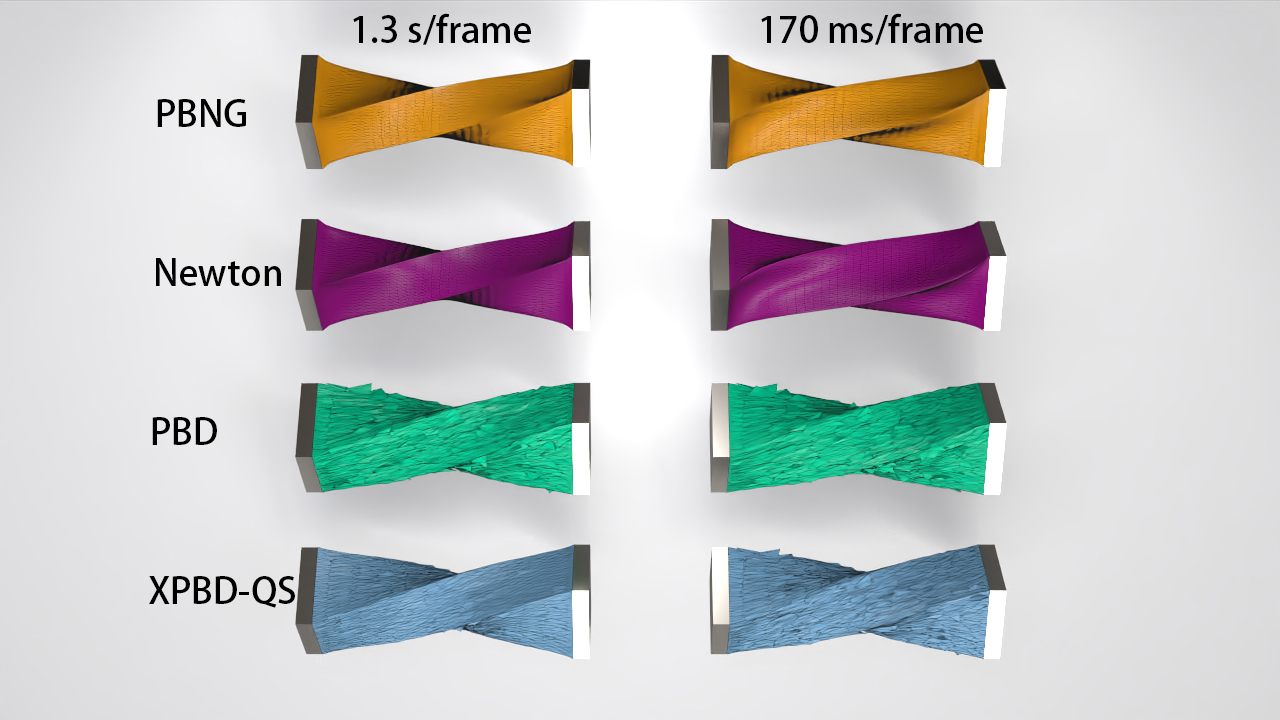}
	\includegraphics[draft=\mydraft,width=0.49\columnwidth,trim={120px 20px 240px 0px},clip]{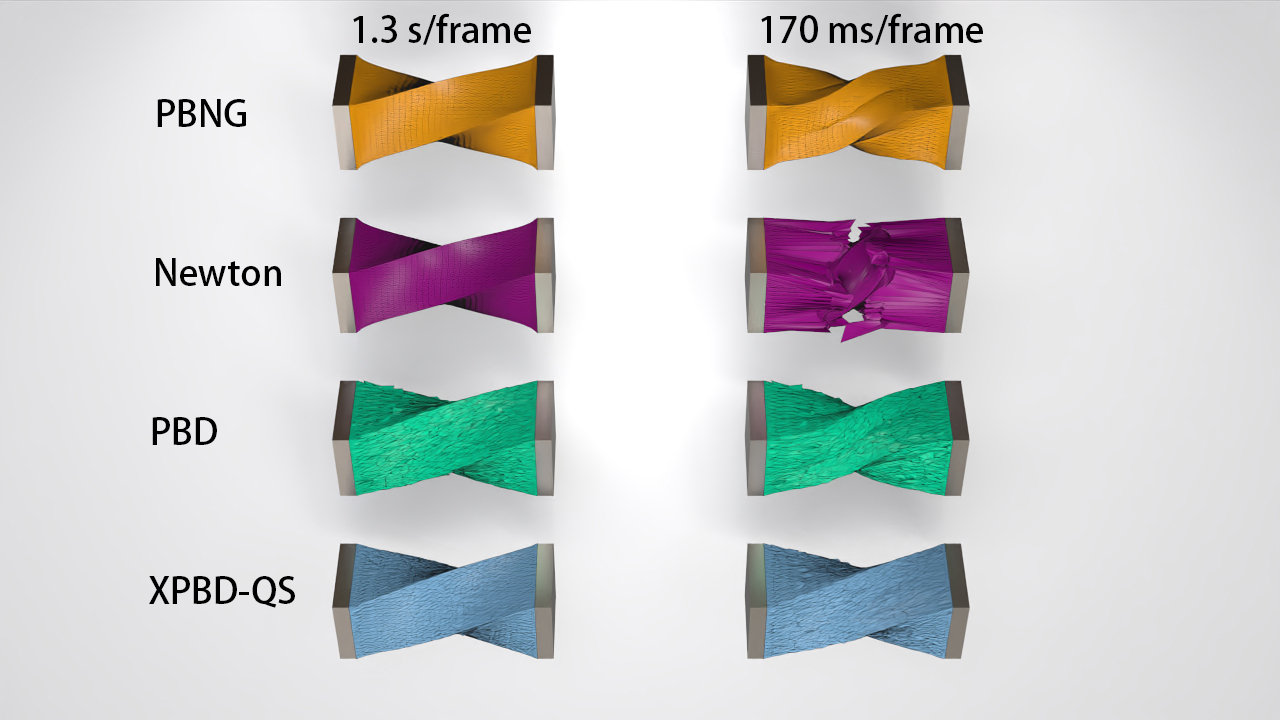}
	\includegraphics[draft=\mydraft,width=0.49\columnwidth,trim={120px 20px 240px 0px},clip]{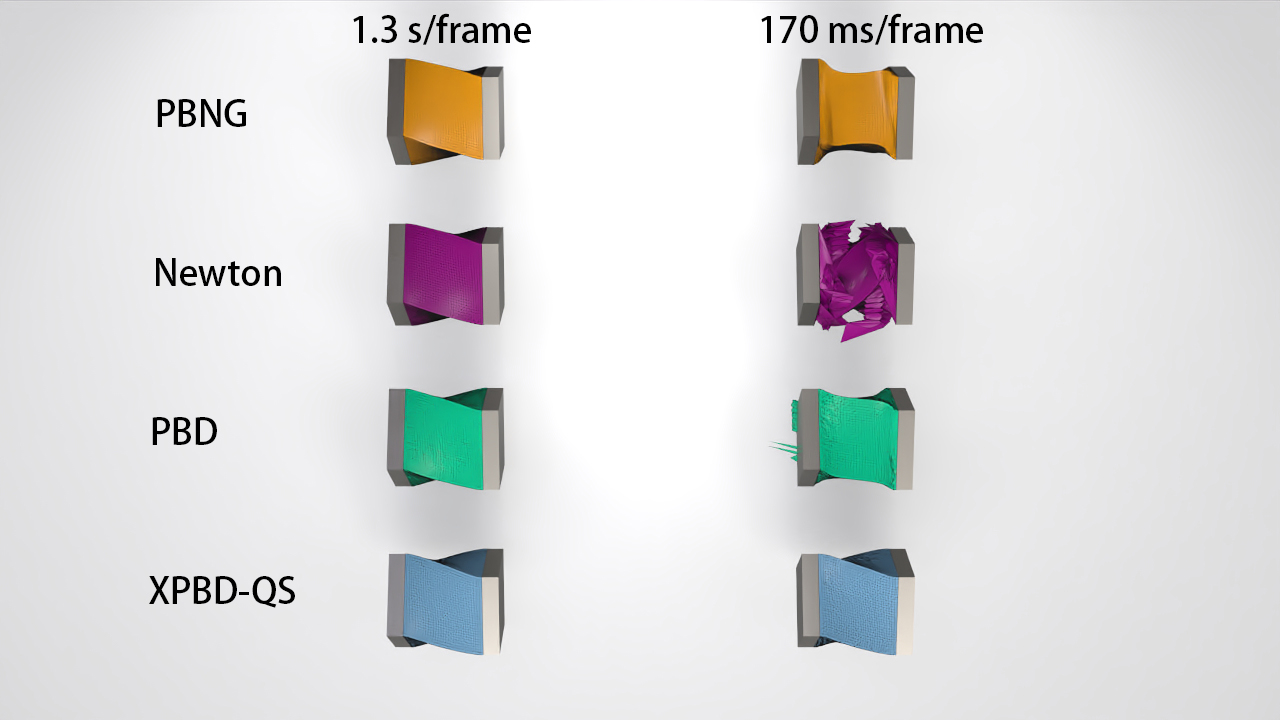}
	\includegraphics[draft=\mydraft,width=0.49\columnwidth,trim={10px 0px 40px 10px},clip]{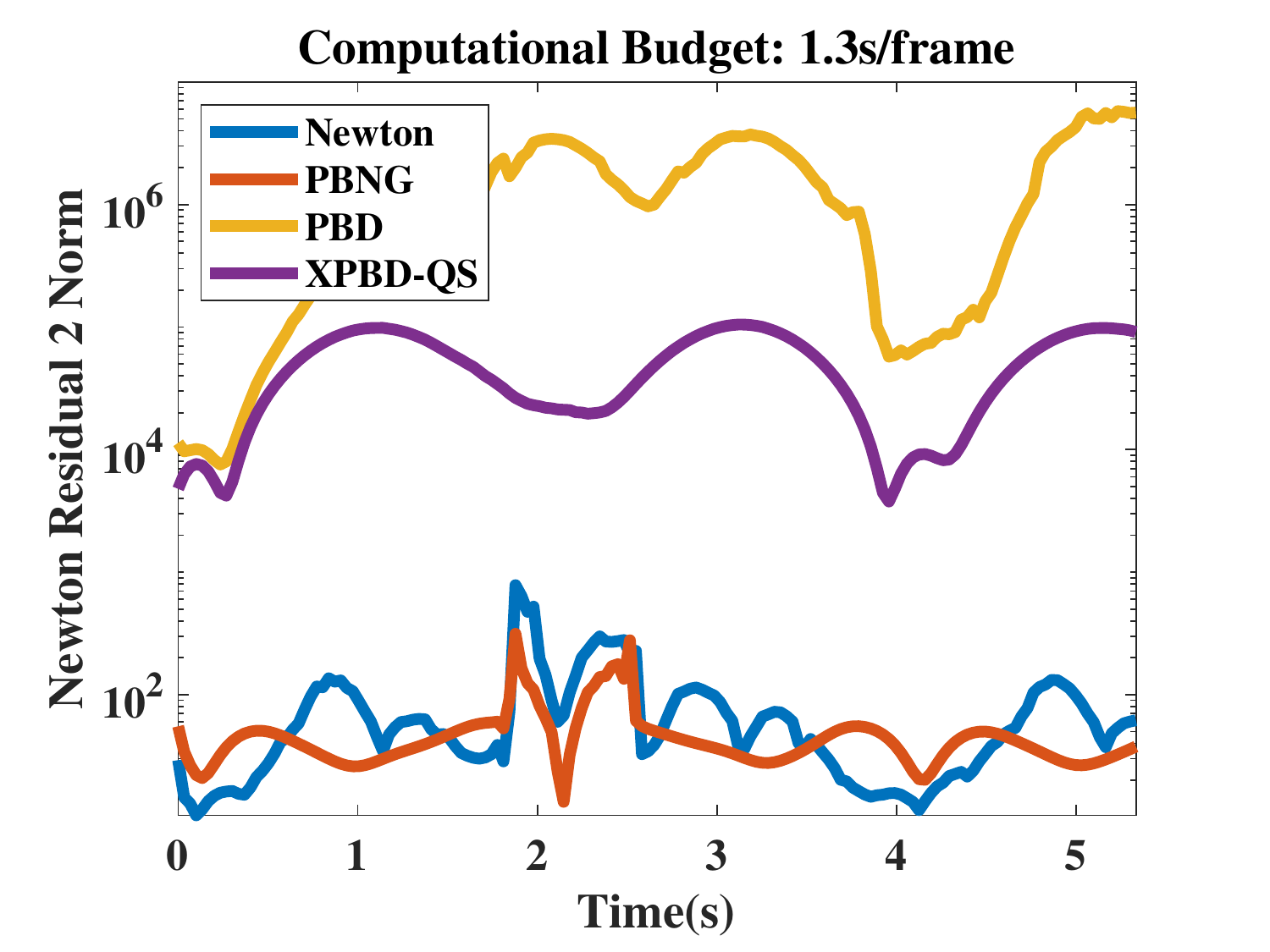}
	\includegraphics[draft=\mydraft,width=0.49\columnwidth,trim={10px 0px 40px 10px},clip]{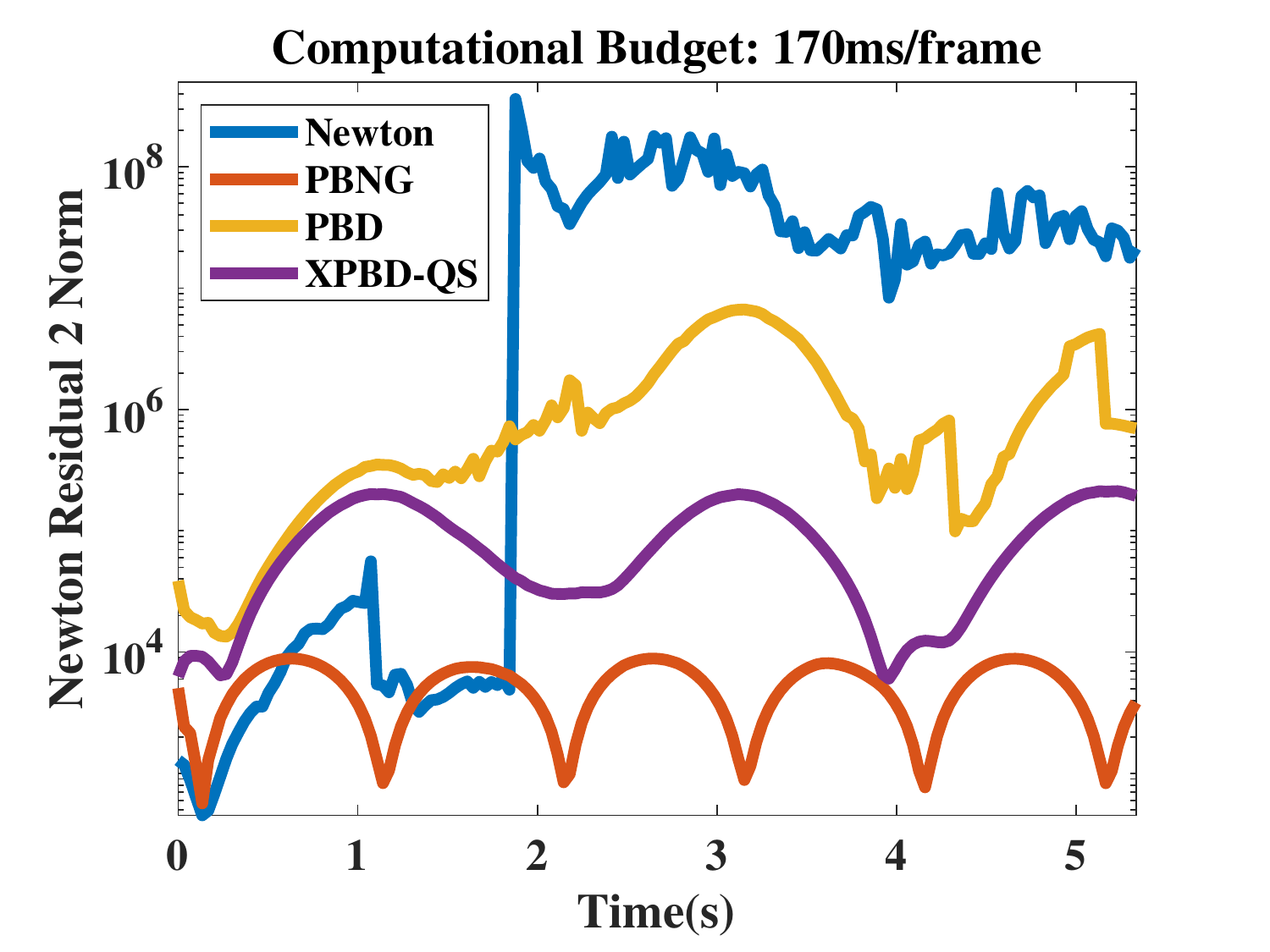}
	\caption{\textbf{Comparisons with Different Computational Budget}. A block is stretched/compressed while being twisted.
	With a sufficiently large computational budget, Newton's method is stable, but it becomes unstable when the computational budget is small.
	PBD and XPBD-QS do not significantly reduce the residual in the given computational time, resulting in noisy artifacts on the mesh.
	PBNG maintains relatively small residuals and generates visually plausible results of the deformable block even if the budget is limited.}
	\label{fig:method_comparison}
\end{figure}

\section{Previous work}
Baraff and Witkin first demonstrated that implicit time stepping with elasticity is essential for efficiency \cite{baraff:1998:cloth}.
Many approaches characterize implicit time stepping with hyperelasticity as a minimization of an incremental potential \cite{martin:2011:ebem,gast:2015:tvcg,liu:2013:fsm,stern:2006:int,bouaziz:2014:pdf,narain:2016:admm}. 
This is often referred to as variational implicit Euler \cite{stern:2006:int,martin:2011:ebem} or optimization implicit Euler \cite{liu:2013:fsm}.
Quasistatic time stepping is an extreme case where inertia terms are ignored and only the strain energy is minimized \cite{teran:2005:robust,sorkine:2007:as-rigid-as-possible,liu:2008:arap,rabinovich:2017:injective,kovalsky:2016:aqp}.
Minimizers are usually found by setting the gradient of the energy to zero and solving the associated nonlinear system of equations with Newton's method.
While Newton's method \cite{nocedal:2006:cg} generally requires the fewest iterations to reach a desired tolerance (often achieving quadratic convergence), each iteration can be costly and a line search is typically required for stability \cite{gast:2015:tvcg}.
There are many techniques that are less costly than Newton, but that can only reduce the system residual by a few orders of magnitude.
However, many are satisfactory for visual accuracy. See discussion in Liu et al. \shortcite{liu:2013:fsm}, Bouaziz et al. \cite{bouaziz:2014:pdf} and Zhu et al. \shortcite{zhu:2018:qn}.\\
\\
Hyperelastic potentials must be rotationally invariant, non-negative and have global minima equal to zero at rotations.
These considerations make the energy minimization non-convex with potentially non-unique solutions in quasistatic problems \cite{bonet:2008:continuum}.
The non-convexity yields indefinite energy Hessians that can prevent convergence.
Quasi-Newton methods can be used to approximate the Hessian with a symmetric semi-definite counterpart \cite{teran:2005:robust,nocedal:2006:cg,zhu:2018:qn,smith:2019:analytic,li:2019:dot}.
Many methods avoid the indefiniteness issue with the inclusion of auxiliary (or secondary) variables.
Narain et al. \shortcite{narain:2016:admm}, Bouaziz et al. \cite{bouaziz:2014:pdf}, Liu et al. \shortcite{liu:2013:fsm} are recent examples of this, but similar approaches have been used in graphics since the local/global approach with ARAP by Sorkine et al. \shortcite{sorkine:2007:as-rigid-as-possible}.
Rabinovich et al. \shortcite{rabinovich:2017:injective} generalize this approach to a wider range of distortion energies.\\
\\
Methods like the Alternating Direction Method of Multipliers (ADMM) \cite{boyd:2011:distributed,narain:2016:admm}, the limited-memory Broyden-Fletcher-Goldfarb-Shanno algorithm (L-BFGS) \cite{bertsekas:1997:nonlinear,zhu:2018:qn,liu:2017:lbfgs,witemeyer:2021:qlb} and Sobolev preconditioned gradient descent (SGD) \cite{neuberger:1985:sgd,bouaziz:2014:pdf,liu:2013:fsm,sorkine:2007:as-rigid-as-possible} require the inversion of a constant discrete elliptic operator (component wise-Laplacian) which can be pre-factored for efficiency.
While this discrete operator does not suffer from indefiniteness issues, various authors note that SGD approaches may converge initially faster than Newton but will often taper off \cite{zhu:2018:qn,liu:2013:fsm,bouaziz:2014:pdf,wang:2015:cheby}.
Zhu et al. \shortcite{zhu:2018:qn} tailor their approach to this observation and use SGD initially and then combine with L-BFGS to incorporate more second-order information.
Liu et al. \shortcite{liu:2017:lbfgs} and Witemeyer et al. \shortcite{witemeyer:2021:qlb} also use combinations of SGD and L-BFGS.
Kovalsky et al. \cite{kovalsky:2016:aqp} add Nesterov acceleration to SGD. 
Hecht et al. \shortcite{hecht:2012:chol} develop efficient updates for a pre-factored Hessian with corotated materials.
Wang \shortcite{wang:2015:cheby} discusses the challenges of using direct solution/pre-factoring in the SGD-style approaches of Narain et al. \shortcite{narain:2016:admm}, Bouaziz et al. \cite{bouaziz:2014:pdf} and Liu et al. \shortcite{liu:2013:fsm} and develops a Chebyshev acceleration technique as an alternative.
In particular, they show that pre-factored discrete elliptic operators are memory-intensive (particularly at high-resolution) and limited since forward and backward substitutions do not parallelize. 
Moreover, \cite{wang:2015:cheby} show that simply replacing the direct solver with an iterative solver with reduced iteration count can lead to visually implausible or even unstable behaviors.
However, Fratarcangeli et al. \shortcite{fratarcangeli:2016:vivace} show that Gauss-Seidel iteration does not suffer from the same limitations in this context, although it does require degree of freedom coloring to facilitate parallel computation.
\\
\\
Tournier et al. \shortcite{tournier:2015:stable_dynamics} develop a technique for bridging elasticity and constraint-based approaches that is robust to large stiffness.
They use a similar primal/dual setup to XPBD. However, unlike XPBD, their approach solves the entire system at once, rather than iterating over individual constraints.
Wang and Yang \shortcite{Wang:2016:des_gpu} use a Chebyshev accelerated gradient descent approach for general hyperelasticity and FEM.
\begin{figure}[h]
	\begin{subfigure}{0.33\columnwidth}
		\includegraphics[draft=\mydraft,width=\linewidth,trim={70px 50px 0px 0px},clip]{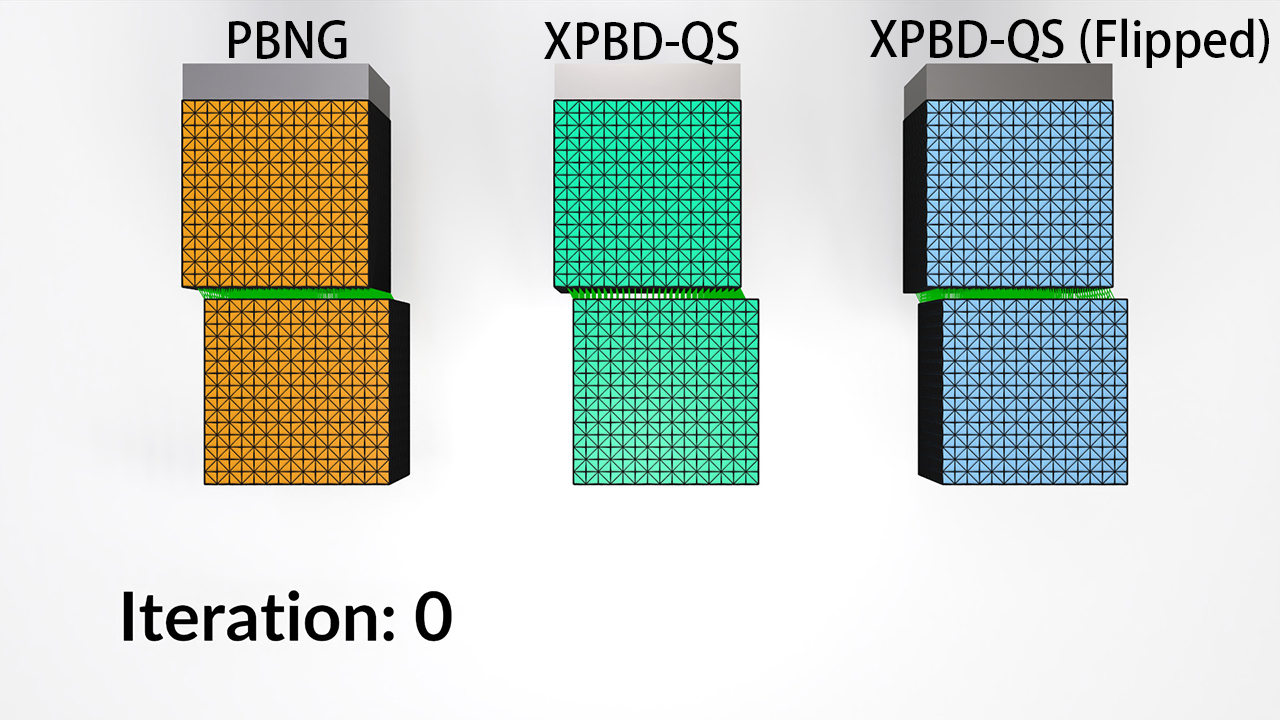}
		\includegraphics[draft=\mydraft,width=\linewidth,trim={70px 50px 0px 0px},clip]{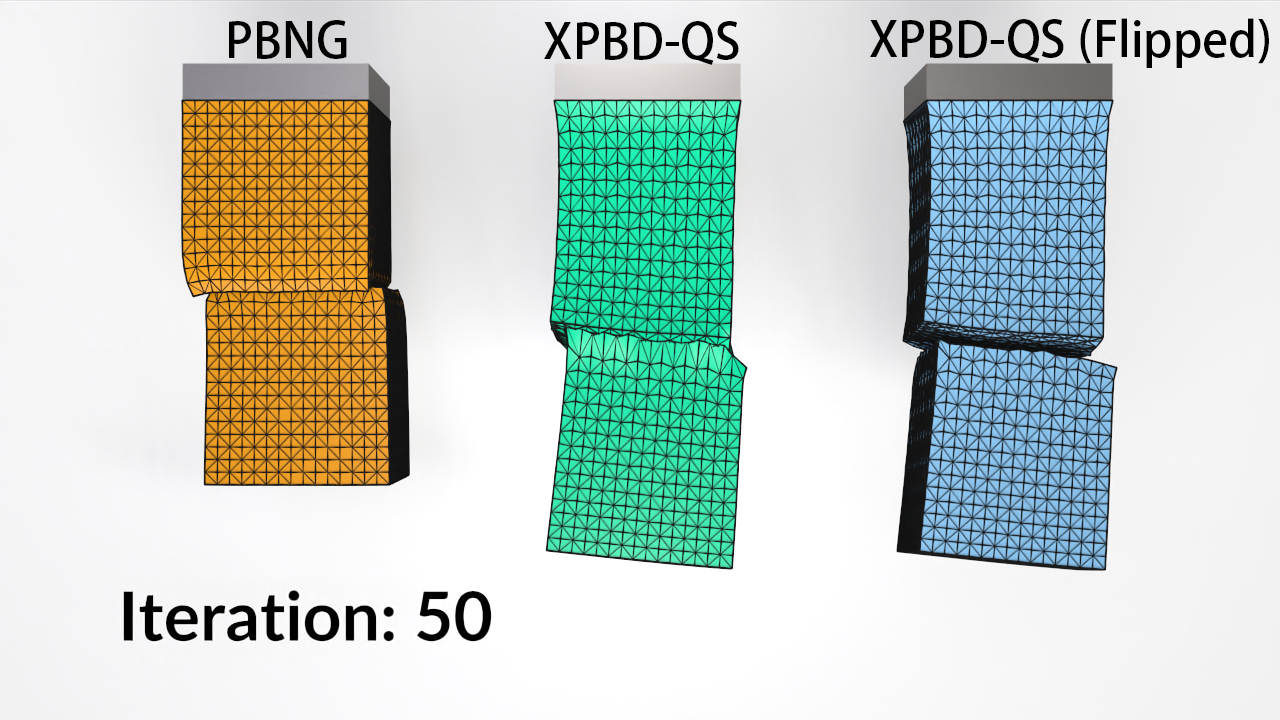}
		\includegraphics[draft=\mydraft,width=\linewidth,trim={70px 50px 0px 0px},clip]{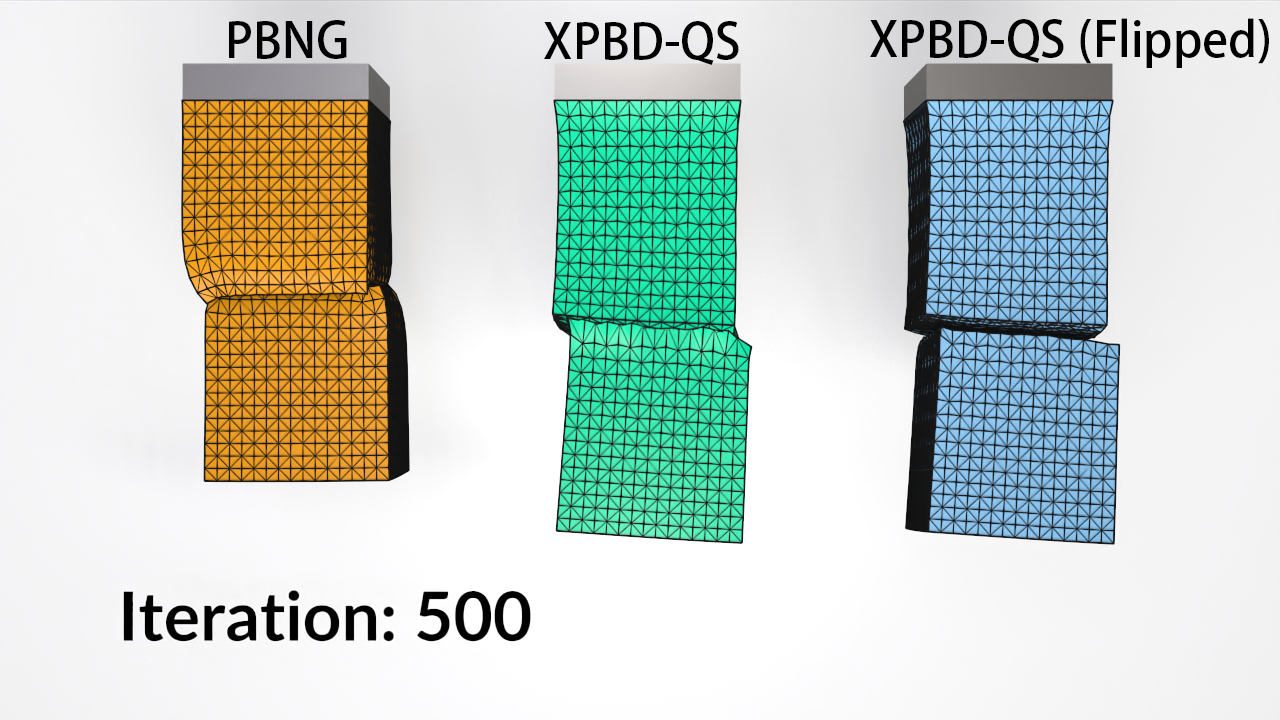}
	\end{subfigure}
	\begin{subfigure}{0.65\columnwidth}
		\includegraphics[width = \linewidth]{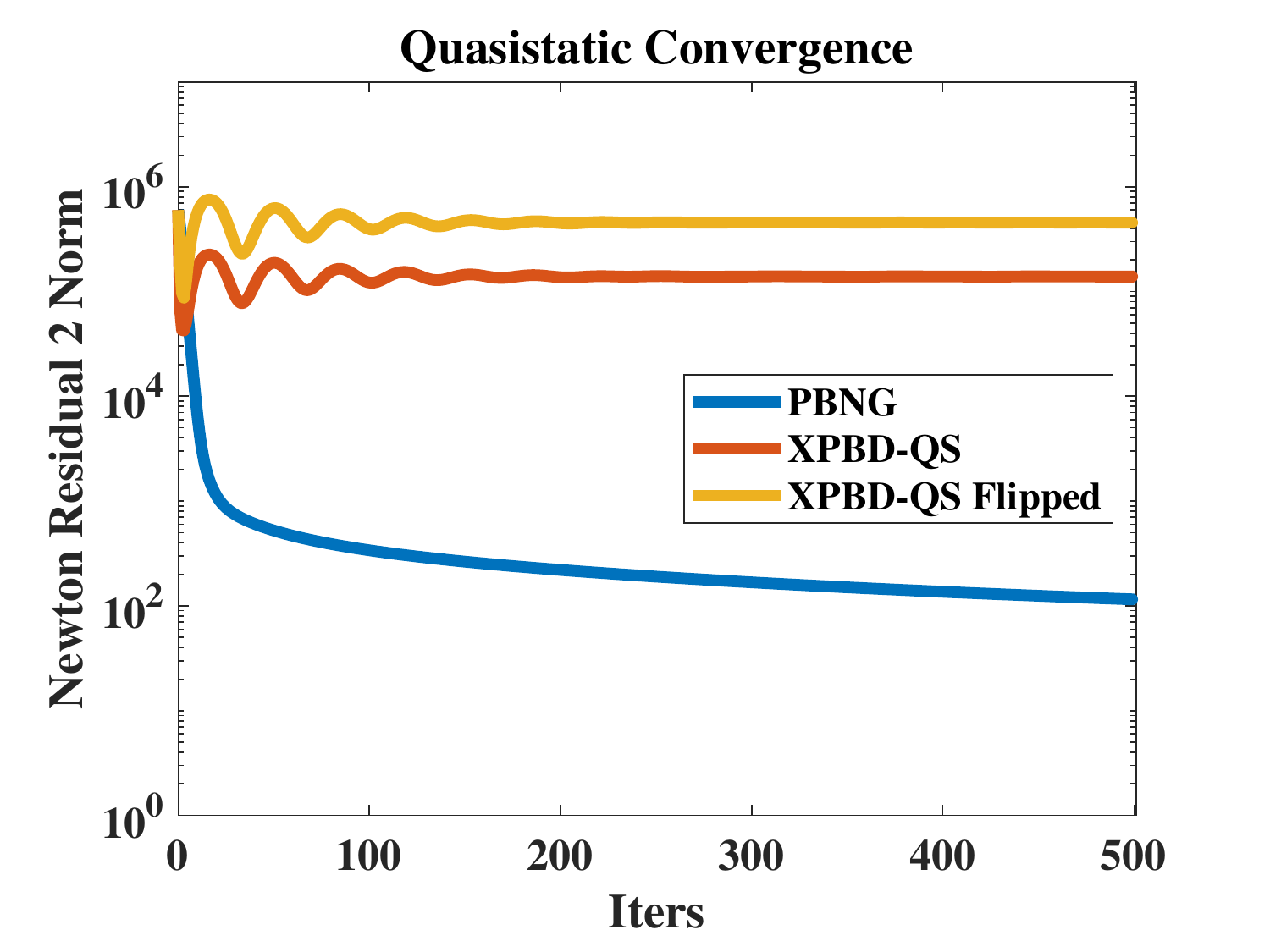}
	\end{subfigure}
	
	\caption{
		\textbf{Two Blocks Hanging}. Two identical blocks are bound together through weak constraints. 
		Green line segments in iteration 0 indciate weak constraint springs.
		PBNG is able to reduce the residual by a few orders of magnitude and converges quickly.
		XPBD-QS methods demonstrate iteration-order-dependent behavior. Residuals oscillate and produce visually incorrect results.
	}
	\label{fig:two_blocks_hanging}
\end{figure}
\section{Equations}
We consider continuum mechanics conceptions of the governing physics where a flow map $\fm:\Omega^0\times[0,T]\rightarrow\mathbb{R}^d$, $d=2$ or $d=3$, describes the motion of the material.
Here the time $t\in[0,T]$ location of the particle $\XX\in\Omega^0\subset\mathbb{R}^d$ is given by $\fm(\XX,t)\in\Omega^t\subset\mathbb{R}^d$ where $\Omega^0$ and $\Omega^t$ are the initial and time $t$ configurations of material respectively.
The flow map $\fm$ obeys the partial differential equation associated with momentum balance
\begin{align}
R^0\frac{\partial^2 \fm}{\partial t^2}=\nabla^\XX \cdot \PP + \ff^\textrm{ext} \label{eq:mom}
\end{align}
where $R^0$ is the initial mass density of the material, $\PP$ is the first Piola-Kirchhoff stress and $\ff^\textrm{ext} $ is external force density.
This is also subject to boundary conditions 
\begin{align}
\fm(\XX,t)&=\xx_D(\XX,t), \ \XX\in\partial\Omega^0_D\label{eq:bC_c}\\
\PP(\XX,t)\NN(\XX,t)&=\TT_N(\XX,t), \ \XX\in\partial\Omega^0_N\label{eq:bc_n}
\end{align}
where $\partial\Omega^0$ is split into Dirichlet ($\partial\Omega^0_D$) and Neumann ($\partial\Omega^0_N$) regions where the deformation and applied traction respectively are specified.
Here $\TT_N$ denotes externally applied traction boundary conditions.
For hyperelastic materials, the first Piola-Kirchhoff stress is related to a notion of potential energy density $\Psi:\mathbb{R}^{d\times d}\rightarrow\mathbb{R}$ as 
\begin{align}
\PP(\XX,t)=\frac{\partial \Psi}{\partial \FF}(\frac{\partial \fm}{\partial \XX}(\XX,t)), \ \PE(\fm(\cdot,t))=\int_{\Omega^0} \Psi(\frac{\partial \fm}{\partial \XX})d\XX
\end{align}
where $\PE(\fm(\cdot,t))$ is the potential energy of the material when it is in the configuration defined by the flow map at time $t$.
Note that we will typically use $\FF=\frac{\partial \fm}{\partial \XX}$ to denote the spatial derivative of the flow map (or deformation gradient).
We refer the reader to \cite{gonzalez:2008:continuum,bonet:2008:continuum} for more continuum mechanics detail.
\\
\\
In quasistatic problems, the inertial terms in the momentum balance (Equation~\eqref{eq:mom}) can be neglected and the material motion is defined by a sequence of equilibrium problems
\begin{align}
\mathbf{0}=\nabla^\XX \cdot \PP + \ff^\textrm{ext} \label{eq:qs}
\end{align}
subject to the boundary conditions in Equations~\eqref{eq:bC_c}-\eqref{eq:bc_n}.
This is equivalent to the minimization problems
\begin{align}
\fm(\cdot,t)=
\begin{array}{c}
\textrm{argmin}\\
\trial\in\mathcal{W}^t
\end{array}
\PE(\trial) - \int_{\Omega_0} {\ff^\textrm{ext}} \cdot \trial d\XX - \int_{\partial\Omega^0_N} \TT_N \cdot \trial ds(\XX)
\end{align}
where $\mathcal{W}^t=\left\{\trial:\Omega_0\rightarrow\mathbb{R}^d\left|\trial(\XX)=\xx_D(\XX,t), \ \XX\in\partial\Omega^0_D \right.\right\}$.
\subsection{Constitutive Models}
We demonstrate our approach with a number of different hyperelastic potentials commonly used in computer graphics applications.
The ``corotated" or ``warped stiffness" model \cite{muller:2002:stable,etzmuss:2003:cont,muller:2004:ivm,schmedding:2008:inversion,chao:2010:geom} has been used for many years with a few variations.
We use the version with the fix to the volume term developed by Stomakhin et al. \shortcite{stomakhin:2012:invertible}
\begin{align}
\Psi^{\textrm{cor}}(\FF)=\mu|\FF-\RR(\FF)|_F^2 + \frac{\lambda}{2}(\det(\FF)-1)^2\label{eq:cor}.
\end{align}
Here $\FF=\RR(\FF)\SS(\FF)$ is the polar decomposition of $\FF$.
Neo-Hookean models \cite{bonet:2008:continuum} have also been used since they do not require polar decomposition and since recently some have been shown to have favorable behavior with nearly incompressible materials \cite{smith:2018:stable}.
We use the Macklin and M\"ueller \shortcite{Macklin:2021:neohookean_xpbd} formulation due to its simplicity and natural use with XPBD
\begin{align}
\Psi^{\textrm{nh}}(\FF)=\frac{1}{2}\mu|\FF|_F^2 + \frac{\hat{\lambda}}{2}(\det(\FF)-1-\frac{\mu}{\hat{\lambda}})^2\label{eq:nh}.
\end{align}
Here $\hat{\lambda}=\mu+\lambda$.
$\lambda$ and $\mu$ are the Lam\'e parameters and are related to the Young's modulus ($E$) and Poisson's ratio ($\nu$) as
\begin{align}
\mu&=\frac{E}{2(1+\nu)}, \ \lambda=\frac{E\nu}{(1+\nu)(1 -2\nu)}\label{eq:lame}.
\end{align}
Note that we distinguish between the $\hat{\lambda}$ used in Macklin and M\"ueller \shortcite{Macklin:2021:neohookean_xpbd} and the Lam\'e parameter $\lambda$, we discuss the reason for this in more detail in Section~\ref{sec:lame}.
We also support the stable Neo-Hookean model proposed in \cite{smith:2018:stable}
\begin{align}
	\Psi^{\textrm{snh}}(\FF)= \frac{1}{2}\mu(|\FF|_F^2 - d) + \frac{1}{2}(\det(\FF)-1-\frac{3\mu}{4\lambda})^2 - \frac{1}{2}\mu\log(1 + |\FF|_F^2)\label{eq:snh}.
\end{align}
\begin{figure}[h]
	\includegraphics[draft=\mydraft,width=0.32\columnwidth,trim={70px 0px 340px 0px},clip]{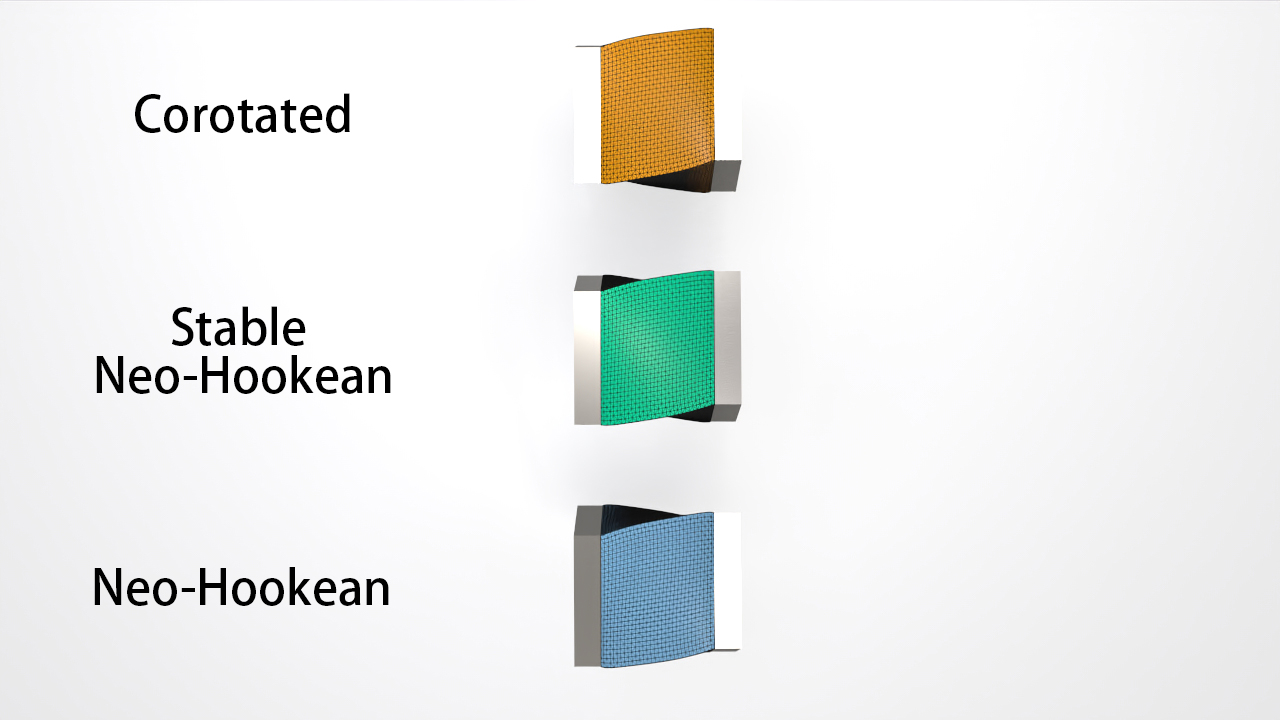}
	\includegraphics[draft=\mydraft,width=0.32\columnwidth,trim={70px 0px 340px 0px},clip]{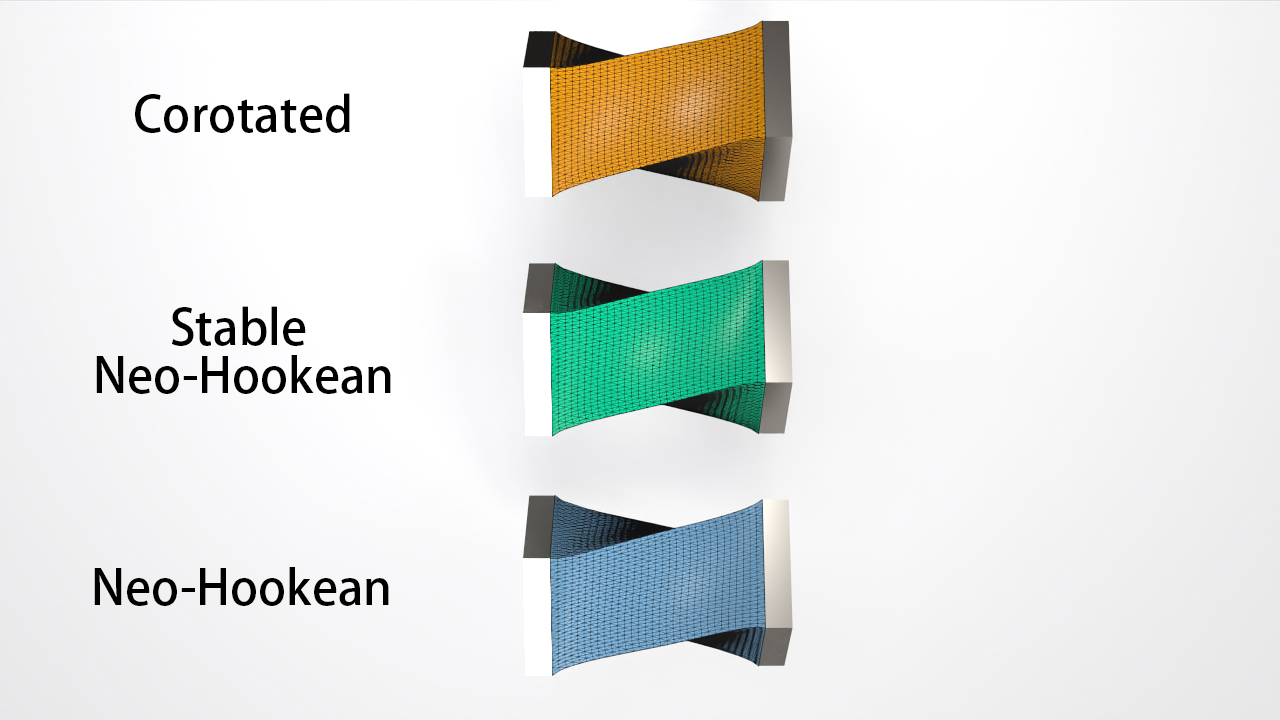}
	\includegraphics[draft=\mydraft,width=0.32\columnwidth,trim={70px 0px 340px 0px},clip]{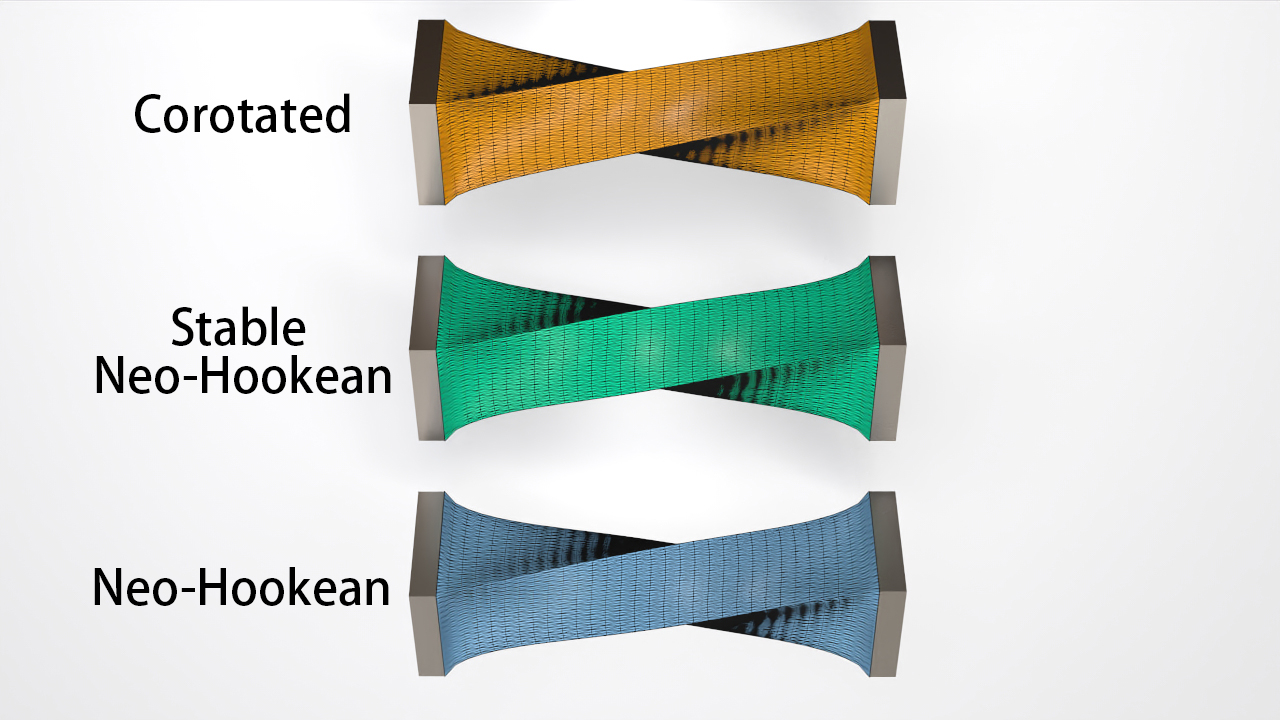}
	\caption{\textbf{Different Constitutive Models}. PBNG works with various constitutive models. We showcase the corotated, Neo-Hookean, and stable Neo-Hoookean models through a block twisting and stretching example.}
	\label{fig:twist_different_models}
\end{figure}

\section{Discretization}
We use the FEM discretization of the quasistatic problem in Equation~\eqref{eq:qs}
\begin{align}
\ff_i(\xx^{n+1})+\hat{\ff}^\textrm{ext}_i&=\mathbf{0}, \ \XX_i\notin\Omega^0_D\label{eq:fem_qs}\\
\xx^{n+1}_i&=\xx_D(\XX_i,t^{n+1}), \ \XX_i\in\Omega^0_D \label{eq:fem_d}.
\end{align}
Here the flow map is discretized as $\fm(\XX,t^{n+1})=\sum_{j=0}^{N^N-1} \xx_j^{n+1}N_j(\XX)$ where the $N_j(\XX)$ are piecewise linear interpolating functions defined over a tetrahedron mesh ($d=3$) or triangle mesh ($d=2$) and the $\xx_j^{n+1}\in\mathbb{R}^d$, $0\leq j<N^N$ are the locations of the vertices of the mesh at time $t^{n+1}$.
Note that we use $\xx^{n+1}\in\mathbb{R}^{dN^N}$ to denote the vector of all vertex locations and $x^{n+1}_{i\beta}$ to denote the $
0\leq\beta<d$ components of the position of vertex $i$ in the mesh.
The forces are given as 
\begin{align}
%\ff_i(\xx^{n+1})&=-\frac{\partial \hat{\textrm{PE}}}{\partial \yy_i}(\xx^{n+1})=-\int_{\Omega^0} \frac{\partial \Psi}{\partial \FF}(\sum_{j=0}^{M-1}  \xx_j^{n+1}\frac{\partial N_j}{\partial \XX})\frac{\partial N_i}{\partial \XX}d\XX\\
\ff_i(\yy)&=-\frac{\partial \hat{\textrm{PE}}}{\partial \yy_i}(\yy)\\
\hat{\textrm{PE}}(\yy)&=\hat{\textrm{PE}}^{\Psi}(\yy)+\hat{\textrm{PE}}^{\textrm{wc}}(\yy)\label{eq:pe_hat}\\
\hat{\textrm{PE}}^{\Psi}(\yy)&=\sum_{e=0}^{N^E-1} \Psi(\sum_{j=0}^{N^N-1}\yy_j \frac{\partial N^e_j}{\partial \XX})V^0_e\label{eq:disc_pe}\\
\hat{\ff}^{\textrm{ext}}_i&=\int_{\Omega^0} \ff^\textrm{ext}N_i d\XX + \int_{\partial \Omega^0_N} \TT_N N_i ds(\XX)
\end{align}
where $\hat{\textrm{PE}}^{\Psi}:\mathbb{R}^{dN^N}\rightarrow\mathbb{R}$ is the discretization of the potential energy, $\sum_{j=0}^{N^N-1}\yy_j \frac{\partial N^e_j}{\partial \XX}$ is the deformation gradient induced by nodal positions $\yy\in\mathbb{R}^{dN^N}$ in tetrahedron ($d=3$) or triangle ($d=2$) element $e$ with $0\leq e<N^E$, $\frac{\partial N^e_i}{\partial \XX}$ is the derivative of the interpolating function in element $e$ (which is constant since we use piecewise linear interpolation) and $V^0_e$ is the measure of the element.
We refer the reader to \cite{bonet:2008:continuum,sifakis:2012:course} for more detail on the FEM derivation of potential energy terms in a hyperelastic formulation.
Also, note that we add another term to the discrete potential energy $\hat{\textrm{PE}}^{\textrm{wc}}:\mathbb{R}^{dN^N}\rightarrow\mathbb{R}$ in Equation~\eqref{eq:pe_hat} to account for self-collisions and similar weak constraints (see Section~\ref{sec:wc}).
Similar to the non-discrete case, the constrained minimization problem
\begin{align}
\xx^{n+1}=&
\begin{array}{c}
\textrm{argmin}\\
\yy\in\mathcal{W}^{n+1}_{\Delta x}
\end{array}
\hat{\textrm{PE}}(\yy) - \yy\cdot\hat{\ff}^{\textrm{ext}}
\end{align}
where $\mathcal{W}^{n+1}_{\Delta x}=\left\{\yy\in\mathbb{R}^{dN^N}\left| \yy_i=\xx_D(\XX_i,t^{n+1}), \ \XX_i\in\partial\Omega^0_D \right.\right\}$ is equivalent to Equations~\eqref{eq:fem_qs}-\eqref{eq:fem_d}.
\subsection{Weak Constraints}\label{sec:wc}
We support weak constraints for self-collision and other similar purposes (as in \cite{mcadams:2011:mge}).
These are terms added to the potential energy in the form
\begin{align}
\hat{\textrm{PE}}^{\textrm{wc}}(\yy)&=\frac{1}{2}\sum_{c=0}^{N^{\textrm{wc}}-1}\CC_c(\yy)^T\KK_c\CC_c(\yy) \\
\CC_c(\yy)&=\sum_{j=0}^{N^N-1} w^c_{0j}\yy_j-w^c_{1j}\yy_j.
\end{align}
Here the $w^c_{0j},w^c_{1j}$ are interpolation weights that sum to one and are non-negative.
This creates constraints between the interpolated points $\sum_{j=0}^{N^N-1} w^c_{0j}\yy_j$ and $\sum_{j=0}^{N^N-1}w^c_{1j}\yy_j$.
The stiffness of the constraint is represented in the matrix $\KK_c$.
This can allow for anisotropic responses where $\KK_c=k_n\nn\nn^T + k_\tau \left( \boldsymbol\tau_0\boldsymbol\tau_0^T +\boldsymbol\tau_1\boldsymbol\tau_1^T\right)$.
Here $\nn^T\boldsymbol\tau_i=0$, $i=0,1$ and $k_n$ is the stiffness in the $\nn$ direction while $k_\tau$ is the stiffness in response to the motion in the plane normal to $\nn$.
In the case of an isotropic constraint ($k_c=k_n=k_\tau$), we use the scalar $k_c$ in place of $\KK_c$ since $\KK_c=k_c\II$ is diagonal.
We note that, in most of our examples, the anisotropic model is used for collision constraints where $\nn$ is the collision constrain direction (see Section~\ref{sec:col}).
\section{Gauss-Seidel Notation}
Our approach, PBD and XPBD all use nonlinear Gauss-Seidel to iteratively improve an approximation to the solution $\xx^{n+1}\in\mathbb{R}^{dN^N}$ of Equation~\eqref{eq:fem_qs}.
We use $l$ to denote the $l^\textrm{th}$ Gauss-Seidel iteration $\xx^{n+1,l}\approx\xx^{n+1}$.
During the course of one iteration, degrees of freedom in the approximate solution will be updated in sub-iterates which we denote as $\xx^{n+1,l}_{(k)}$ with $0\leq k< N^\textrm{GS}$.
Here $\xx^{n+1,l}_{(0)}=\xx^{n+1,l}$ and $\xx^{n+1,l}_{(N^\textrm{GS}-1)}=\xx^{n+1,l+1}$.
For example, with PBD/XPBD, in the $k^\textrm{th}$ sub-iterate, the nodes in the $k^\textrm{th}$ constraint will be projected/solved for.
In our position-based approach, in the $k^\textrm{th}$ sub-iterate, only a single node $i_k$ will be updated.
It is important to introduce this notation, since unlike with Jacobi-based approaches, the update of the $k^\textrm{th}$ sub-iterate will depend on the contents of the $k-1^\textrm{th}$ sub-iterate.
\section{Position-Based Dynamics: Constraint-Based Nonlinear Gauss-Seidel}\label{sec:pbd}
Macklin et al. \shortcite{macklin:2016:xpbd} show that PBD \cite{muller:2007:pbd} can be seen to be the extreme case of a numerical method for the approximation of the backward Euler temporal discretization of the FEM spatial discretization of Equation~\eqref{eq:mom}
\begin{align}
\sum_{j=0}^{N^N-1} m_{ij}\left(\frac{\xx_j^{n+1}-2\xx_j^n+\xx_j^{n-1}}{{\Delta t}^2}\right)=\ff_i(\xx^{n+1}) + \ff^{\textrm{ext}}_i, \ \XX_i\notin\Omega^0_D.\label{eq:fem_be}
\end{align}
Here $m_{ii}=\int_{\Omega^0} R^ 0N_i d\XX$ and $m_{ij}=0$, $j\neq i$ are entries in the mass matrix.
However, they require that the discrete potential energy in Equation~\eqref{eq:disc_pe} is of the form
\begin{align}
\hat{PE}^\Psi(\yy)=\sum_{c=0}^{2N^E-1} \frac{1}{2\alpha_c} C^2_c(\yy), \ \yy\in\mathbb{R}^{dN^E}.
\end{align}
For example, this can be done with the energy densities in Equations~\eqref{eq:cor} and \eqref{eq:nh} using two constraints $c=2e$ and $c=2e+1$ per element $e$
\begin{align}
C^\textrm{cor}_{2e}(\yy)&=|\FF^e(\yy)-\RR(\FF^e(\yy))|_F, \ C^\textrm{cor}_{2e+1}(\yy)=\det(\FF^e(\yy))-1\\
C^\textrm{nh}_{2e}(\yy)&=|\FF^e(\yy)|_F, \ C^\textrm{nh}_{2e+1}(\yy)=\det(\FF^e(\yy))-1-\frac{\mu}{\hat\lambda}.
\end{align}
In this case, $\alpha^{\textrm{cor}}_{2e}=\frac{1}{2\mu V^0_e}$, $\alpha^{\textrm{cor}}_{2e+1}=\frac{1}{\lambda V^0_e}$, $\alpha^{\textrm{nh}}_{2e}=\frac{1}{\mu V^0_e}$, $\alpha^{\textrm{nh}}_{2e+1}=\frac{1}{\hat\lambda V^0_e}$
\\
\\
To demonstrate the connection between Equation~\eqref{eq:fem_be} and PBD, Macklin et al. \shortcite{macklin:2016:xpbd} develop XPBD.
It is based on the total Lagrange multiplier formulation
\begin{align}
\sum_{j=0}^{N^N-1} m_{ij}\left(\xx_j^{n+1}-\hat{\xx}_j\right)-\sum_{c=0}^{P-1}\frac{\partial C_c}{\partial \xx_i}(\xx^{n+1})\lambda_c^{n+1}&=0, \ \XX_i\notin\Omega^0_D\label{eq:xpbd_prim}\\
C_c(\xx^{n+1})+\frac{\alpha_{c}}{\Delta t^2}\lambda^{n+1}_c&=0, \ 0\leq c<P \label{eq:xpbd_dual}
\end{align}
where $\hat{\xx}_j=2\xx_j^n-\xx_j^{n-1}-\frac{\Delta t^2}{m_{jj}} \ff^{\textrm{ext}}_j$ and $\boldsymbol\lambda^{n+1}\in\mathbb{R}^{P}$ is introduced as an additional unknown.
The $\xx^{n+1}\in\mathbb{R}^{dN^N}$ in Equations~\eqref{eq:xpbd_prim}-\eqref{eq:xpbd_dual} is the same in the solution to Equation~\eqref{eq:fem_be}.
Macklin et al. \shortcite{macklin:2016:xpbd} use a per-constraint Gauss-Seidel update of Equations~\eqref{eq:xpbd_prim}-\eqref{eq:xpbd_dual}
\begin{align}
\xx^{n+1,l}_{i(k+1)}&=\xx^{n+1,l}_{i(k)}+\Delta \xx^{n+1,l}_{i(k+1)}, \ \XX_i\notin\Omega^0_D\label{eq:xpbd_prim_gn}\\
\Delta \xx^{n+1,l}_{i(k+1)}&=\frac{\Delta \lambda^{n+1,l}_{(k+1)c_k}}{m_{ii}}\frac{\partial C_{c_k}}{\partial \xx_i}(\xx^{n+1,l}_{(k)})\label{eq:xpbd_dx}\\
\Delta \lambda^{n+1,l}_{(k+1)c_k}&=\frac{-C_{c_k}(\xx^{n+1,l}_{(k)}) + \frac{\alpha_{c_k}}{\Delta t^2}C_{c_k}(\xx^{n+1,l}_{(k)})}{\sum_{j=0}^{N^N-1} \frac{1}{m_{jj}}\sum_{\beta=0}^{d-1}\left(\frac{\partial C_{c_k}}{\partial x_{j\beta}}(\xx^{n+1,l}_{(k)})\right)^2 + \frac{\alpha_{c_k}}{\Delta t^2}}\label{eq:xpbd_dual_gn}.
\end{align}
Here the $k+1^\textrm{th}$ sub-iterate in iteration $l$ is generated by solving for the change in a single Lagrange multiplier $\Delta \lambda^{n+1,l}_{(k+1)c_k}$ associated with a constraint $c_k$ that varies from sub-iteration to sub-iteration.
\subsection{Quasistatics}
As noted by Macklin et al. \shortcite{macklin:2016:xpbd}, the XPBD update in Equations~\eqref{eq:xpbd_prim_gn}-\eqref{eq:xpbd_dual_gn} is the same as in the original PBD \cite{muller:2007:pbd} in the limit $\alpha_c\rightarrow0$.
By choosing a stiffness inversely proportionate to a parameter $s \geq 0$ and examining the limiting behavior of the equations being approximated, we see that the original PBD approach generates an approximation to the quasistatic problem (Equations~\eqref{eq:qs}), albeit with the external forcing terms omitted.
More precisely, define $\fm_s$ to be a solution of the problem
\begin{align}
s R^0\frac{\partial^2 \fm_s}{\partial t^2}=\nabla^\XX \cdot \PP + s\ff^\textrm{ext}. \label{eq:mom_s}
\end{align}
subject to the same boundary conditions in Equations~\eqref{eq:bC_c}-\eqref{eq:bc_n}.
This is equivalent to scaling the $\alpha_c$ that would appear in Equation~\eqref{eq:mom} (through $\PP)$ by $s$.
The $\alpha_c$ are inversely proportionate to the Lam{\'e} parameters, so as $s\rightarrow 0$, the material stiffness increases.
Since the inertia and external force terms in Equation~\eqref{eq:mom_s} vanish as $s\rightarrow 0$, it is clear then that the original PBD formulation generates an approximation to the solution of a quasistatic problem with the external forcing $\ff^\textrm{ext}$ omitted.
Note that PBD does include the external forcing term in its initial guess $\xx^{n+1}_i=\xx^n_i+\Delta t \vv^n_i + \frac{\Delta t^2}{m_{ii}}\ff^\textrm{ext}_i$.
However, the effect of the initial guess vanishes as the iteration count is increased.
We demonstrate this in Section~\ref{sec:ex_pbd}
Also, note that this is not the case in the XPBD formulation where $\alpha_c>0$.
\\
\begin{figure}[h]
	\includegraphics[draft=\mydraft,width=0.49\columnwidth,trim={0px 70px 0px 120px},clip]{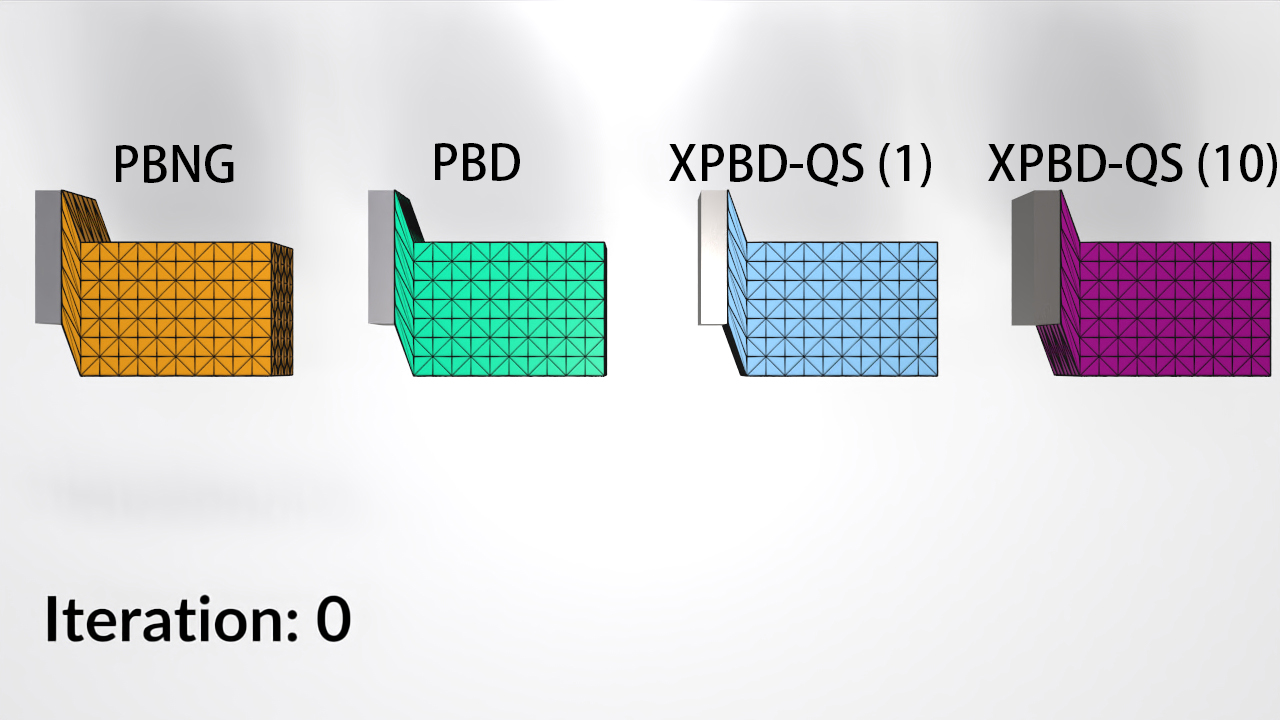}
	\includegraphics[draft=\mydraft,width=0.49\columnwidth,trim={0px 70px 0px 120px},clip]{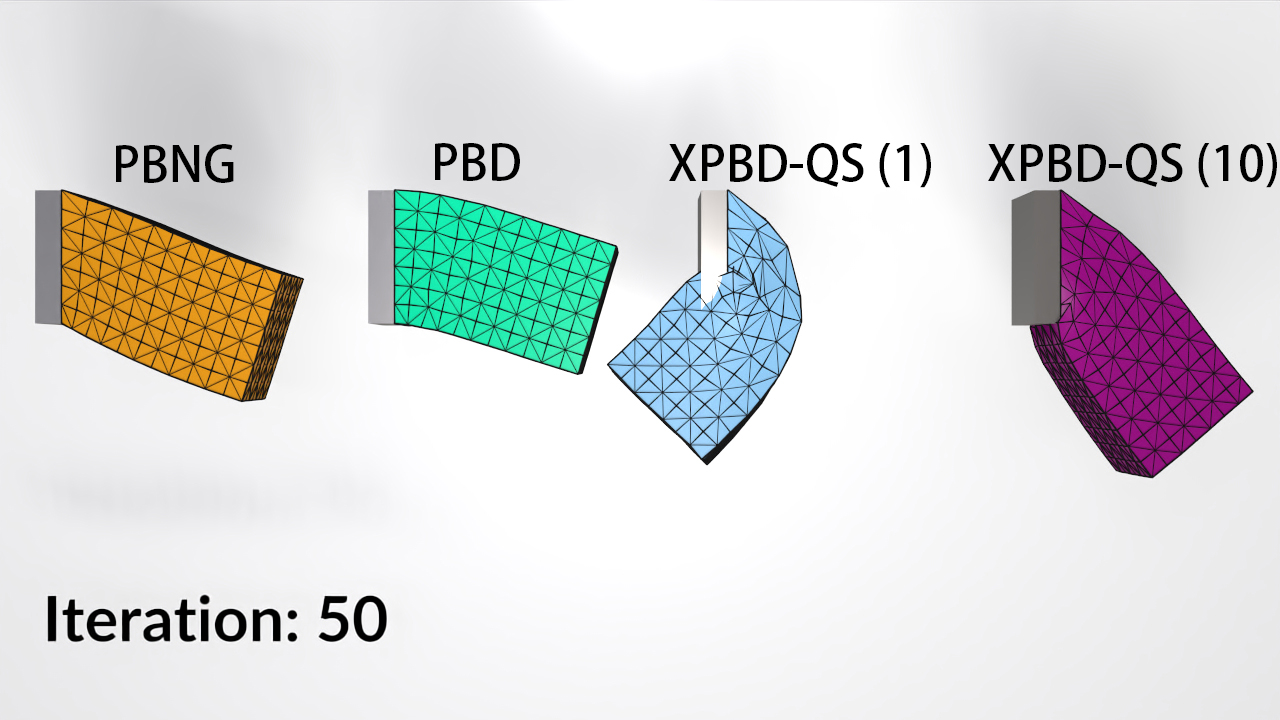}
	\includegraphics[draft=\mydraft,width=0.49\columnwidth,trim={0px 70px 0px 120px},clip]{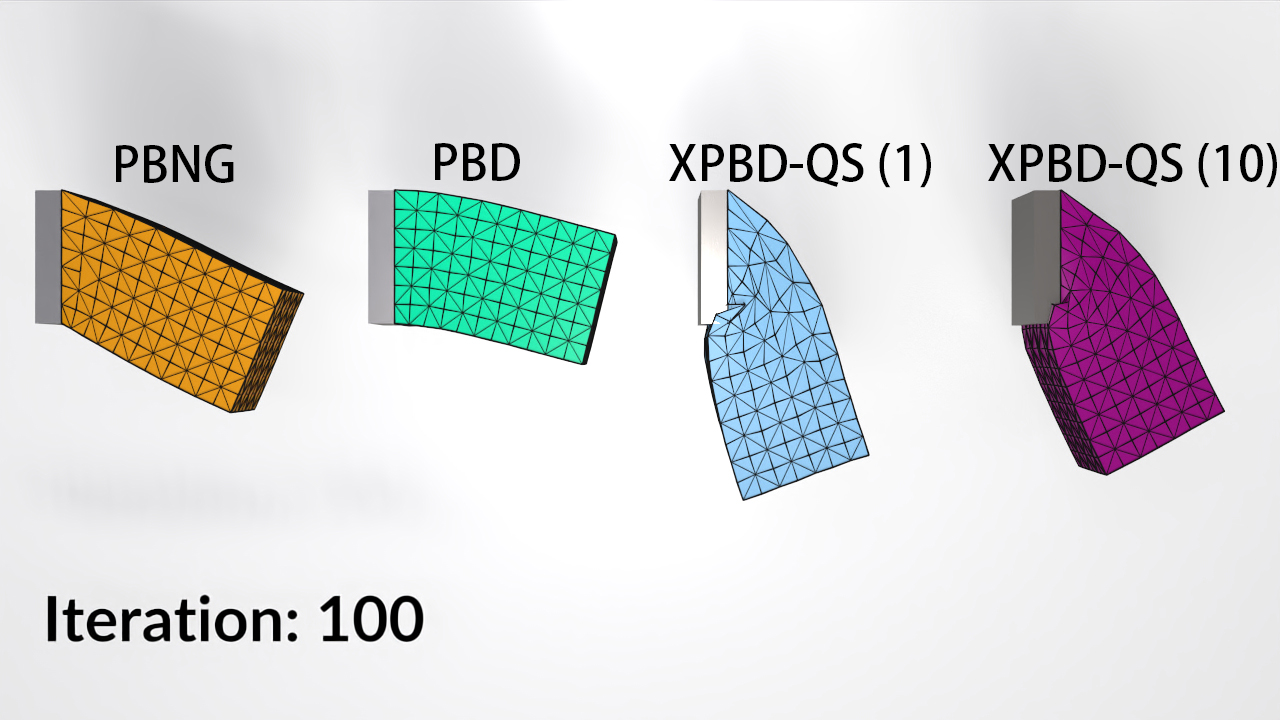}
	\includegraphics[draft=\mydraft,width=0.49\columnwidth,trim={0px 70px 0px 120px},clip]{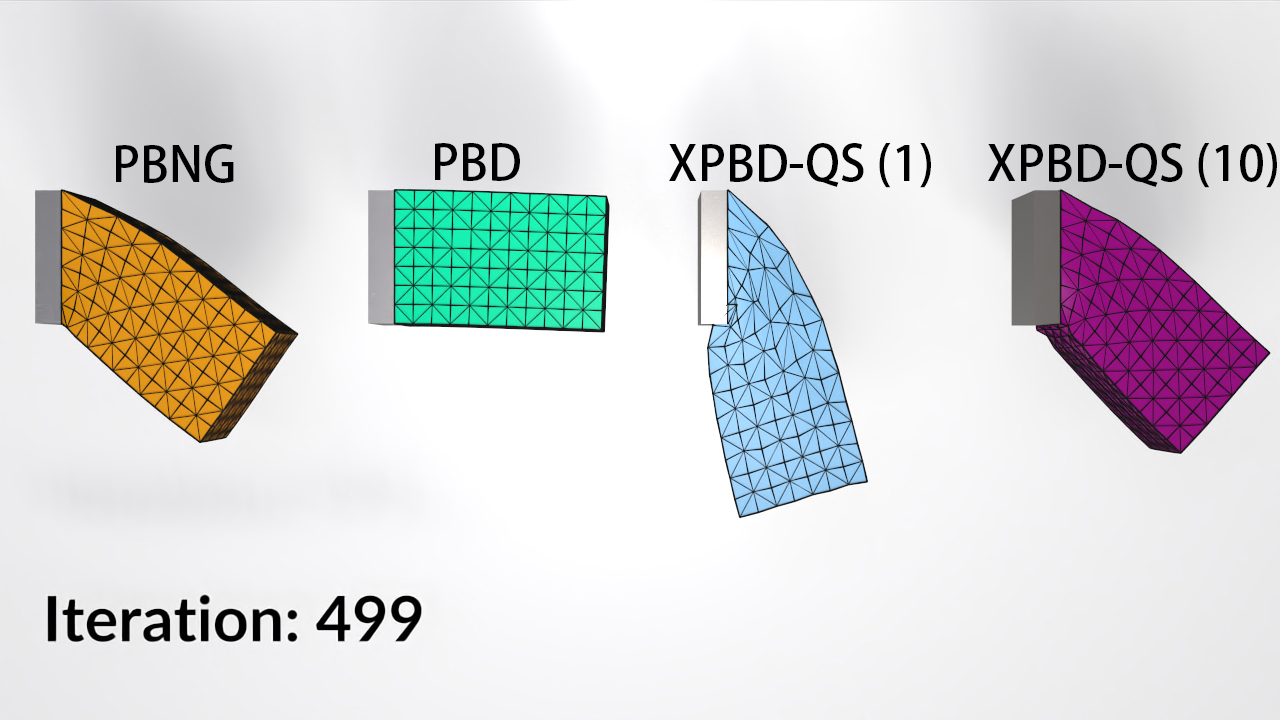}
	\caption{\textbf{Bar under Gravity}. A quasistatic simulation of a bar bending under gravity using different methods. 
	The effect of external forcing vanishes in the PBD example as the number of iterations increases. 
	More local iterations of XPBD-QS produces better results.
	PBNG converges to visually plausible results within fewer iterations than XPBD-QS.}
	\label{fig:bar_under_gravity}
\end{figure}\\
Unfortunately, XPBD cannot be trivially modified to run quasistatic problems.
For example, omitting the mass terms on the left-hand side of Equation~\eqref{eq:xpbd_prim} makes the Gauss-Seidel update in Equations\eqref{eq:xpbd_prim_gn}-\eqref{eq:xpbd_dual_gn} impossible since there would be a division by zero.
The simplest fix for quasistatic problems we can conceive of in the PBD framework is to use XPBD run to steady state using a pseudo-time iteration. 
This prevents the need for scaling the $\alpha_c$ which inherently removes the external forcing terms and does not introduce a divide by zero in Equation~\eqref{eq:xpbd_dx}.
However, this is very costly since each quasistatic time step is essentially the cost of an entire XPBD simulation.
Nevertheless, we compare our approach against this option (see Section~\ref{sec:ex_pbd}) since it will at least allow for the correct representation of the forcing terms.
We refer to this technique as XPBD-QS.
\subsection{XPBD Convergence}\label{sec:xpbd_conv}
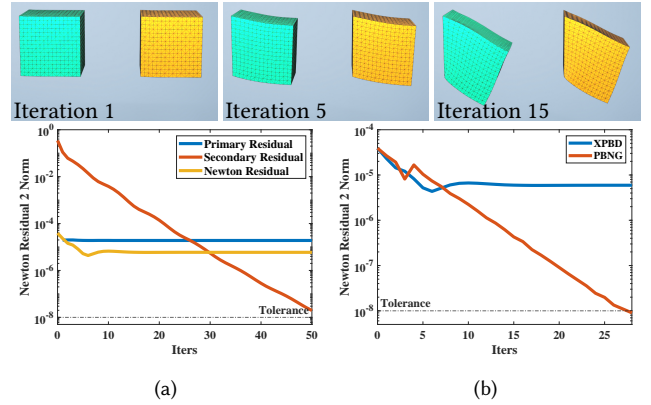
\begin{figure}[h]
	\begin{tikzpicture}
		\node [anchor=south west, inner sep=0pt] (image1) at (0,0) {
			\includegraphics[draft=\mydraft,width=0.32\columnwidth,trim={290px 150px 260px 150px},clip]{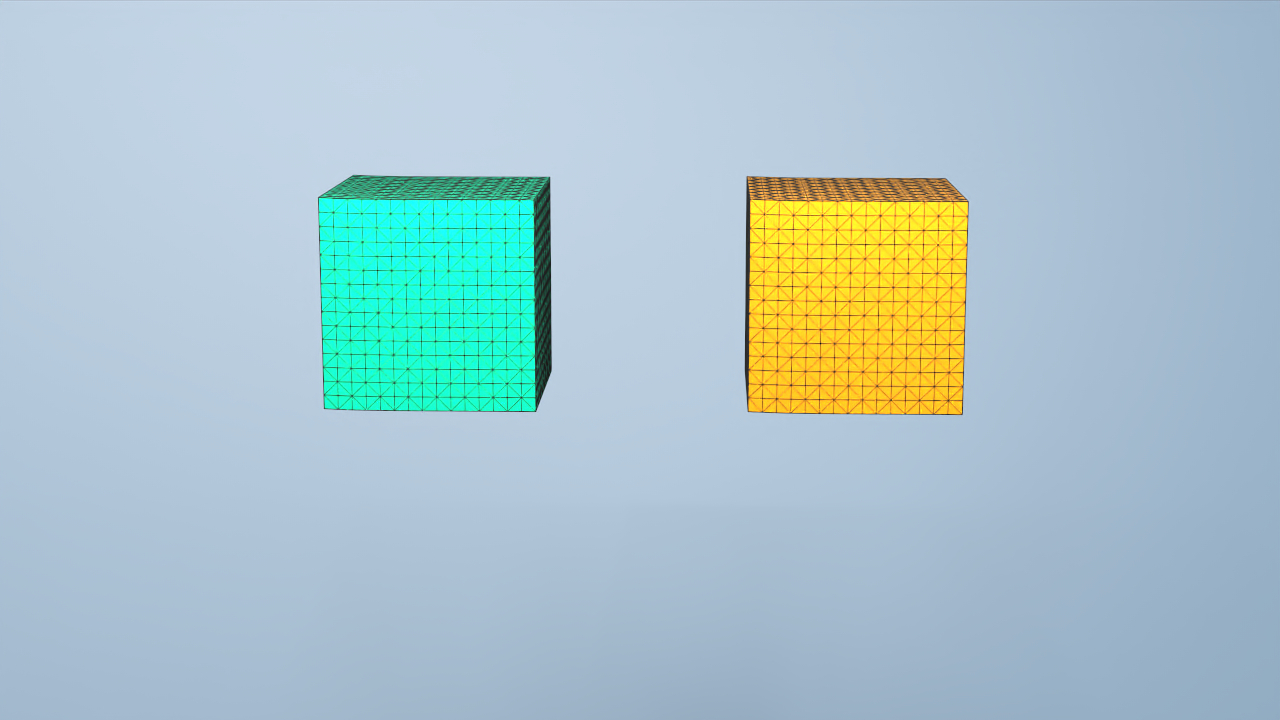}
		};
		\node [anchor=south west, inner sep=1pt] at (image1.south west) {Iteration 1};
	\end{tikzpicture}
	\begin{tikzpicture}
		\node [anchor=south west, inner sep=0pt] (image1) at (0,0) {
			\includegraphics[draft=\mydraft,width=0.32\columnwidth,trim={290px 150px 260px 150px},clip]{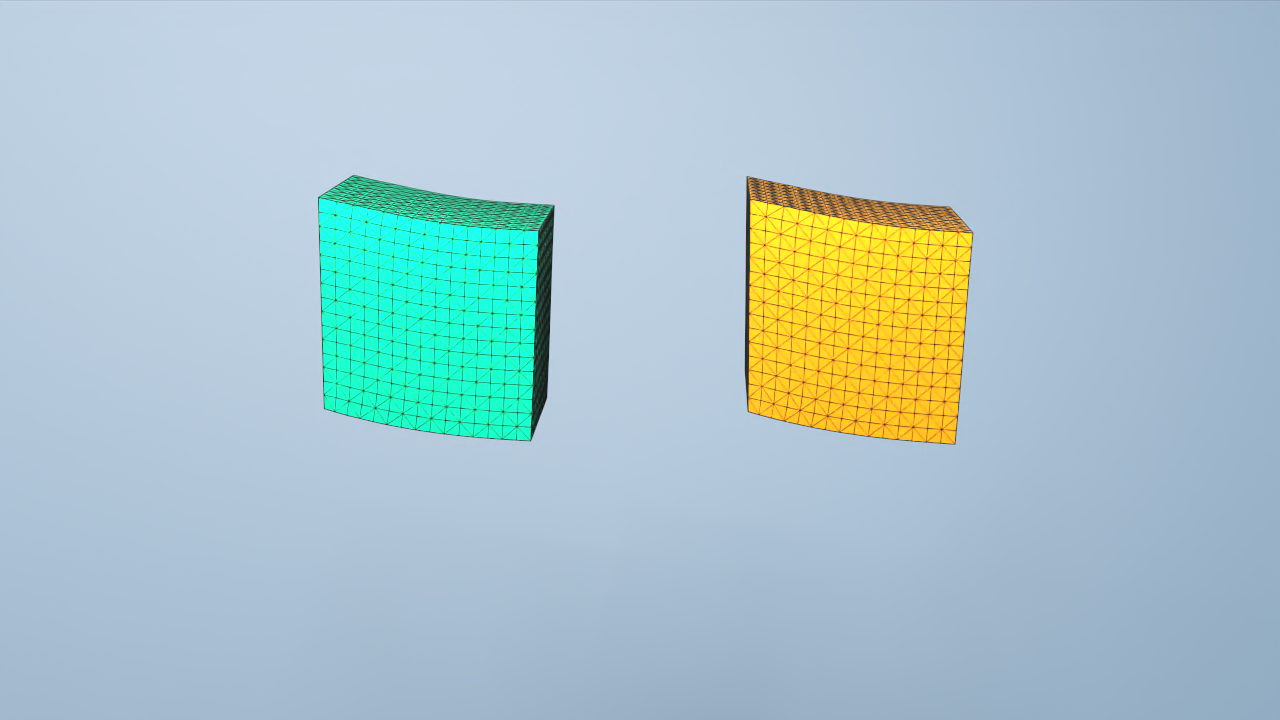}
		};
		\node [anchor=south west, inner sep=1pt] at (image1.south west) {Iteration 5};
	\end{tikzpicture}
	\begin{tikzpicture}
		\node [anchor=south west, inner sep=0pt] (image1) at (0,0) {
			\includegraphics[draft=\mydraft,width=0.32\columnwidth,trim={290px 150px 260px 150px},clip]{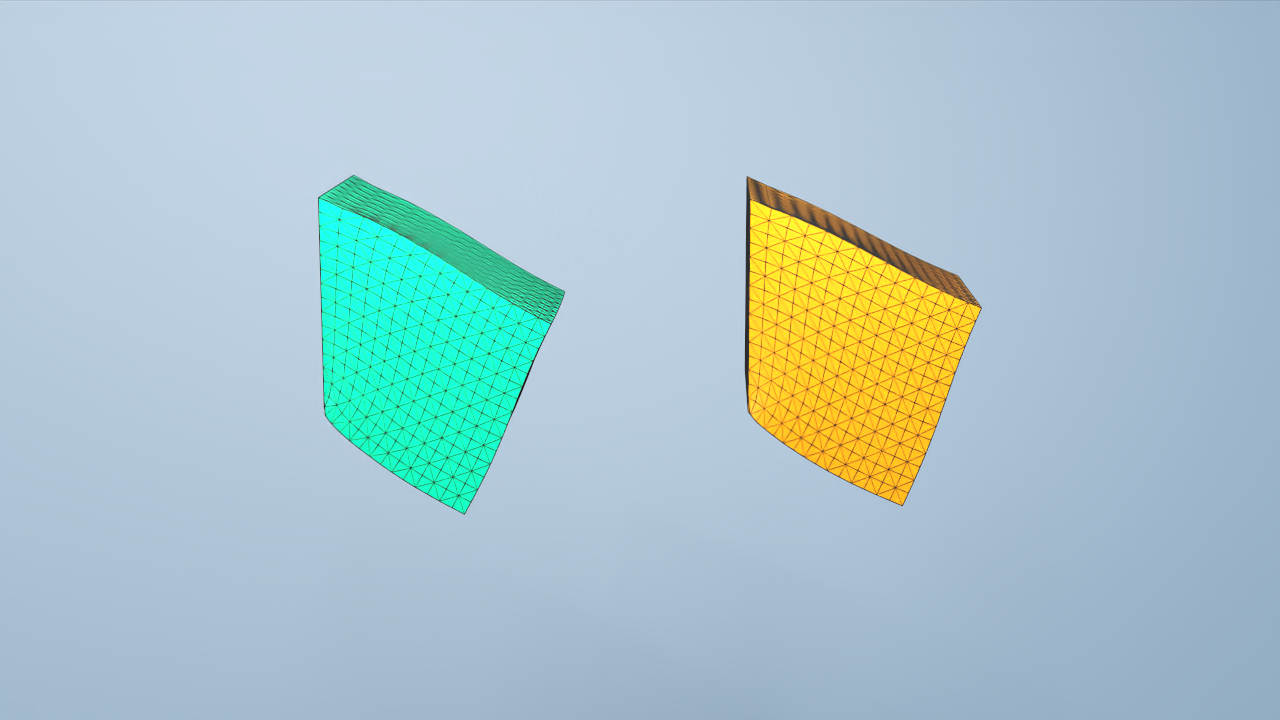}
		};
		\node [anchor=south west, inner sep=1pt] at (image1.south west) {Iteration 15};
	\end{tikzpicture}

	\subcaptionbox{}{
		\includegraphics[width = 0.47\columnwidth,trim={10px 0 35px 20px},clip]{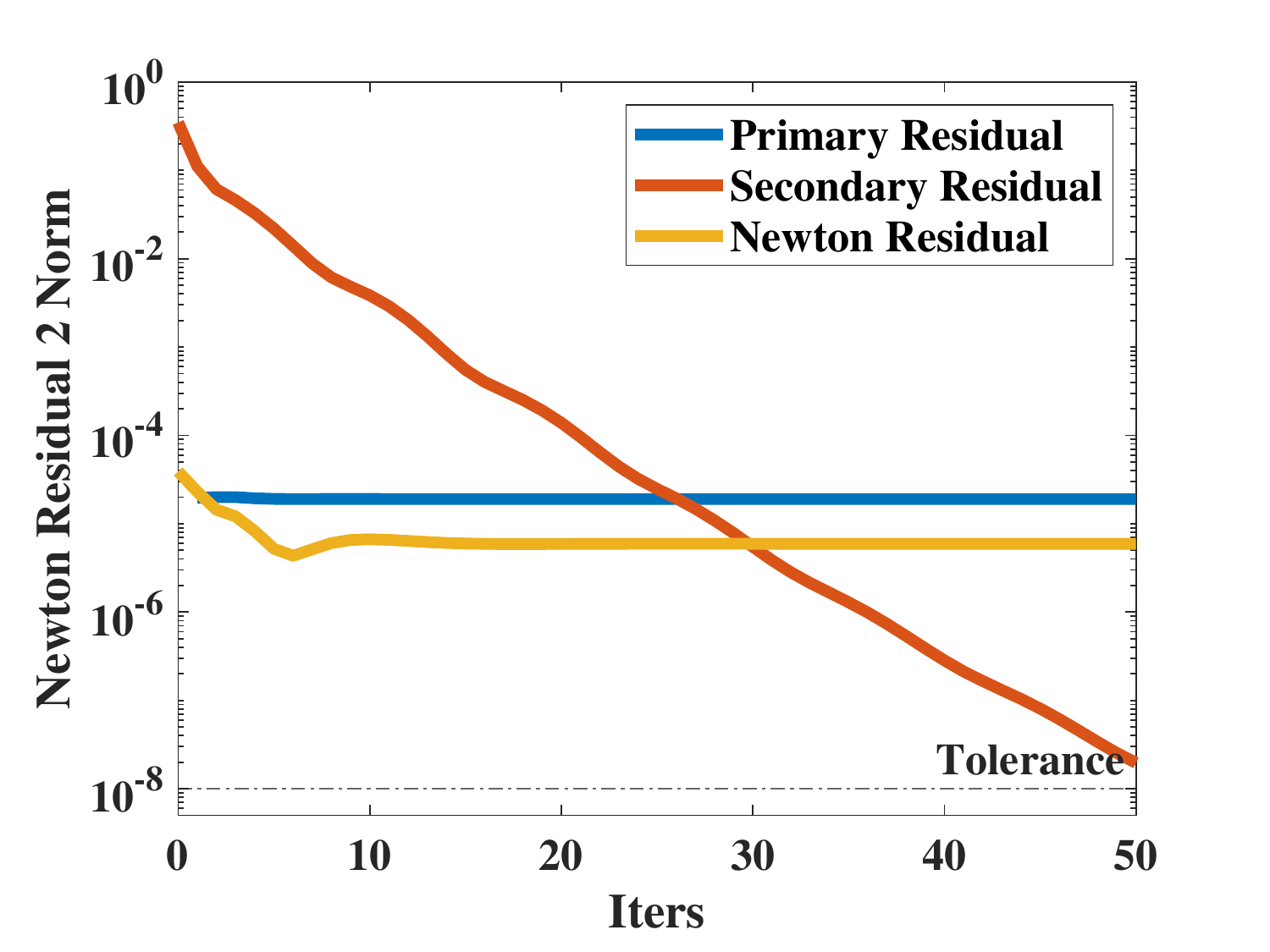}
	}
	\subcaptionbox{}{
		\includegraphics[width = 0.47\columnwidth,trim={10px 0 35px 20px},clip]{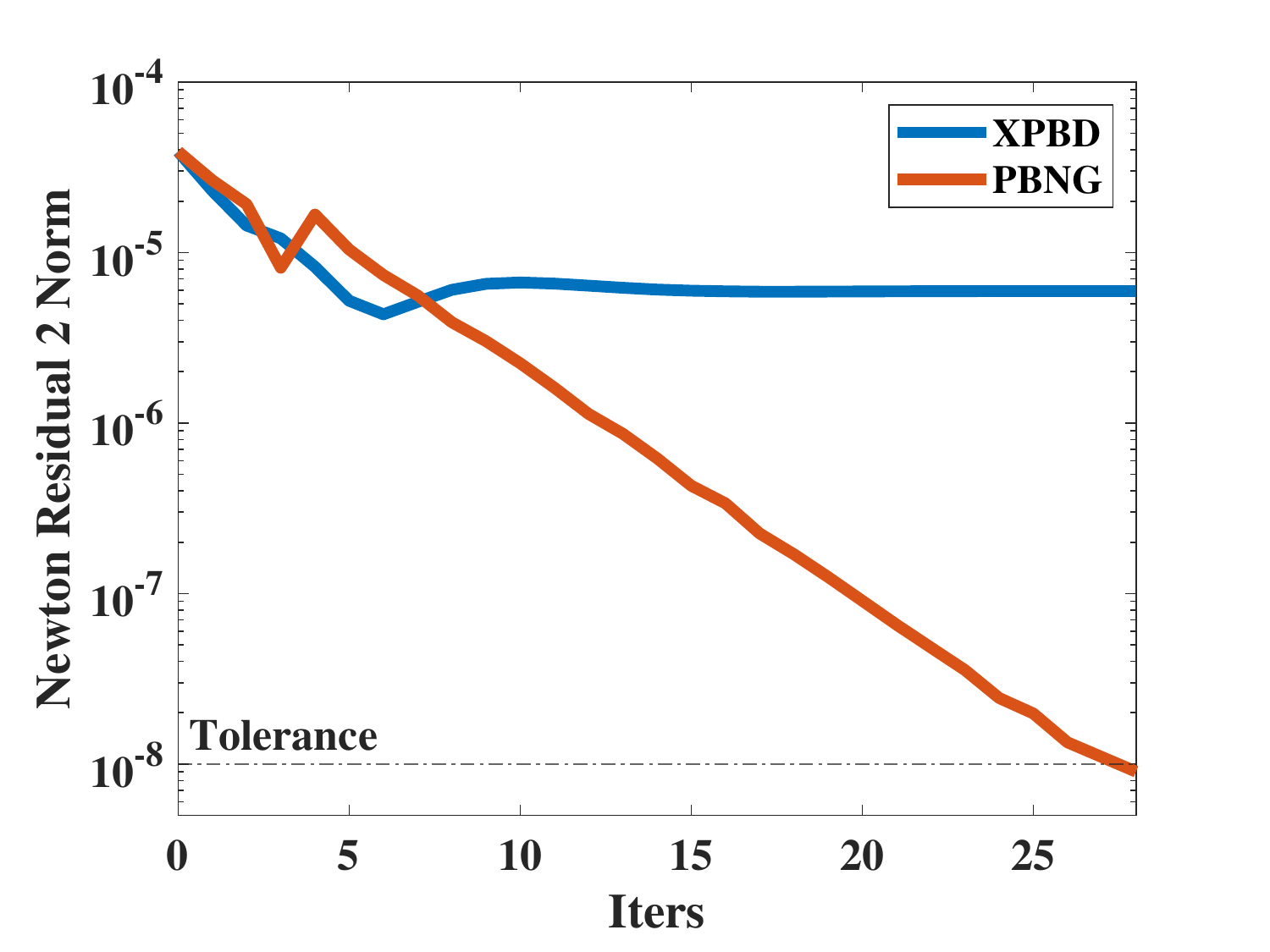}
	}
	\caption{
		\textbf{Top}. Clamped blocks under gravity. The green block is XPBD, and the yellow one is PBNG.
		\textbf{(a) Primary Residual Comparison: Stagnation}. While XPBD reliably reduces the secondary residual, its omission of the primary residual in the linearization causes its primary residual to stagnate, making its true (Newton) residual stagnate as well.
		\textbf{(b) Convergence}. PBNG is able to reduce the Newton residual to the tolerance, whereas XPBD's residual stagnates.
	}
	\label{fig:xpbd_stagnation}
\end{figure}

The linear system in Equations~\eqref{eq:xpbd_prim}-\eqref{eq:xpbd_dual} that forms the basis for the XPBD Gauss-Seidel update is generated from the omission of two terms in their Lagrange multiplier formulation (shown in red)
\small
\begin{align}
\left(
\begin{array}{cc}
\MM+{\color{red}\sum_c \lam_{ck}\frac{\partial^2 C_c}{\partial \xx^2}(\xx^{n+1}_k)}&-\nabla C^T_c(\xx^{n+1}_k)\\
\nabla \CC(\xx^{n+1}_k)&\frac{\AA}{\dt^2}
\end{array}
\right)
\left(
\begin{array}{c}
\Delta \xx_{k+1}\\
\Delta \boldsymbol\lam_{k+1}
\end{array}
\right)=
\left(
\begin{array}{c}
{\color{red}\gg(\xx^{n+1}_k,\boldsymbol\lam_k)}\\
\hh(\xx^{n+1}_k,\boldsymbol\lam_k)
\end{array}
\right).\label{eq:xlin}
\end{align}
\normalsize
The left-hand side term is second-order, but its omission is essential for the decoupling of position (primary) and Lagrange multipliers.
However, the omission on the right-hand side is of the residual in the position (primary) equations.
Without this term, residual reduction with XPBD stagnates after an iteration or two. 
However, the inclusion of this term leads to unstable behavior.
We demonstrate this in Section~\ref{sec:ex_xpbd} and Figure~\ref{fig:xpbd_stagnation}.
We believe that the omission of the primary residual in the update causes iteration-dependent behavior and generally degraded convergence with XPBD since information about adjacent constraints would be included in this term if it could be stably added.
Furthermore, the effect is more visually pronounced in quasistatic problems.

\begin{figure*}[ht]
	
	\begin{tikzpicture}
		\node [anchor=south west, inner sep=0pt] (image3) at (0,0) {
			\includegraphics[draft=\mydraft,width=0.17\textwidth,trim={0 0 0 0},clip]{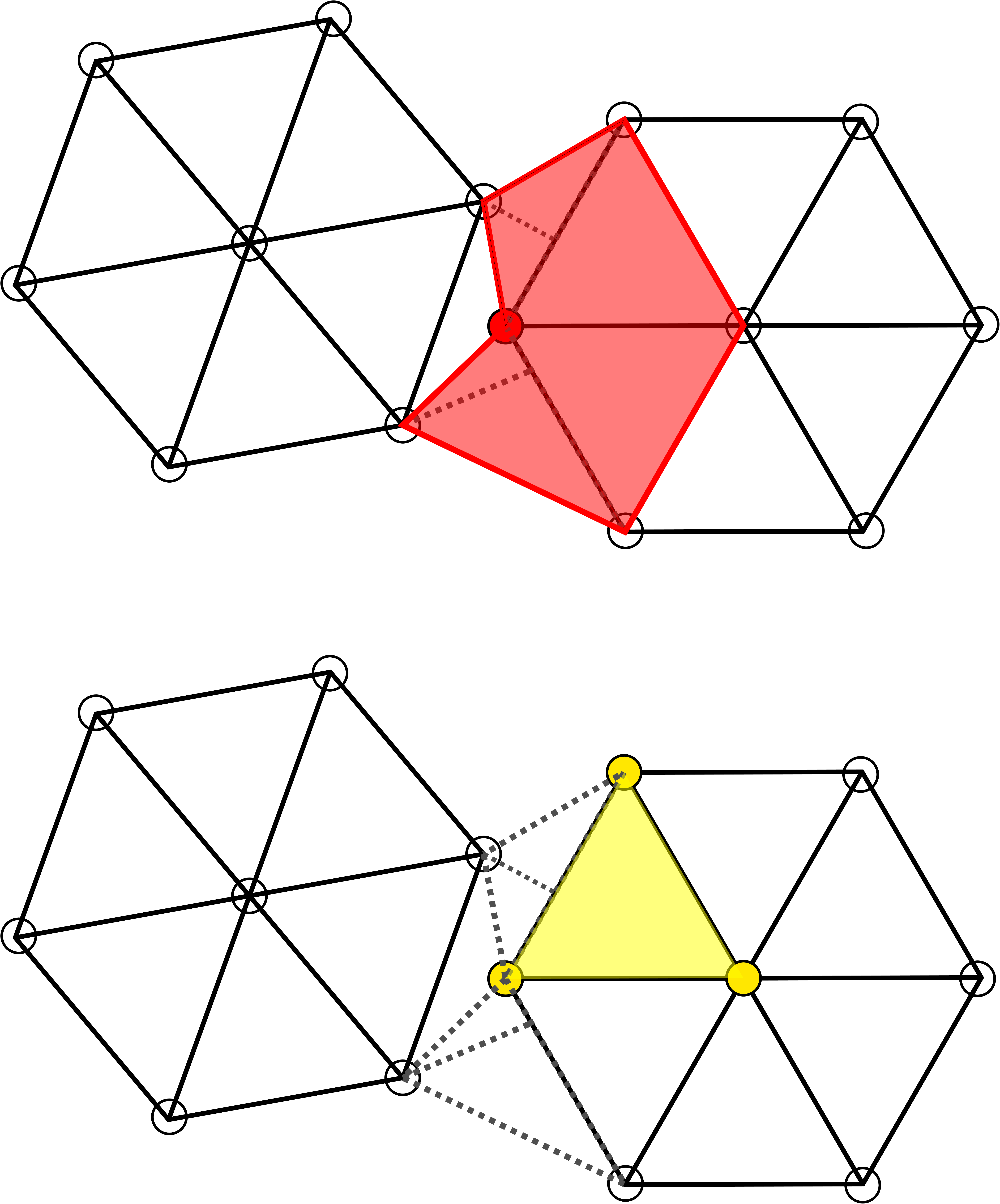}
		};
		\node [anchor=south west, inner sep=0pt] at (image3.south west) {(a)};
	\end{tikzpicture}
	\quad
	\unskip\vrule
	\quad
	\begin{tikzpicture}
		\node [anchor=south west, inner sep=0pt] (image1) at (0,0) {
			\includegraphics[draft=\mydraft,width=0.38\textwidth,trim={0 0 0 0},clip]{figures/constraints_coloring.pdf}
		};
		\node [anchor=south west, inner sep=0pt] at (image1.south west) {(b)};
	\end{tikzpicture}
	\quad
	\unskip\vrule
	\quad
	\begin{tikzpicture}
		\node [anchor=south west, inner sep=0pt] (image2) at (0,0) {
			\includegraphics[draft=\mydraft,width=0.38\textwidth,trim={0 0 0 0},clip]{figures/particle_coloring.pdf}
		};
		\node [anchor=south west, inner sep=0pt] at (image2.south west) {(c)};
	\end{tikzpicture}

	\caption{
		\textbf{(a) Dual Coloring }. Node based coloring (top) is contrasted with constraint based coloring (bottom). When a node is colored as red, its incident elements register red as used colors. When a constraint is colored yellow, its incident particles register yellow as used colors.  
		{\textbf{(b) Constraints-Based Coloring}}. A step-by-step constraint mesh coloring scheme is shown. The dotted line indicates two weak constraints between the elements. The first constraint is colored red, all its incident points will register red as a used color. Other constraints incident to the first constraint have to choose other colors.
		{\textbf{(c) Node-Based Coloring}}. A step-by-step node coloring scheme is shown. The constraint register the colors used by its incident particles. The first particle is colored red, so all its incident constraints will register red as used. Other particles incident to the constraints have to choose other colors.
	}
	\label{fig:constraints_coloring}
\end{figure*}

\section{Position-Based Nonlinear Gauss-Seidel} \label{sec:ngs}
To fix the issues with PBD/XPBD and quasistatics, we abandon the Lagrange-multiplier formulation and approximate the solution of Equation~\eqref{eq:fem_qs} using position-centric, rather than constraint-centric nonlinear Gauss-Seidel.,
This update takes into account each constraint that the position participates in.
Visual intuition for this is illustrated in top of Figure~\ref{fig:constraints_coloring}(a).
More specifically, in the $k^\textrm{th}$ sub-iterate of iteration $l$, we modify a single node $i_k$ with $\XX_{i_k}\notin\partial\Omega^0_D$ as 
\begin{align}
\xx^{n+1,l}_{(k+1)i_k}&=\xx^{n+1,l}_{(k)i_k}+\Delta \xx^{n+1,l}_{(k+1)i_k} \label{eq:pbgn_update}\\
\Delta \xx^{n+1,l}_{(k+1)i_k}&=
\begin{array}{c}
\textrm{argmin}\\
\Delta \yy\in\mathbb{R}^d
\end{array}
\hat{PE}(\xx^{n+1,l}_{(k)} + \tilde{\CC}^{i_k}\Delta \yy)-\Delta\yy\cdot\hat{\ff}^\textrm{ext}_{i_k}\notag.
\end{align}  
Here $\tilde{\CC}^{i_k}\in\mathbb{R}^{dN^E \times d}$ is a selection matrix that isolates the degrees of freedom on the node $i_k$ and has entries $\tilde{C}^{i_k}_{j\alpha\beta}=\delta_{j{i_k}}\delta_{\alpha\beta}$.
We solve this minimization by setting the gradient to zero
\begin{align}
\mathbf{0}&=\ff_{i_k}(\xx^{n+1,l}_{(k)} + \tilde{\CC}^{i_k}\Delta \xx^{n+1,l}_{(k+1)i_k})+\hat{\ff}^\textrm{ext}_{i_k}\label{eq:pbgn_newt}.
\end{align}
We use a single step of a modified Newton's method to approximate the solution of Equation~\eqref{eq:pbgn_newt} for $\Delta \xx^{n+1,l}_{(k+1)i_k}\in\mathbb{R}^{d}$.
We use $\Delta \xx^{n+1,l}_{(k+1)i_k}=\mathbf{0}$ as the initial guess.
We found that using more than one iteration did not significantly improve robustness or convergence.
Our update is of the form
\begin{align}
\Delta \xx^{n+1,l}_{(k+1)i_k}=\left(\AA^{n+1,l}_{(k+1)i_k}\right)^{-1}\left(\ff_{i_k}(\xx^{n+1,l}_{(k)})+\hat{\ff}^\textrm{ext}_{i_k}\right). \label{eq:pbgn_dx}
\end{align}
Here $\AA^{n+1,l}_{(k+1)i_k}\approx-\frac{\partial \ff_{i_k}}{\partial \yy_{i_k}}(\xx^{n+1,l}_{(k)})\in\mathbb{R}^{d\times d}$ is an approximation to the potential energy Hessian/negative force gradient.
\subsection{Modified Hessian}
We choose the modified energy Hessian $\AA^{n+1,l}_{(k+1)i_k}$ to minimize its computational cost.
The true Hessian $\frac{\partial \ff_{i_k}}{\partial \yy_{i_k}}\in\mathbb{R}^{d\times d}$  has entries
\begin{align}
\frac{\partial f_{i_k \alpha}}{\partial y_{i_k \beta}}(\yy)&=-\sum_{e=0}^{N^E-1}\sum_{\delta,\gamma=0}^{d-1}\mathcal{C}^e_{\alpha\gamma\beta\delta}(\yy)\frac{\partial N_{i_k}^e}{\partial X_\gamma}\frac{\partial N^e_{i_k}}{\partial X_\delta}V^0_e-\label{eq:hess}\\
&\sum_{c=0}^{N^{\textrm{wc}}-1} \left(w^c_{0i_k}-w^c_{1i_k}\right)^2 K_{c\alpha\beta}, \ 0\leq \alpha,\beta< d\notag
\end{align}
where $\mathcal{C}^e_{\alpha\gamma\beta\delta}(\yy)=\frac{\partial^2\Psi}{\partial F_{\beta\delta}\partial F_{\alpha\gamma}}(\sum_{j=0}^{N^N-1}\yy_j \frac{\partial N^e_j}{\partial \XX})$ is the Hessian of the potential energy density evaluated at the deformation gradient in element $e$.
This follows since the potential force on the node $i_k$ is
\begin{align}
\ff_{i_k}(\yy)&=-\sum_{e=0}^{N^E-1} \hP_e(\yy) \frac{\partial N_{i_k}}{\partial \XX}(\XX^e)V^0_e-\sum_{c=0}^{N^{\textrm{wc}}-1}\left(w^c_{0i_k}-w^c_{1i_k}\right)\KK_c\CC_c(\yy)
\end{align}
where $\hP_e(\yy)=\frac{\partial \Psi}{\partial \FF}(\sum_{j=0}^{N^N-1}\yy_j \frac{\partial N^e_j}{\partial \XX})$ is the first Piola-Kirchhoff stress in the element.\\
\\
The primary cost in Equation~\eqref{eq:hess} is the evaluation of the Hessian of the energy density $\mathcal{C}^e_{\alpha\gamma\beta\delta}(\yy)$ which is a symmetric fourth order tensor with $d^2\times d^2$ entries.
Furthermore, this tensor can be indefinite, which would complicate the convergence of the Newton procedure.
We use a definiteness projection as in \cite{teran:2005:robust} and \cite{smith:2019:analytic}.
However we use a very simple symmetric positive definite approximation instead of their approaches which require the singular value decomposition of the element deformation gradient $\sum_{j=0}^{N^N-1}\yy_j \frac{\partial N^e_j}{\partial \XX}$.
Teran et al. \shortcite{teran:2005:robust} also require the solution of a $3\times 3$ and three $2\times 2$ symmetric eigenvalue problems, our approach does not require this.
Our the simple approximation is $\tilde{\mathcal{C}}^e_{\alpha\gamma\beta\delta}(\yy)\approx\mathcal{C}^e_{\alpha\gamma\beta\delta}(\yy)$ with
\begin{align}
\tilde{\mathcal{C}}^e_{\alpha\gamma\beta\delta}(\yy)=2\mu\delta_{\alpha\beta}\delta_{\gamma\delta} + \lambda J{F^{e^{-1}}}_{\alpha\gamma}(\yy)J{F^{e^{-1}}}_{\beta\delta}(\yy) \label{eq:mod_hess_dens}.
\end{align}
Here $J{\FF^e}(\yy)=\det(\FF^e(\yy)){\FF^e}^{-T}(\yy)$ is the cofactor matrix of the element deformation gradient $\FF^e(\yy)=\sum_{j=0}^{N^N-1}\yy_j \frac{\partial N^e_j}{\partial \XX}$. We note that the cofactor matrix is defined for all deformation gradients $\FF^e$, singular, inverted (negative determinant) or otherwise.
This is essential for robustness to large deformation. 
We discuss the motivation for this simplification in Section~\ref{sec:lame}, but note here that it is clearly positive definite since it is a scaled version of the identity with a rank-one update from the cofactor matrix (with positive $\lambda>0$ scaling).
With this convention, our symmetric positive definite modified nodal Hessian is of the form
\begin{align}
A^{n+1}_{(k+1)i_k\alpha\beta}&=\sum_{e=0}^{N^E-1}\sum_{\delta,\gamma=0}^{d-1}\tilde{\mathcal{C}}^e_{\alpha\gamma\beta\delta}(\xx^{n+1,l}_{(k)})\frac{\partial N_{i_k}^e}{\partial X_\gamma}\frac{\partial N^e_{i_k}}{\partial X_\delta}V^0_e+\label{eq:mod_hess}\\
&\sum_{c=0}^{N^{\textrm{wc}}-1} \left(w^c_{0i_k}-w^c_{1i_k}\right)^2 K_{c\alpha\beta}, \ 0\leq \alpha,\beta< d 
\end{align}
\begin{figure}[h]
	\includegraphics[draft=\mydraft,width=0.32\columnwidth,trim={180px 20px 350px 0px},clip]{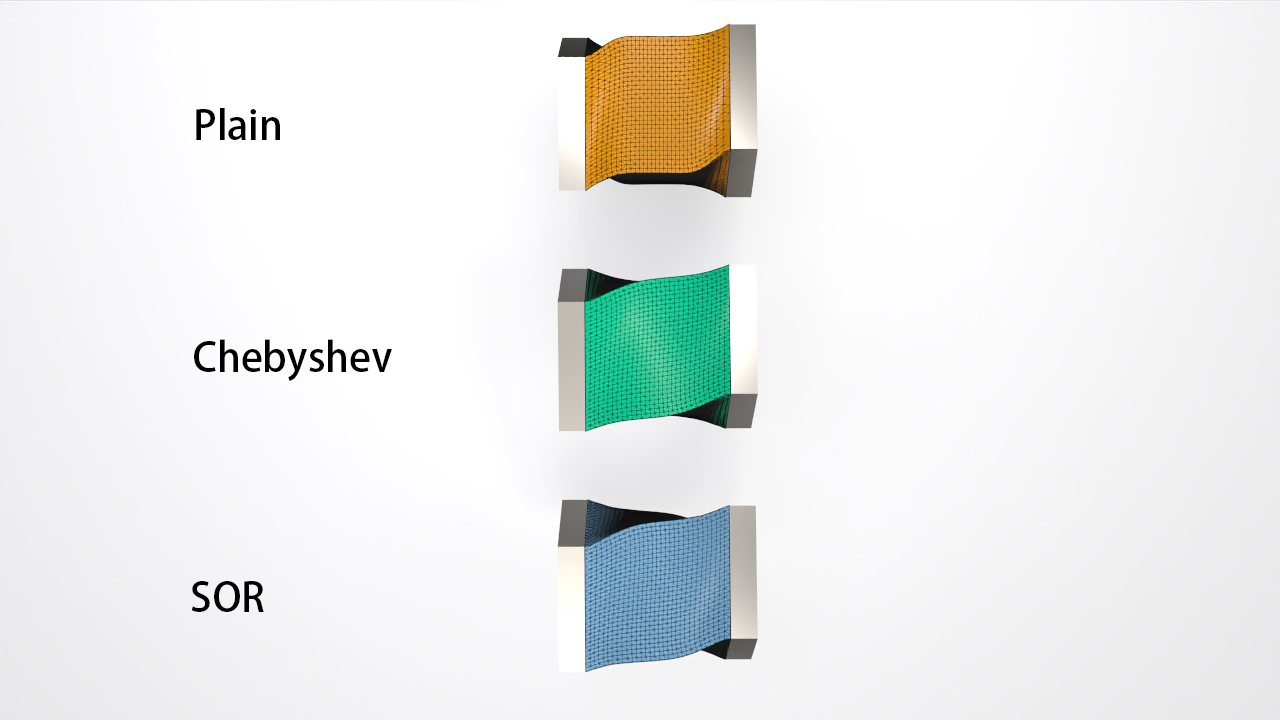}
	\includegraphics[draft=\mydraft,width=0.32\columnwidth,trim={180px 20px 350px 0px},clip]{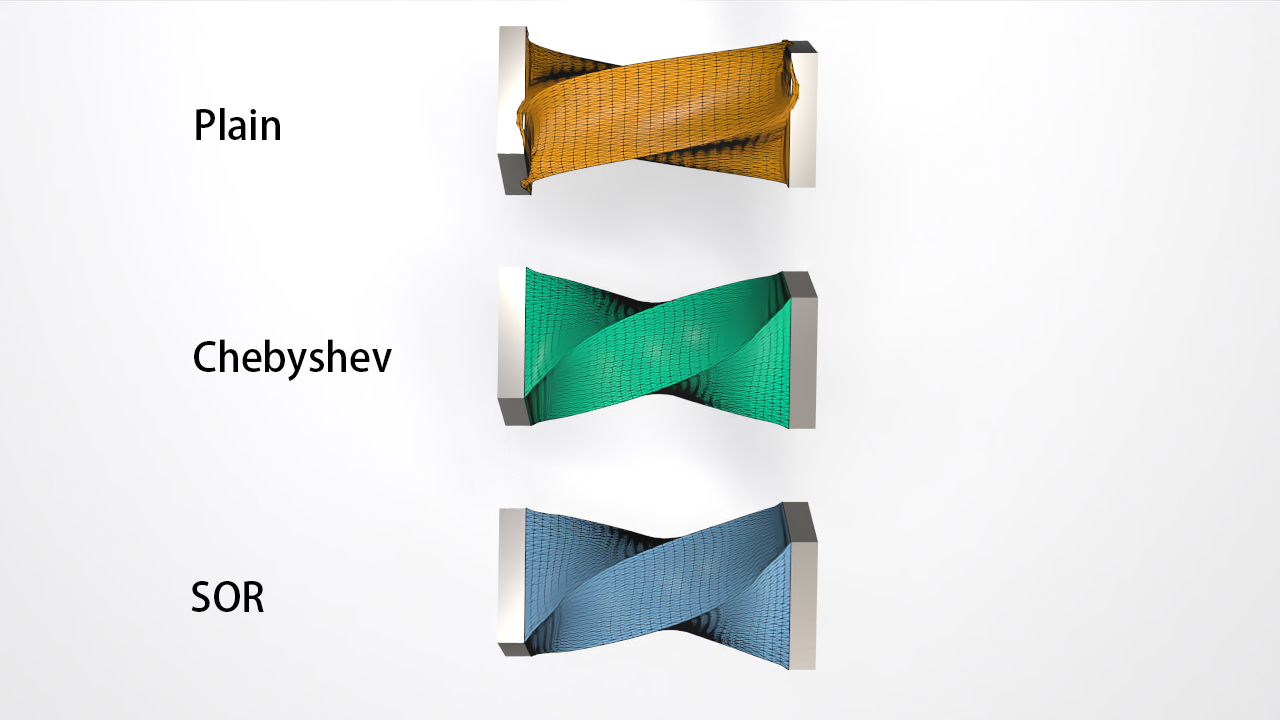}
	\includegraphics[draft=\mydraft,width=0.32\columnwidth,trim={180px 20px 350px 0px},clip]{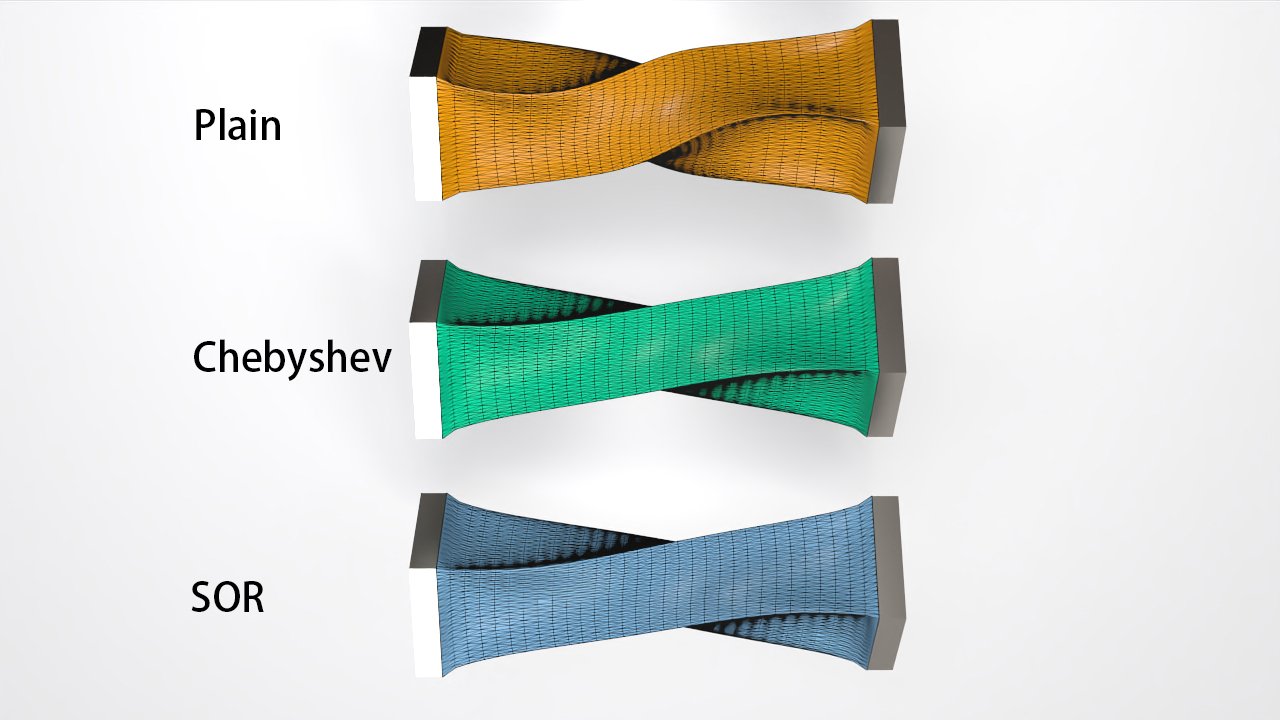}
	\includegraphics[width = \columnwidth,trim={10px 5px 40px 20px},clip]{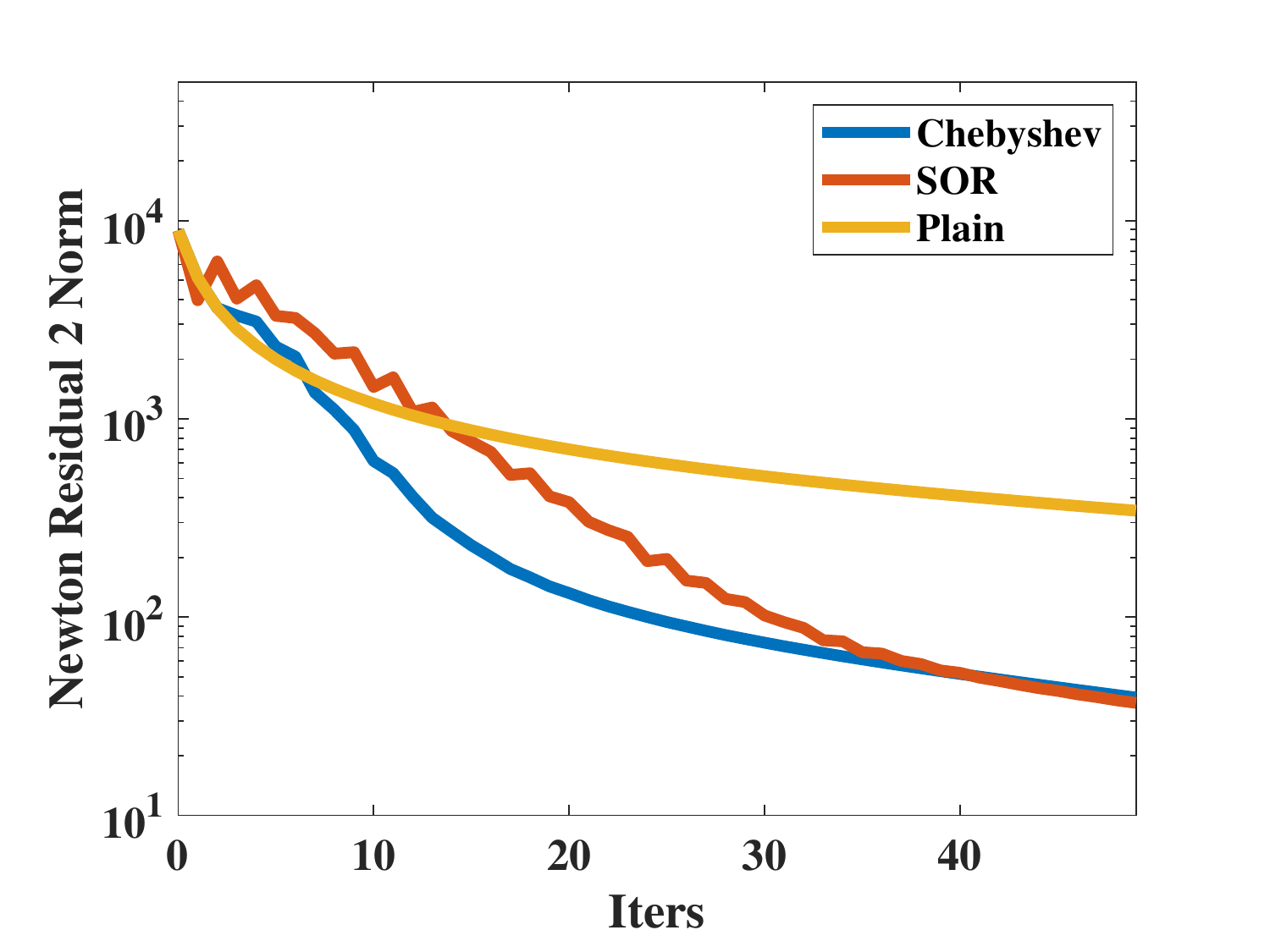}
	\caption{
			\textbf{Acceleration Techniques}. The convergence rate of PBNG may slow down as the iteration count increases. 
			Chebyshev semi-iterative method and SOR effectively accelerate the Newton residual reduction.
	}
	\label{fig:twist_sor}
\end{figure}
\subsection{Acceleration Techniques}
PBNG is remarkably stable and gives visually plausible results when the computational budget is limited, but it is also capable of producing numerically accurate results as the budget is increased.
However, as shown in Figure ~\ref{fig:twist_sor} and as with most Gauss-Seidel approaches, the convergence rate of PBNG may decrease with iteration count. 
We investigated two simple acceleration techniques to help mitigate this:
the Chebyshev semi-iterative method (as in \cite{wang:2015:cheby}) and SOR (as in \cite{fratarcangeli:2016:vivace}). 
The Chebyshev method uses the update
\begin{align}
	{\xx}^{n+1, l+1} &= \omega_{l+1} (\gamma (\xx^{n+1, l+1}_{\textrm{PBNG}} - {\xx}^{n+1, l}) + {\xx}^{n+1, l} - {\xx}^{n+1, l-1}) + {\xx}^{n+1, l-1}
\end{align}
where $\xx^{n+1, l+1}$ denotes the accelerated update and $\xx^{n+1, l+1}_{\textrm{PBNG}}$ denotes the standard PBNG update.
Here $\omega_{l+1}=\frac{4}{4- \rho^2\omega_l}$ for $l > 2$, $\frac{2}{2 - \rho^2}$ for $l=2$ and $1$ for $l<2$.
$\gamma$ is an under-relaxation parameter that stabilizes the algorithm. 
For our examples, we set $\rho = .95$. 
PBNG is very stable, and this allows for the use of over-relaxation as well. 
We set $\gamma = 1.7$.\\
\\
The SOR method uses a similar, but simpler update
\begin{align}
	{\xx}^{n+1, l+1} &= \omega (\xx^{n+1, l+1}_{\textrm{PBNG}} - {\xx}^{n+1, l-1}) + {\xx}^{n+1, l-1}.
\end{align}
We use $\omega=1.7$ for this under-relaxation parameter. 
As shown in Figure ~\ref{fig:twist_sor}, Chebyshev and SOR behave similarly in terms of residual reduction and visual appearance. 
%We illustrate PBNG with SOR in Algorithm ~\ref{alg:pbng}.
%
%\begin{algorithm}[ht]
%	\textbf{Initialize:} $tol = 10^{-8}$. \\
%	\While{not reached maximal iterations}{
%		\For{unconstrained particle $i_k$}{
%			\Begin(One-Step Quasi-Newton Solve){
%				1. Compute Newton residual on node $i_k$ via right hand side of Equation~\eqref{eq:pbgn_newt}\;
%				\If{Newton residual norm > $tol$}{
%					2. Compute the modified hessian using Equation~\ref{eq:mod_hess}\;
%					3. Compute $\Delta \xx^{n+1, l}_{(k+1), i_k}$ via Equation ~\ref{eq:pbgn_dx}\;
%					4. Update $\xx^{n+1, l}_{(k+1), i_k}$ with Equation ~\ref{eq:pbgn_update} \;
%				}
%			}
%		}
%		$\tilde{\xx}^{n+1, l+1} \gets \omega (\xx^{n+1, l+1} - \tilde{\xx}^{n+1, l-1}) + \tilde{\xx}^{n+1, l-1} $
%	}
%	\caption{PBNG with SOR Simulation Loop} 
%	\label{alg:pbng}
%\end{algorithm}  
%
%
%
\section{Lam\'e Coefficients}\label{sec:lame}
The parameters of an isotropic constitutive model are often based on Lam\'e coefficients $\mu$ and $\lambda$ which are themselves set from Young's modulus $E$ and Poisson's ratio $\nu$ according to Equation~\eqref{eq:lame}.
This relationship is based on the assumption of linear dependence of stress on strain, or quadratic potential energy density 
\begin{align}
\Psi^\textrm{le}(\FF)&=\mu\textrm{tr}(\eps^2(\FF))+\frac{\lambda}{2}\textrm{tr}(\eps(\FF))^2\label{eq:le}\\
\eps&=\frac{1}{2}(\FF+\FF^T)-\II.
\end{align}
Furthermore, Equation~\eqref{eq:lame} is derived from the model in Equation~\eqref{eq:le} by holding one end of a cuboidal domain fixed and applying a displacement at its opposite end. 
The remaining faces of the domain are assumed to be traction-free.
Young's modulus is the scaling in a linear relationship between the traction exerted by the material in resistance to the displacement.
The Poisson's ratio correlates with the degree of volume preservation via deformation in the directions orthogonal to the applied displacement.\\
\\
The use of Lam\'e coefficients with nonlinear models is not directly analogous since the relation between displacement and traction is not a linear scaling in the cuboid example.
When using Lam\'e coefficients with nonlinear problems, the cuboid derivation should hold if the model were linearized around $\FF=\II$.
All isotropic hyperelastic constitutive models can be written in terms of the isotropic invariants $I_\alpha:\mathbb{R}^{d\times d}\rightarrow\mathbb{R}$, $0\leq \alpha<d$
\begin{align}
I_0(\FF)&=\textrm{tr}(\FF^T\FF), \ I_1(\FF)=\textrm{tr}((\FF^T\FF)^2), \ I_2(\FF)=\det(\FF)\\
\Psi(\FF)&=\hat{\Psi}(I_0(\FF),I_1(\FF), I_2(\FF)).
\end{align}
See \cite{gonzalez:2008:continuum} for more detailed derivation. 
Note, when $d=2$, $I_1(\FF)=\textrm{tr}((\FF^T\FF)^2)$ is not used.
With this convention, the Hessian of the potential energy density is of the form
\begin{align}
\frac{\partial^2 \Psi}{\partial \FF^2}=\sum_{\alpha=0}^{d-1} \frac{\partial \hat{\Psi}}{\partial I_\alpha} \frac{\partial^2 I_\alpha}{\partial \FF^2}+\sum_{\alpha,\beta=0}^{d-1} \frac{\partial^2 \hat{\Psi}}{\partial I_\alpha\partial I_\beta} \frac{\partial I_\alpha}{\partial \FF}\otimes\frac{\partial I_\beta}{\partial \FF} \label{eq:inv_hess}.
\end{align}
If Lam\'e parameters are to be used with a nonlinear model, the Hessian $\frac{\partial^2 \Psi}{\partial \FF^2}(\FF)$ should match that of linear elasticity when evaluated at $\FF=\II$.
For example, this is why we adjust the Lam\'e parameters used in \cite{Macklin:2021:neohookean_xpbd} in Equation~\eqref{eq:nh}.
See the supplementary technical document for derivation details \cite{pbng:techdoc}.
\\
\\
We choose our approximate Hessian in Equation~\eqref{eq:mod_hess_dens} based on this fact.
That is, by omitting all but the first and last terms in Equation~\eqref{eq:inv_hess}, our approximate Hessian is both symmetric positive definite and consistent with any model that is set from Lam\'e coefficients (e.g. from Young's modulus and Poisson's ratio)
\begin{align}
\tilde{\mathbf{\mathcal{C}}}=\mu \frac{\partial^2 I_0}{\partial \FF^2}+\lambda \frac{\partial I_{d-1}}{\partial \FF}\otimes\frac{\partial I_{d-1}}{\partial \FF}.
\end{align}
Again, see the supplementary technical document for more details \cite{pbng:techdoc}.
\begin{figure}[h]
	\begin{tikzpicture}
		\node [anchor=south west, inner sep=0pt] (image2) at (0,0) {
			\includegraphics[draft=\mydraft,width=0.48\columnwidth,trim={400px 200px 400px 200px},clip]{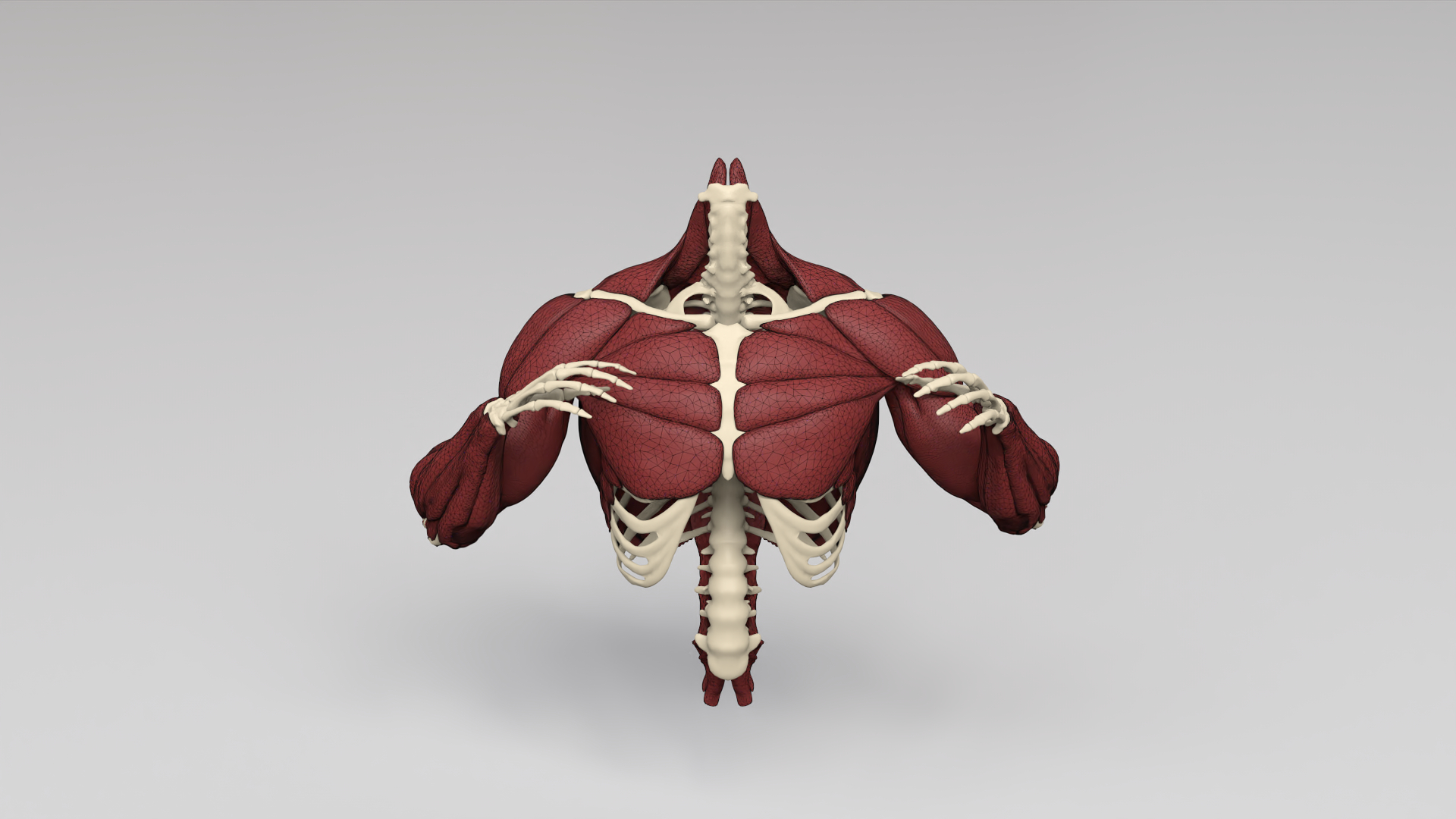}
		};
		\node [anchor=south west, inner sep=1pt] at (image2.south west) {Frame 387};
		\node [anchor=north east, inner sep=1pt, font =\tiny] at (image2.north east) {\textbf{\textcopyright 2023 Epic Games, Inc}};
	\end{tikzpicture}
	\begin{tikzpicture}
		\node [anchor=south west, inner sep=0pt] (image3) at (0,0) {
			\includegraphics[draft=\mydraft,width=0.48\columnwidth,trim={400px 200px 400px 200px},clip]{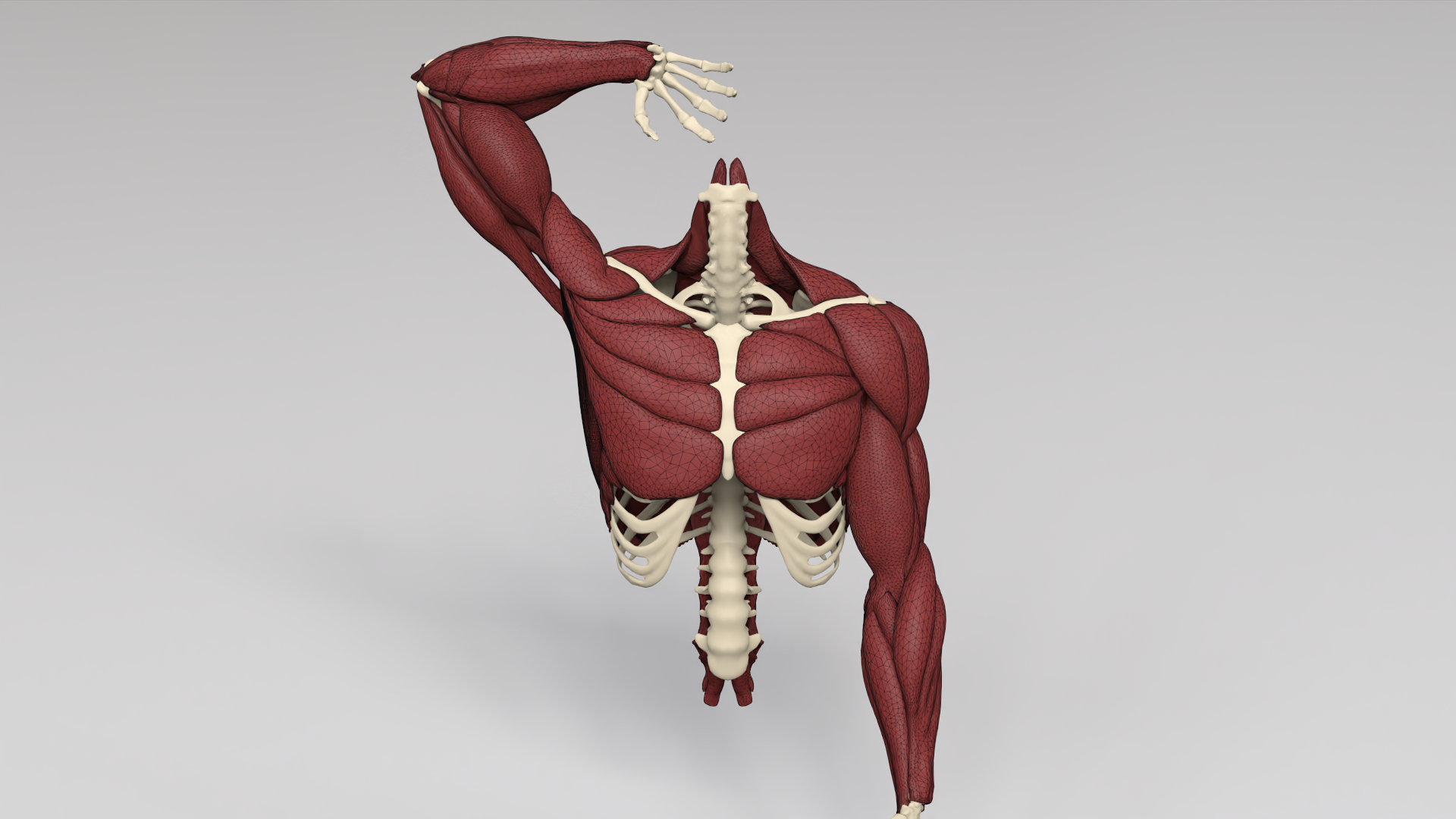}
		};
		\node [anchor=south west, inner sep=1pt] at (image3.south west) {Frame 650};
		\node [anchor=north east, inner sep=1pt, font =\tiny] at (image3.north east) {\textbf{\textcopyright 2023 Epic Games, Inc}};
	\end{tikzpicture}

	\begin{tikzpicture}
		\node [anchor=south west, inner sep=0pt] (image5) at (0,0) {
			\includegraphics[draft=\mydraft,width=0.48\columnwidth,trim={0 0 0 0},clip]{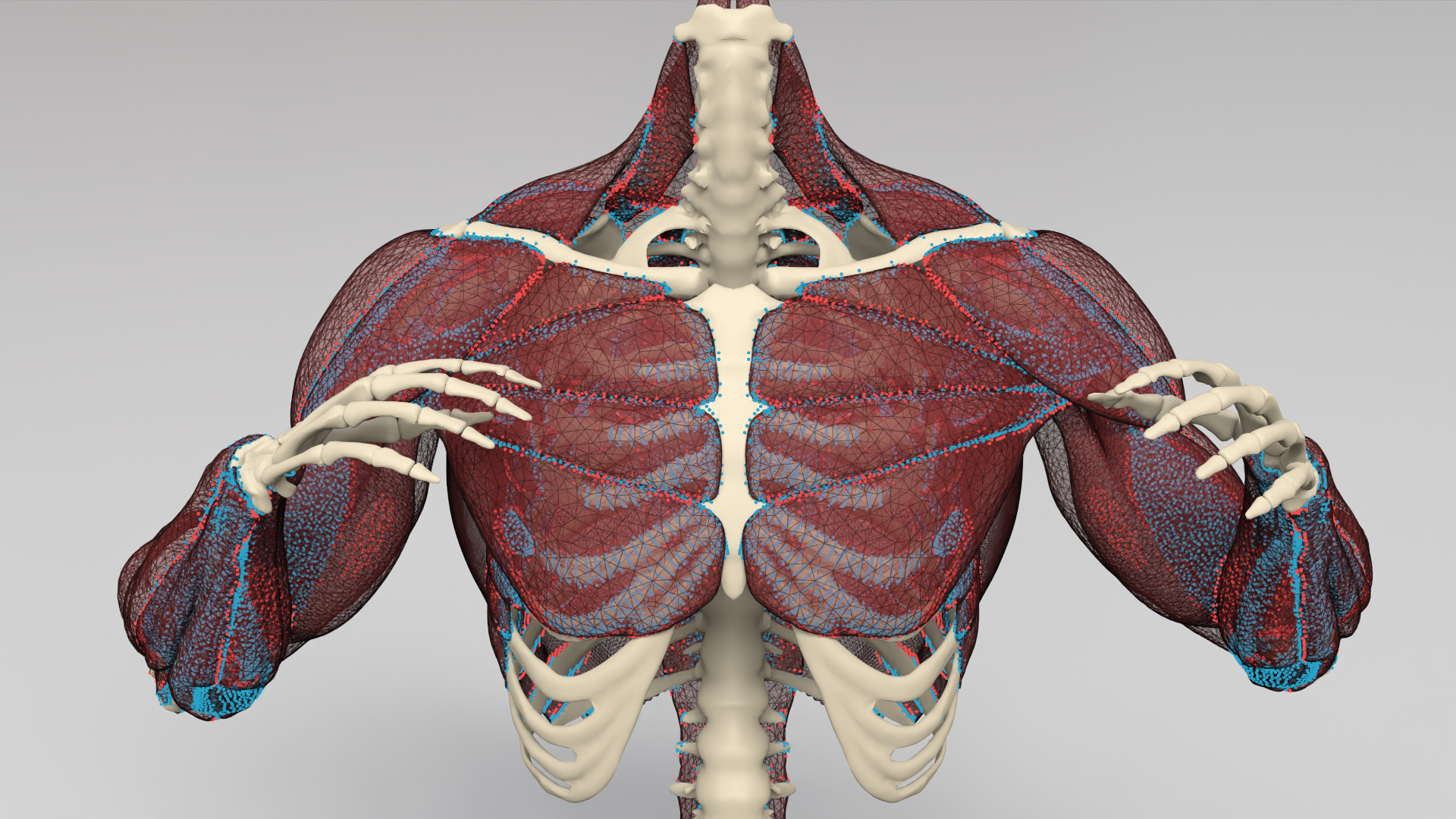}
		};
		\node [anchor=south west, inner sep=1pt] at (image5.south west) {Frame 387};
		\node [anchor=north east, inner sep=1pt, font =\tiny] at (image5.north east) {\textbf{\textcopyright 2023 Epic Games, Inc}};
	\end{tikzpicture}
	\begin{tikzpicture}
		\node [anchor=south west, inner sep=0pt] (image6) at (0,0) {
			\includegraphics[draft=\mydraft,width=0.48\columnwidth,trim={0 0 0 0},clip]{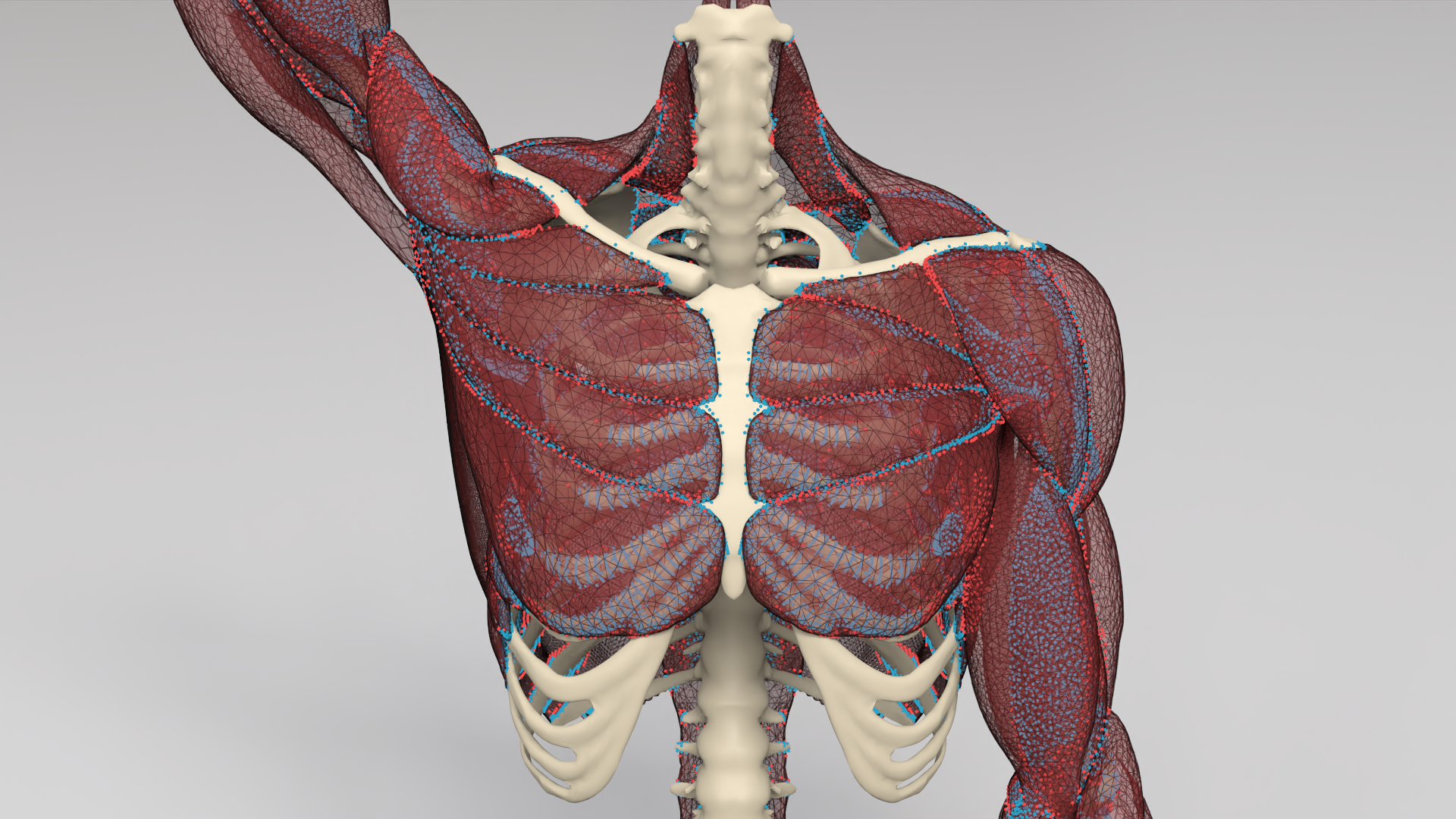}
		};
		\node [anchor=south west, inner sep=1pt] at (image6.south west) {Frame 650};
		\node [anchor=north east, inner sep=1pt, font =\tiny] at (image6.north east) {\textbf{\textcopyright 2023 Epic Games, Inc}};
	\end{tikzpicture}
	\caption{\textbf{PBNG Muscle Simulation}. 
	The top row shows simulation results while the bottom row visualizes the vertex constraint status.
	\textit{Red} indicates a vertex involved in contact, weak constraints are dynamically built to resolve the collisions. 
  \textit{Blue} represents the vertex positions of connective tissue bindings.
	}
	\label{fig:pbgn_muscle}
\end{figure}
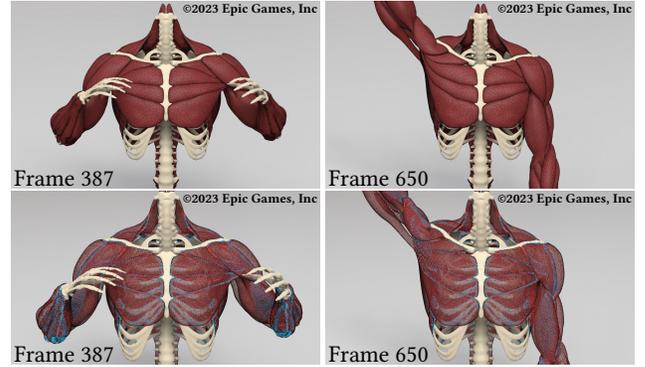
\section{Coloring and Parallelism}\label{sec:coloring}
Parallel implementation of Gauss-Seidel techniques is complicated by data dependencies in the updates.
This can be alleviated by careful ordering of sub-iterate position updates.
We provide simple color-based orderings for both PBD and PBNG techniques.
For PBD, colors are assigned to constraints so that those in the same color do not share incident nodes. 
Constraints in the same color can then be solved at the same time with no race conditions. 
For each vertex $\xx_i$ in the mesh, we maintain a set $S_{\xx_i}$ that stores the colors used by its incident constraints.
For each constraint $c$, we find the minimal color as the least integer that is not contained in the set $\cup_{\xx_{i} \in c} S_{\xx_{i}}$. 
We then register the color by adding it into $S_{\xx_{i}}$ for each $\xx_{i}$ in constraint $c$.
With PBNG, we color the nodes so that those in the same color do not share any mesh element or weak constraint. 
For each element or weak constraint $c$, we maintain a set $S_c$ that stores the colors used by its incident nodes. 
For a position $\xx_i$, we compute its color as the minimal one not contained in the set $\cup_{\xx_{i} \in c} S_{c}$. 
Then we register the color by adding it into $S_c$ for each element or weak constraint $\xx_i$ is incident to. 
The coloring process is illustrated in Figures~\ref{fig:constraints_coloring}(b) and \ref{fig:constraints_coloring}(c).
We observe that coloring the nodes instead of the constraints gives fewer colors. 
This makes simulations run faster since more work can be done without race conditions. 
In Table ~\ref{tbl:colors_comparison}, we demonstrate this performance observation. 
Note that we use the omp parallel directive from Intel's OpenMP library for parallelizing the updates.  

\begin{table*}[t]
	
	\resizebox{\textwidth}{!}{%
		\begin{tabular}{@{}lllllll@{}}
			\toprule
			Example & \# Vertices & \# Elements. & \# Particle Colors & \# Constraint Colors & PBNG Runtime/Iter & PBD Runtime/Iter \\ \midrule
			Res 32 Box Stretching  & 32K & 150K & 5 & 39 & 28ms  & 26.8ms\\
			Muscles Without Collisions & 284k & 1097K & 13 & 82 & 131ms & 140ms \\
			Res 64 Box Stretching & 260K & 1250K & 5 & 39 & 65ms & 137ms\\
			Res 128 Box Stretching & 2097K & 10242K & 5 & 40 & 1520ms & 1080ms\\
			Dropping Simple Shapes Into Box & 256K & 1069K & 11 & 52 & 270ms & 140ms\\
			Res 16 Box Dropping & 4.1K & 17K & 5 & 39 & 3.6ms & 4.1ms\\
			\bottomrule
		\end{tabular}
	}
	\caption{Number of Colors Comparison: runtime is measured per iteration (averaged over the first 200 iterations). PBNG does more work per-iteration than PBD, but has comparable speed due to improved scaling resulting from a smaller number of colors. }
	\label{tbl:colors_comparison}
\end{table*}

\begin{table*}[t]

	\resizebox{\textwidth}{!}{%
	\begin{tabular}{@{}lllllllll@{}}
	\toprule
	Example & \# Vertices & \# Elements. & PBNG Runtime & Newton Runtime & PBD Runtime  & PBNG \# iter & PBD \# iter & Newton \# iter \\ \midrule
	Box Stretching (low budget) & 32K & 150K & 170ms & 170ms & 170ms & 6 & 6 & 2 (7 CGs) \\
	Box Stretching (big budget) & 32K & 150K & 1.3s & 1.3s & 1.3s & 40 & 40 & 11 (10 CGs) \\
	Muscle with collisions & 284k & 1097K & 67s & 430s & - & 510 & - & 34 (200CGs) \\

\bottomrule
\end{tabular}
}
\caption{Methods Comparisons: We show runtime per frame for different methods for some of the examples. Each frame is run after advancing time .033. }
\label{tbl:method_comp}
\end{table*}
\subsection{Collision Coloring}\label{sec:collision_coloring}
For simulations with static weak constraints, the coloring is a one-time cost. 
Otherwise, the colors have to be updated every time the weak constraint structure changes, e.g. from self-collision (Figures~\ref{fig:pbgn_muscle} and \ref{fig:two_blocks_colliding}).
We propose a simple coloring scheme for this purpose: only nodes incident to the newly added weak constraints need recoloring. 
We first compute all nodes $\xx_i^{\textrm{extra}}$ that are incident to newly added weak constraints. 
For each $\xx_i^{\textrm{extra}}$, we compute the used color set $\cup_{\xx_{i}^{\textrm{extra}} \in c} S_{c}$. 
We use the color of $\xx_{i}^{\textrm{extra}}$ from the previous time step as an initial guess. 
If it already exists in the used color set, then we find the minimal color that is not used. 
This is generally of moderate cost, e.g. in the muscle examples with collisions (Figures~\ref{fig:teaser},  \ref{fig:pbgn_vs_xpbd_muscle} and \ref{fig:pbgn_muscle}), our algorithm takes less than 680ms/frame for recoloring, while the actual simulation takes a total of 67s to run.     

%  \begin{figure}[h]
% 	\includegraphics[width = \linewidth]{figures/constraints_coloring.pdf}
% 	\caption{
% 		{\textbf{Constraints Coloring}}. A step-by-step constraints mesh coloring scheme is shown. The dotted line indicates two weak constraints between the elements. The first constraint is colored red, all its incident points will register red as used color. Other constraints incident to the first constraint have to choose other colors.
% 	}
% 	\label{fig:constraints_coloring}
% \end{figure} 

%  \begin{figure}[h]
% 	\includegraphics[width = \linewidth]{figures/particle_coloring.pdf}
% 	\caption{
% 		{\textbf{Particles Coloring}}. A step-by-step particles coloring scheme is shown. The constraints register the colors used by its incident particles. The first particle is colored red, so all its incident constraints will register red as used. Other particles incident to the constraints have to choose other colors.
% 	}
% 	\label{fig:particle_coloring}
% \end{figure} 

%  \begin{figure}[h]
% 	\includegraphics[width = \linewidth]{figures/dual_coloring.pdf}
% 	\caption{
% 		{\textbf{Dual Coloring}}.\textbf{TODO}
% 	}
% 	\label{fig:dual_coloring}
% \end{figure} 

\section{Examples}

We demonstrate the versatility and robustness of PBNG with a number of representative simulations of quasistatic (and dynamic) hyperelasticity. 
Examples run with the corotated model (Equation~\eqref{eq:cor}) use the algorithm from \cite{gast:2016:svd} for its accuracy and efficiency. 
All the examples use Poisson's ratio $\nu=0.3$. 
We compare PBNG, PBD, XPBD, XPBD-QS and XPBD-QS (Flipped). 
For XPBD-QS we do the hyperelastic constraints first, followed by weak constraints. 
For XPBD-QS (Flipped) the order is swapped. 
All the examples were run on an AMD Ryzen Threadripper PRO 3995WX CPU with 64 cores and 128 threads.
In Table ~\ref{tbl:perf}, we provide comprehensive performance statistics for PBNG. 
In Table ~\ref{tbl:method_comp}, we provide runtime comparisons between PBNG and the other methods. 
\begin{figure}[h]
	\includegraphics[draft=\mydraft,width=0.32\columnwidth,trim={180px 0px 320px 0px},clip]{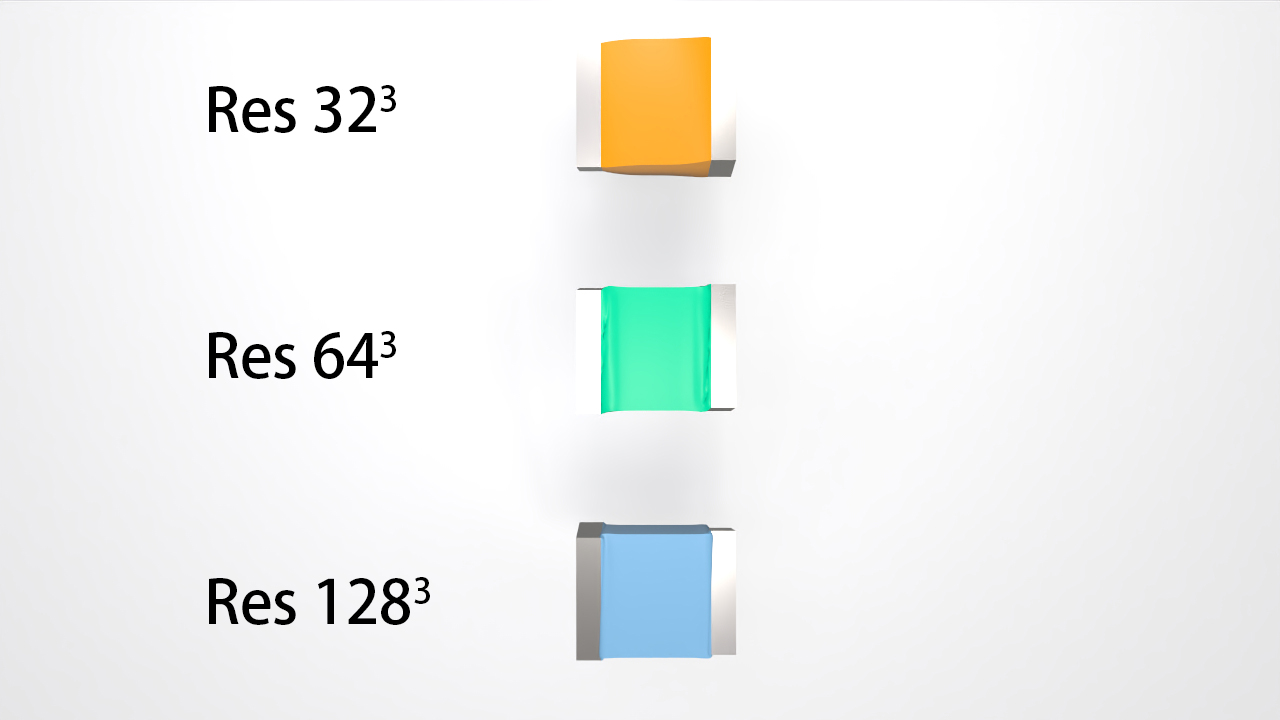}
	\includegraphics[draft=\mydraft,width=0.32\columnwidth,trim={180px 0px 320px 0px},clip]{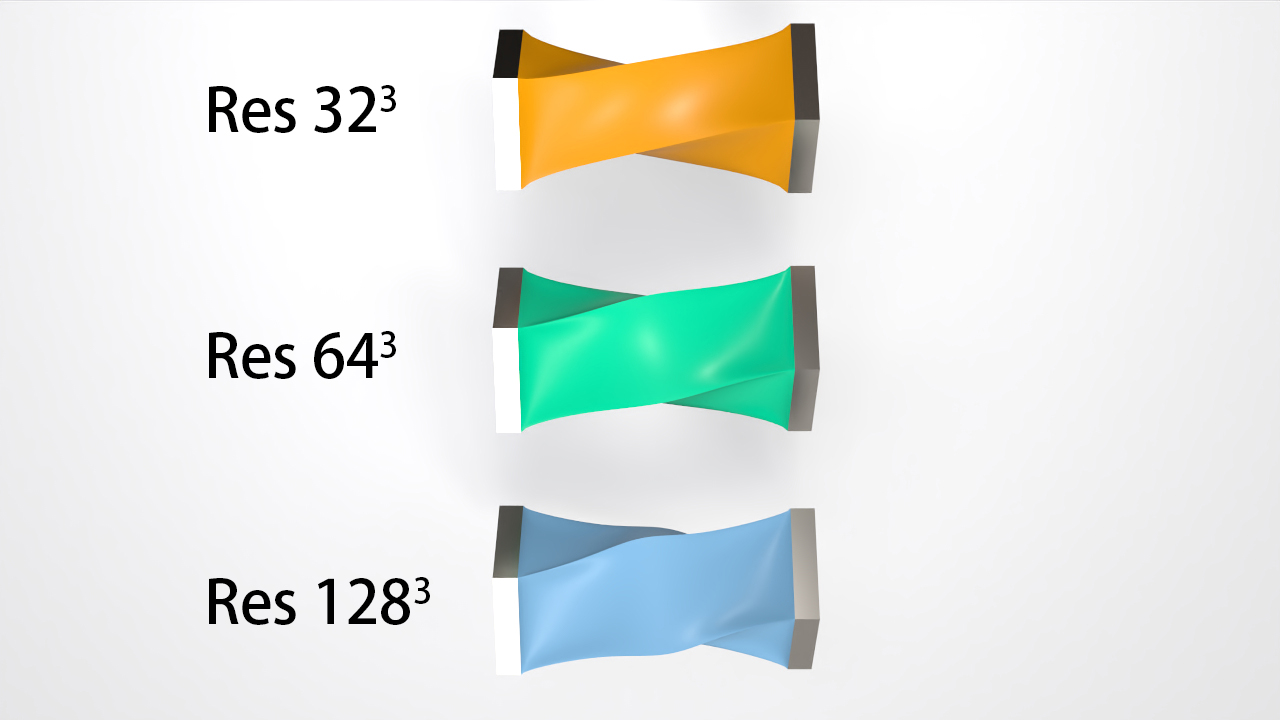}
	\includegraphics[draft=\mydraft,width=0.32\columnwidth,trim={180px 0px 320px 0px},clip]{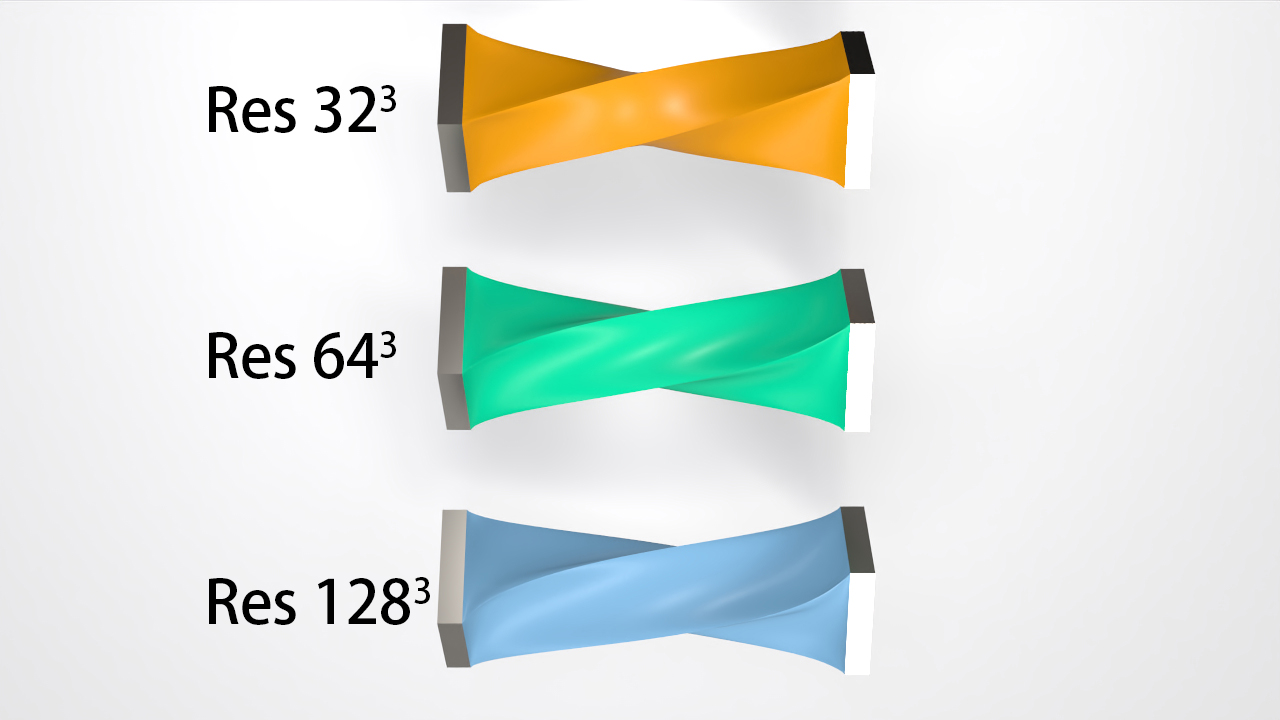}
	\caption{\textbf{Different Mesh Resolution}. PBNG produces consistent results when the mesh is spatially refined. 
	The highest resolution mesh in this comparison has over 2M vertices and only requires 40 iterations to produce visually plausible results.}
	\label{fig:twist_different_res}
\end{figure}
\subsection{Stretching Block}
We stretch and twist a simple block in a simple scenario. 
The block has 32K particles and 150K elements. 
Both ends of the block are clamped. 
They are stretched, squeezed and twisted in opposite directions. 
The block has $R^0=10$ and Young's modulus $E=10^5$. 
There is no gravity. 
The simulation is quasistatic. 
We compare performance between Newton's method, PBD, PBNG and XPBD as described in Section ~\ref{sec:pbd}. 
In Figure ~\ref{fig:method_comparison}, these methods are run under a fixed budget. 
Every method has a runtime of 1.3s/frame. 
With an ample budget, PBNG converges to ground truth, while PBD and XPBD do not.
In Figure ~\ref{fig:method_comparison}, we show a simulation where every method has a runtime of 170ms/frame. 
Newton's method is remarkably unstable. 
PBNG looks visually plausible. 
PBD and XPBD-QS have visual artifacts and fail to converge. 
Residual plots vs. time are shown at the bottom of Figure ~\ref{fig:method_comparison}. 

\subsubsection{Different Resolution}
In this example, we demonstrate PBNG's versatility by running the block stretching and twisting with various resolutions. 
As shown in Figure ~\ref{fig:twist_different_res}, the top block has 32K particles and 150K elements. 
The middle block has 260K particles and 1250K elements. 
The bottom block has 2097K particles and 10242K elements. 
Even at high-resolution (bottom block), PBNG is visually plausible after only 40 iterations and 61 seconds/frame of runtime. 

\subsubsection{Different Constitutive Models}
In this example, we apply PBNG to various constitutive models on the same block examples. 
All three blocks have 32K particles and 150K elements. 
Frames are shown in Figure ~\ref{fig:twist_different_models}. 
The blocks from top to bottom are run with corotated (Equation~\ref{eq:cor}), stable Neo-Hookean (Equation~\ref{eq:snh}) and Neo-Hookean (Equation~\ref{eq:nh}) models respectively. 
With 40 iterations per frame, they are all visually plausible.  

\subsubsection{Acceleration Comparison}
In this example, we compare the effects of the Chebyshev semi-iterative method and the SOR method. 
In Figure ~\ref{fig:twist_sor}, we stretch and twist the same block with 32K particles and 150K elements. 
The top bar is simulated with plain PBNG. 
The middle bar is simulated with PBNG with Chebyshev semi-iterative method with $\gamma = 1.7, \rho = .95$. 
The bottom bar is simulated with PBNG with SOR and $\omega = 1.7$. 
10 iterations are used for each time step. 
With a limited budget, plain PBNG is less converged than accelerated techniques.
Figure ~\ref{fig:twist_sor} shows the convergence rate of the three methods vs. the number of iterations at the first time step. 
We can see that the acceleration techniques boost the convergence rate. 

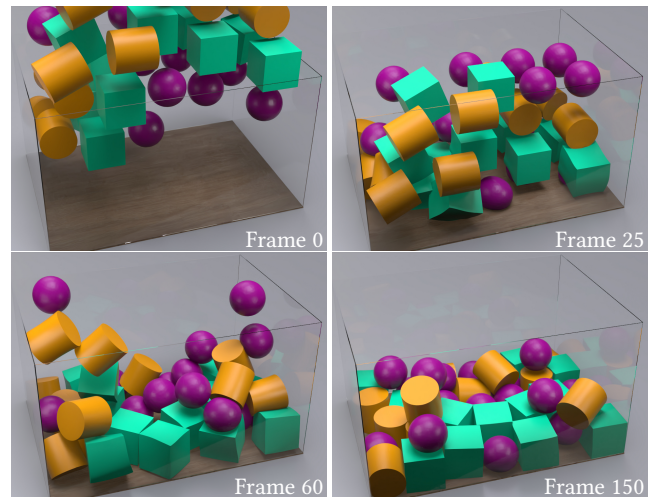
\begin{figure}[h]
	\begin{tikzpicture}
		\node [anchor=south west, inner sep=0pt] (image1) at (0,0) {\includegraphics[draft=\mydraft,width=0.49\columnwidth,trim={200px 0px 150px 0px},clip]{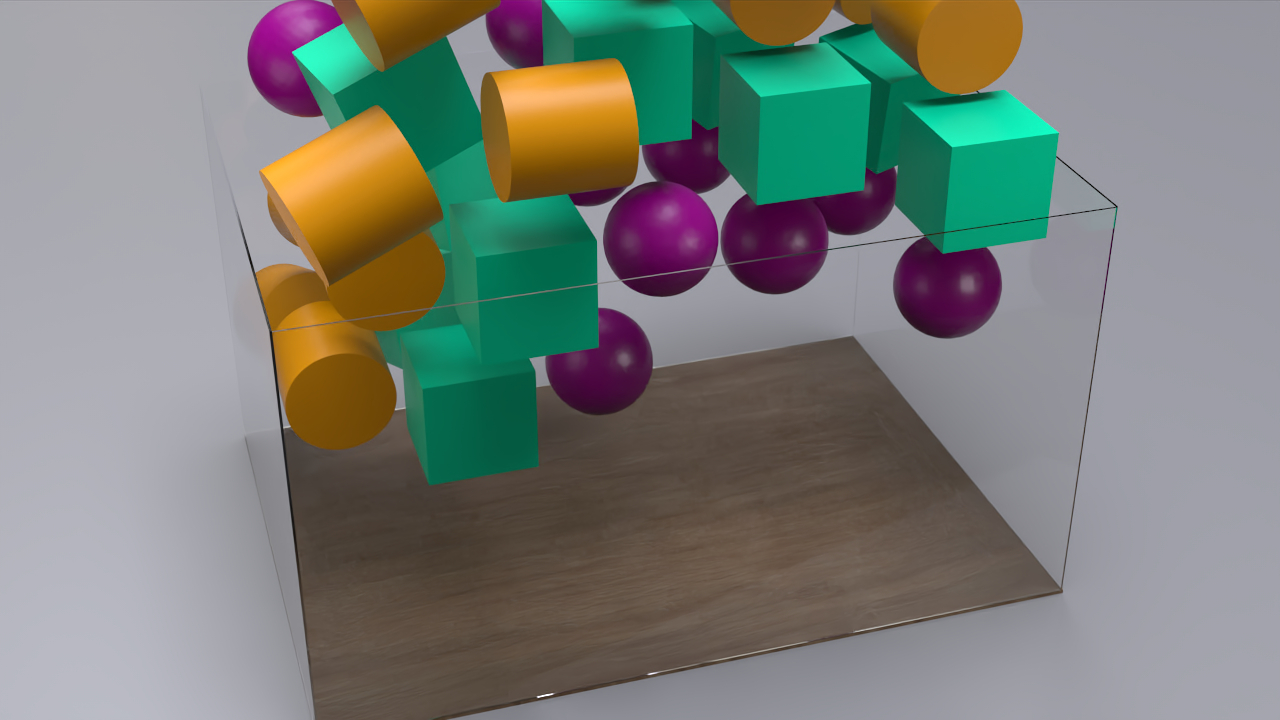}};
		\node [anchor=south east, inner sep=1pt, text=white] at (image1.south east) {Frame 0};
	\end{tikzpicture}
	\begin{tikzpicture}
		\node [anchor=south west, inner sep=0pt] (image2) at (0,0) {\includegraphics[draft=\mydraft,width=0.49\columnwidth,trim={200px 0px 150px 0px},clip]{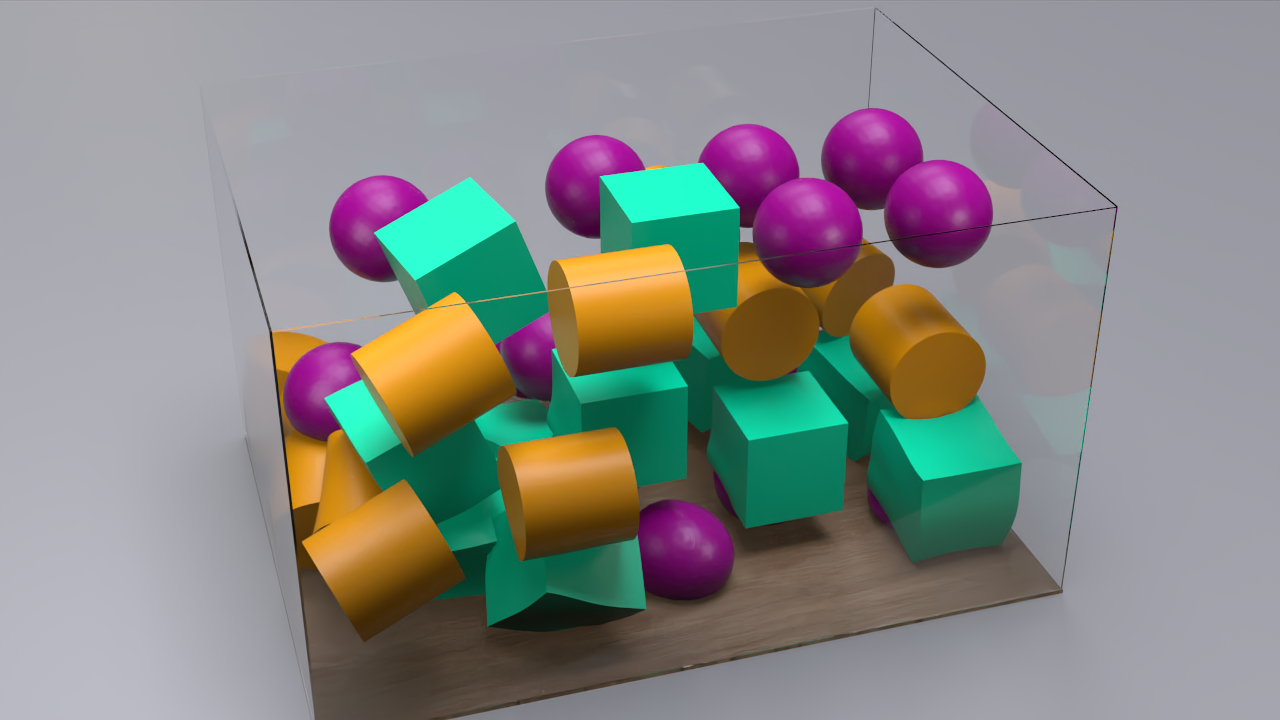}};
		\node [anchor=south east, inner sep=1pt, text=white] at (image2.south east) {Frame 25};
	\end{tikzpicture}
	\begin{tikzpicture}
		\node [anchor=south west, inner sep=0pt] (image3) at (0,0) {\includegraphics[draft=\mydraft,width=0.49\columnwidth,trim={200px 0px 150px 0px},clip]{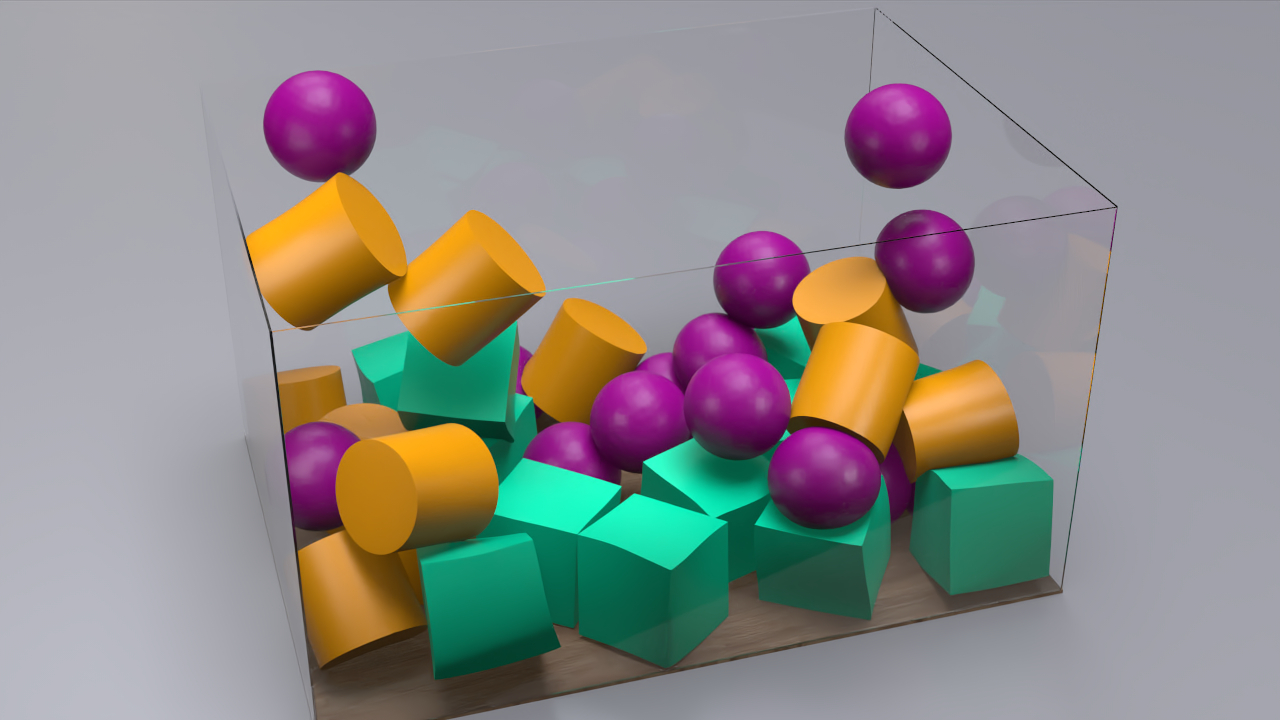}};
		\node [anchor=south east, inner sep=1pt, text=white] at (image3.south east) {Frame 60};
	\end{tikzpicture}
	\begin{tikzpicture}
		\node [anchor=south west, inner sep=0pt] (image4) at (0,0) {\includegraphics[draft=\mydraft,width=0.49\columnwidth,trim={200px 0px 150px 0px},clip]{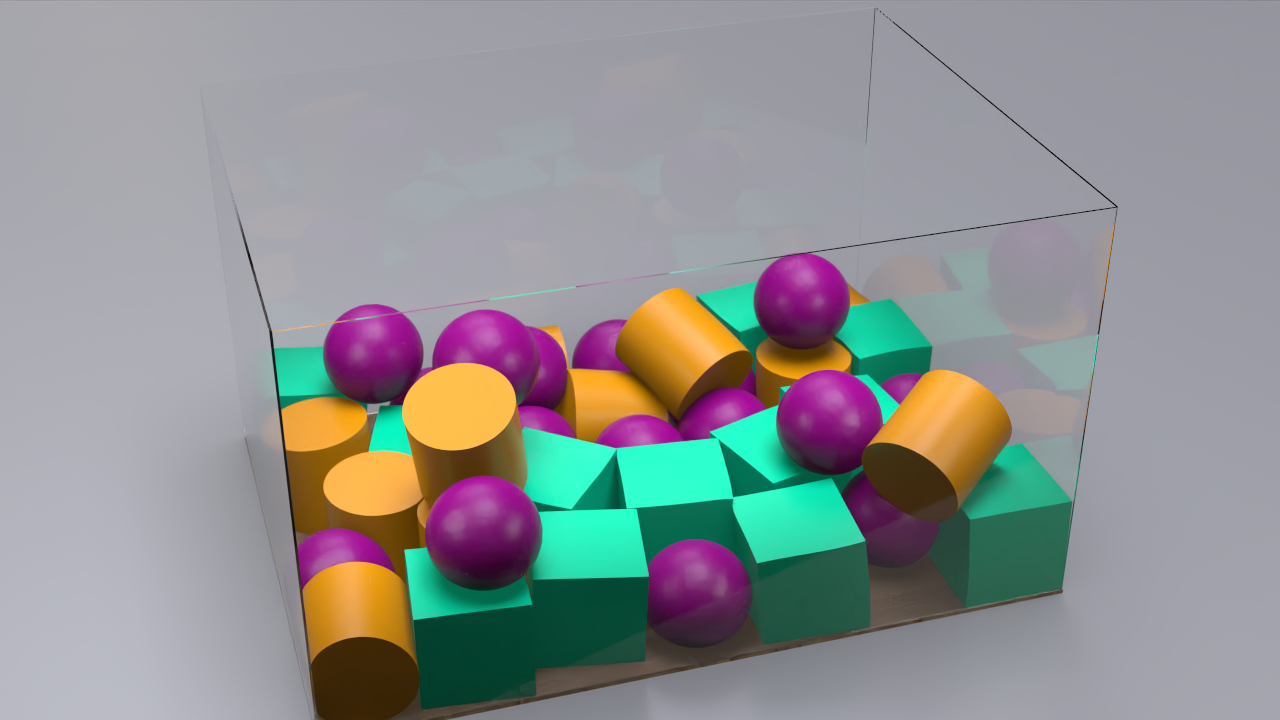}};
		\node [anchor=south east, inner sep=1pt, text=white] at (image4.south east) {Frame 150};
	\end{tikzpicture}	
	\caption{\textbf{Objects Dropping}. A variety of objects drop under gravity. Our method is able to robustly handle collisions between deformable objects through weak constraints.}
	\label{fig:objects_dropping}
\end{figure}
\subsection{Collisions}\label{sec:col}
We support collisions by dynamically adding  weak constraints as discussed in Section ~\ref{sec:wc}. 
We use a time step of $\Delta t=.002$ and detect collision every time step.

\subsubsection{Two Blocks Colliding}
\begin{figure}[h]
	\begin{tikzpicture}
		\node [anchor=south west, inner sep=0pt] (image1) at (0,0) {
			\includegraphics[draft=\mydraft,width=0.49\columnwidth,trim={300px 100px 300px 100px},clip]{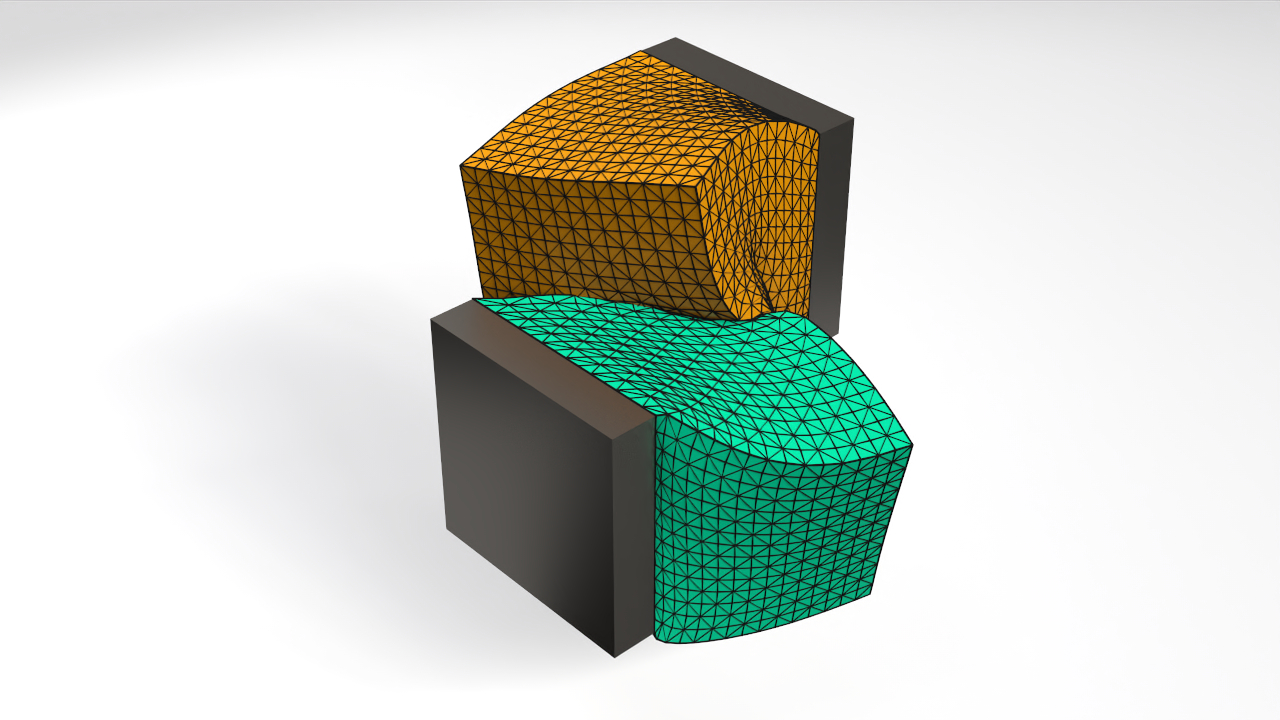}
		};
		\node [anchor=south west, inner sep=1pt] at (image1.south west) {Frame 9};
	\end{tikzpicture}
	\begin{tikzpicture}
		\node [anchor=south west, inner sep=0pt] (image2) at (0,0) {
			\includegraphics[draft=\mydraft,width=0.49\columnwidth,trim={300px 100px 300px 100px},clip]{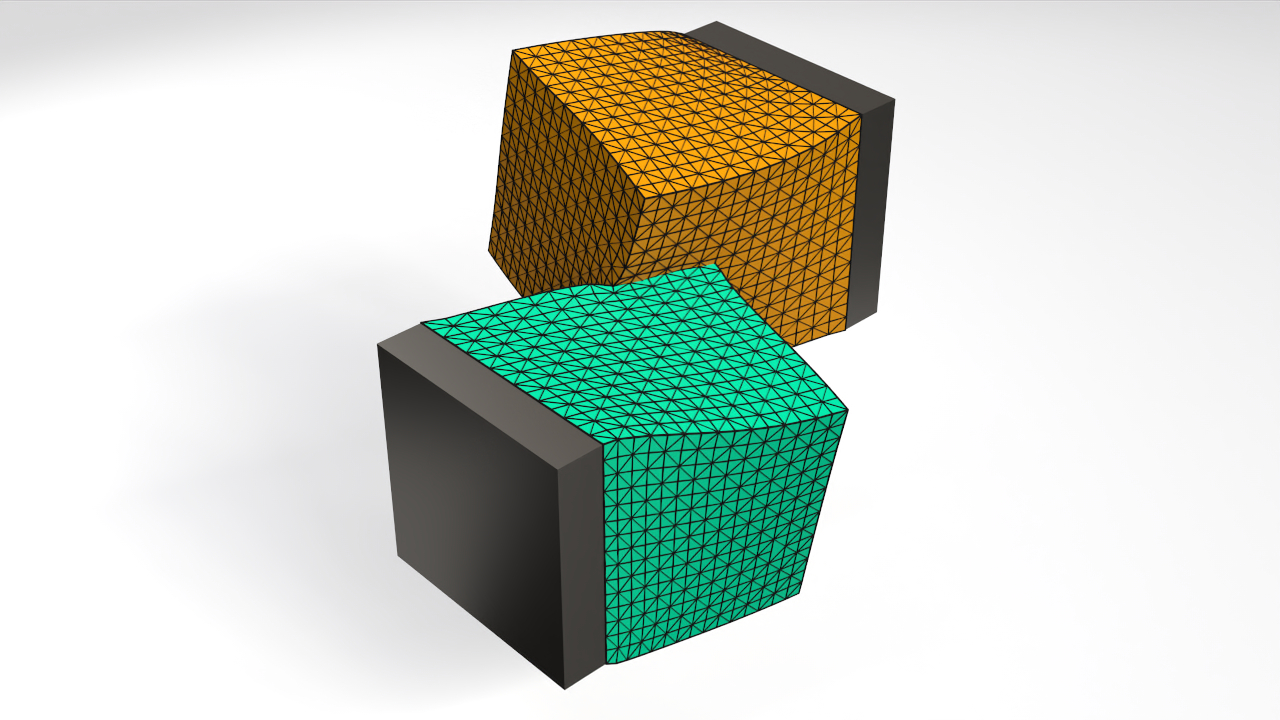}
		};
		\node [anchor=south west, inner sep=1pt] at (image2.south west) {Frame 25};
	\end{tikzpicture}
	\includegraphics[draft=\mydraft,width=0.49\columnwidth,trim={0px 0px 0px 0px},clip]{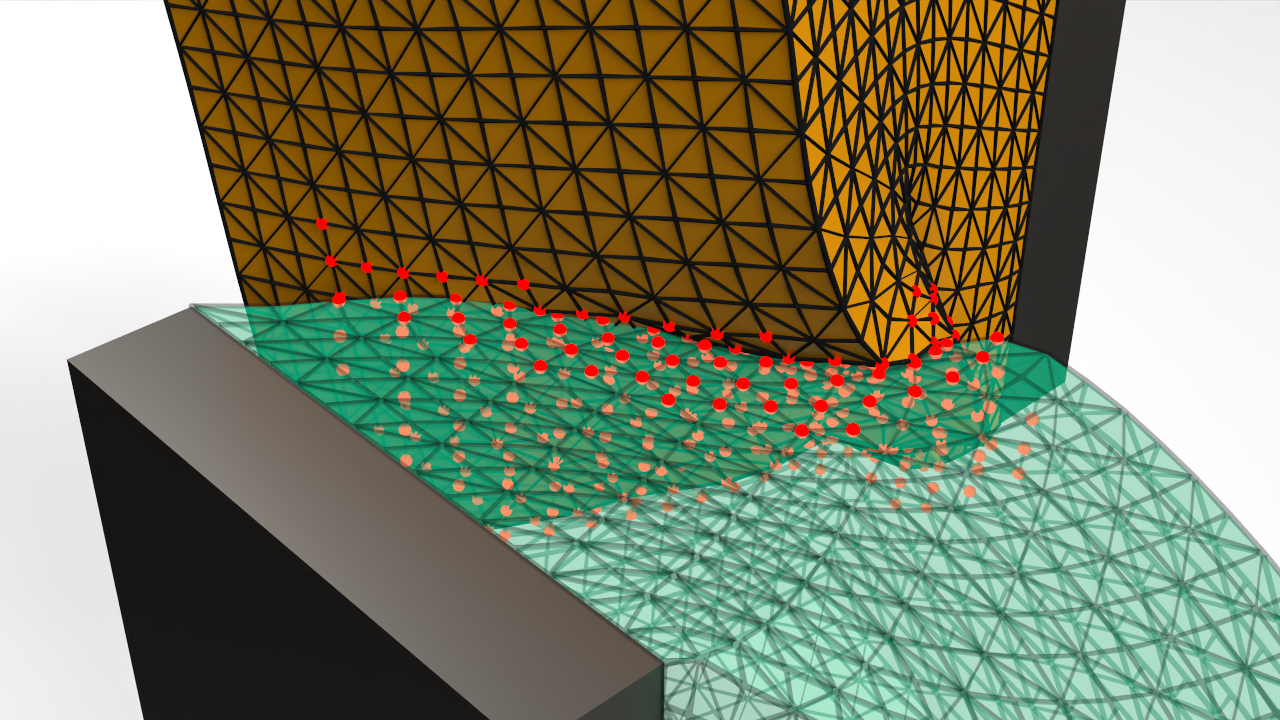}
	\includegraphics[draft=\mydraft,width=0.49\columnwidth,trim={0px 0px 0px 0px},clip]{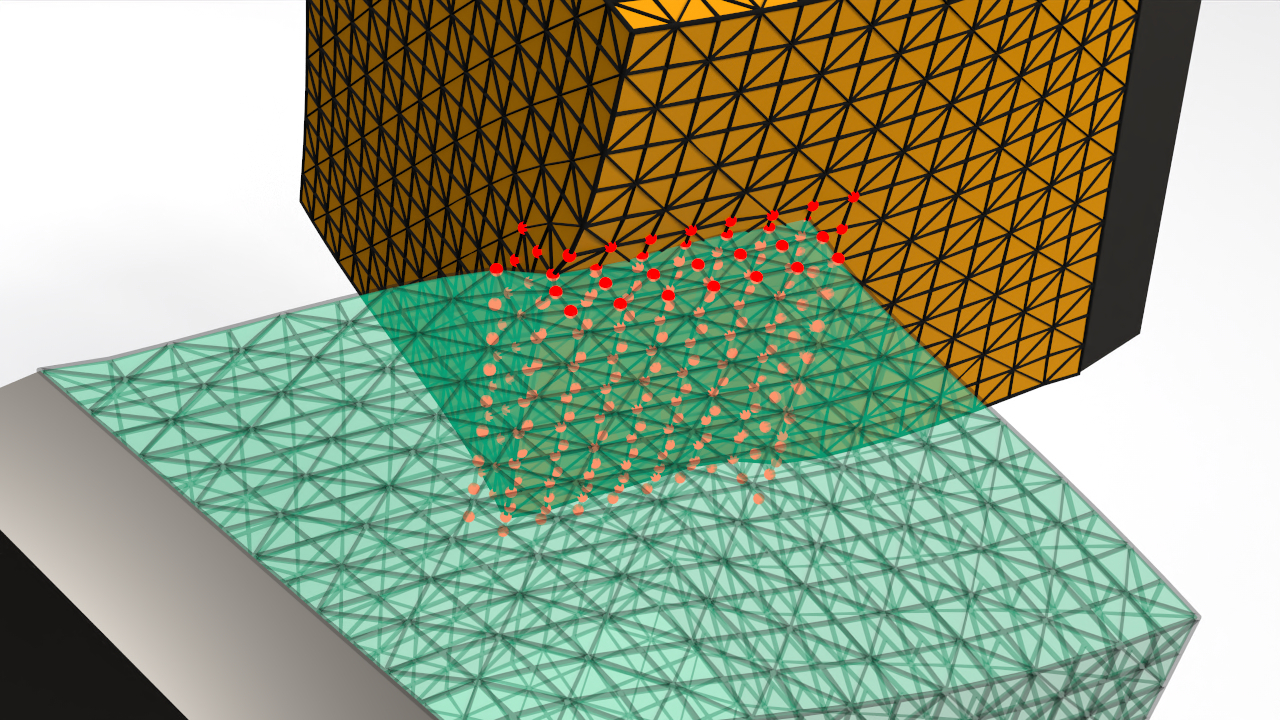}
	\caption{\textbf{Two Blocks Colliding}. Two blocks collide with each other with one face clamped. 
	Red particles indicate that dynamic weak constraints have been built to resolve the collision of corresponding mesh vertices.}
	\label{fig:two_blocks_colliding}
\end{figure}

We demonstrate the generation of dynamic weak constraints with a simple example. 
We take two blocks with one side fixed and drive them toward each other. 
This is a dynamic/backward Euler simulation. 
The blocks have $R^0=10$ and  Young's modulus $E=1000$. The weak constraints have stiffness $k_n=10^8$ and $k_\tau=0$.
The dynamic weak constraints are visualized in Figure ~\ref{fig:two_blocks_colliding} as red nodes in the mesh. 

\subsubsection{Muscles}
We quasistatically simulate a large-scale musculature with collision and connective tissue weak constraints. 
The mesh has a total of 284K particles and 1097K elements. 
The muscles have $R^0=1000$, Young's modulus $E=10^5$, connective tissue (blue) weak constraint stiffness is isotropic $k_n=k_\tau=10^8$. 
Dynamic collision (red) weak constraint stiffness is anisotropic $k_n=10^8$ and $k_\tau=0$.
We show several frames of muscles simulated with PBNG and dynamically generated weak constraints in Figure ~\ref{fig:pbgn_muscle}.
PBNG takes 67 seconds to simulate a frame, while Newton's method takes 430s. 
In figure ~\ref{fig:teaser}, we show that PBNG looks visually the same as Newton, while running 6-7 times faster. 
We also show that PBD and XPBD-QS fail to converge. 
In Figure ~\ref{fig:teaser}, we show PBD becomes unstable. 
In Figure ~\ref{fig:pbgn_vs_xpbd_muscle}, we demonstrate sub-iteration order-dependent behavior with PBD.
XPBD-QS has weak constraints processed last, which leads to excessive stretching of elements.
XPBD-QS (Flipped) has weak constraints processed first, which degrades their enforcement and leaves a gap. 

\subsubsection{Dropping Objects}
40 objects with simple shapes are dropped into a glass box. 
The objects have a total of 256K particles and 1069K elements. 
The simulation is run with dynamic/backward Euler. 
Some frames are shown in Figure ~\ref{fig:objects_dropping}. 
We show PBNG's capability of handling collision-intensive scenarios. 
The example is run with $R^0=10$, Young's modulus $E= 3000$ and weak constraint stiffness $k_n=10^8$ and $k_\tau = 0$.

\subsection{Varying Stiffness}
In this example, we demonstrate that XPBD-QS fails to resolve quasistatic problems with varied stiffness. 
In Figure ~\ref{fig:two_blocks_hanging}, we show the initial setup for the simulation. 
The simulation is quasistatic. 
Both block meshes have $R^0=10$ and Young's modulus $E=1000$. 
The first block mesh has its top boundary constrained. 
The second block is weakly constrained to the first block via weak constraints between them. 
The springs have stiffness $k_n=k_\tau=10^8$. 
There is gravity in the scene with acceleration $-9.8$ in the $y-$direction. 
As we show in Figure ~\ref{fig:two_blocks_hanging}, PBNG converges to a plausible state. 
XPBD-QS and XPBD-QS (Flipped) fail to converge. 
Depending on the order of the constraints, it either leaves a gap between the two blocks or a very stretched top layer of the bottom block. 
This example also serves as a simplified version of the connective bindings on the muscles, which are used in Figure ~\ref{fig:pbgn_vs_xpbd_muscle}. 
The residual plot is shown on the right of Figure ~\ref{fig:two_blocks_hanging}.

\begin{figure}[h]
	\includegraphics[draft=\mydraft,width=0.6\columnwidth,trim={0px 0px 0px 50px},clip]{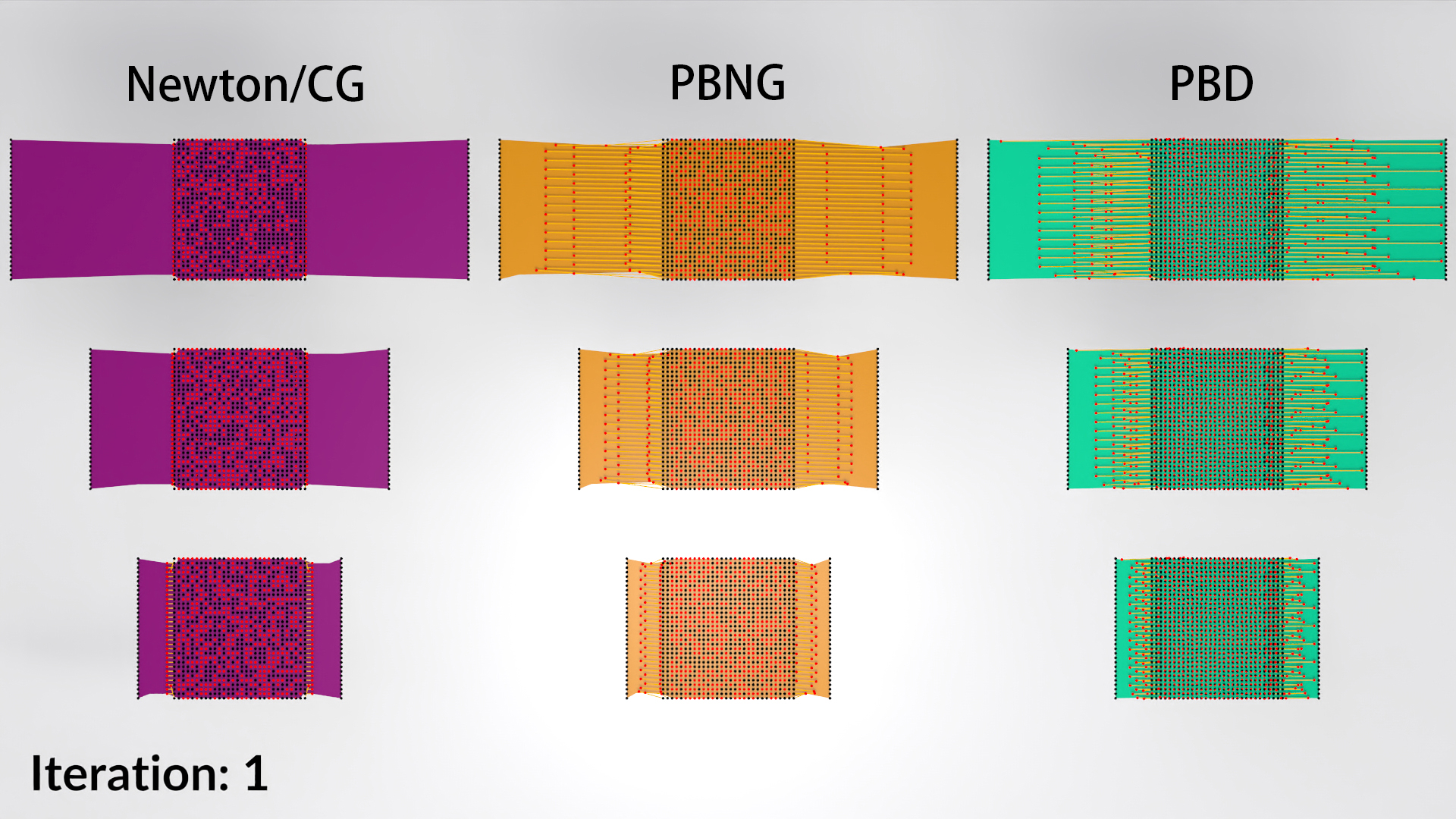}
	\includegraphics[draft=\mydraft,width=0.39\columnwidth,trim={5px 0px 35px 0px},clip]{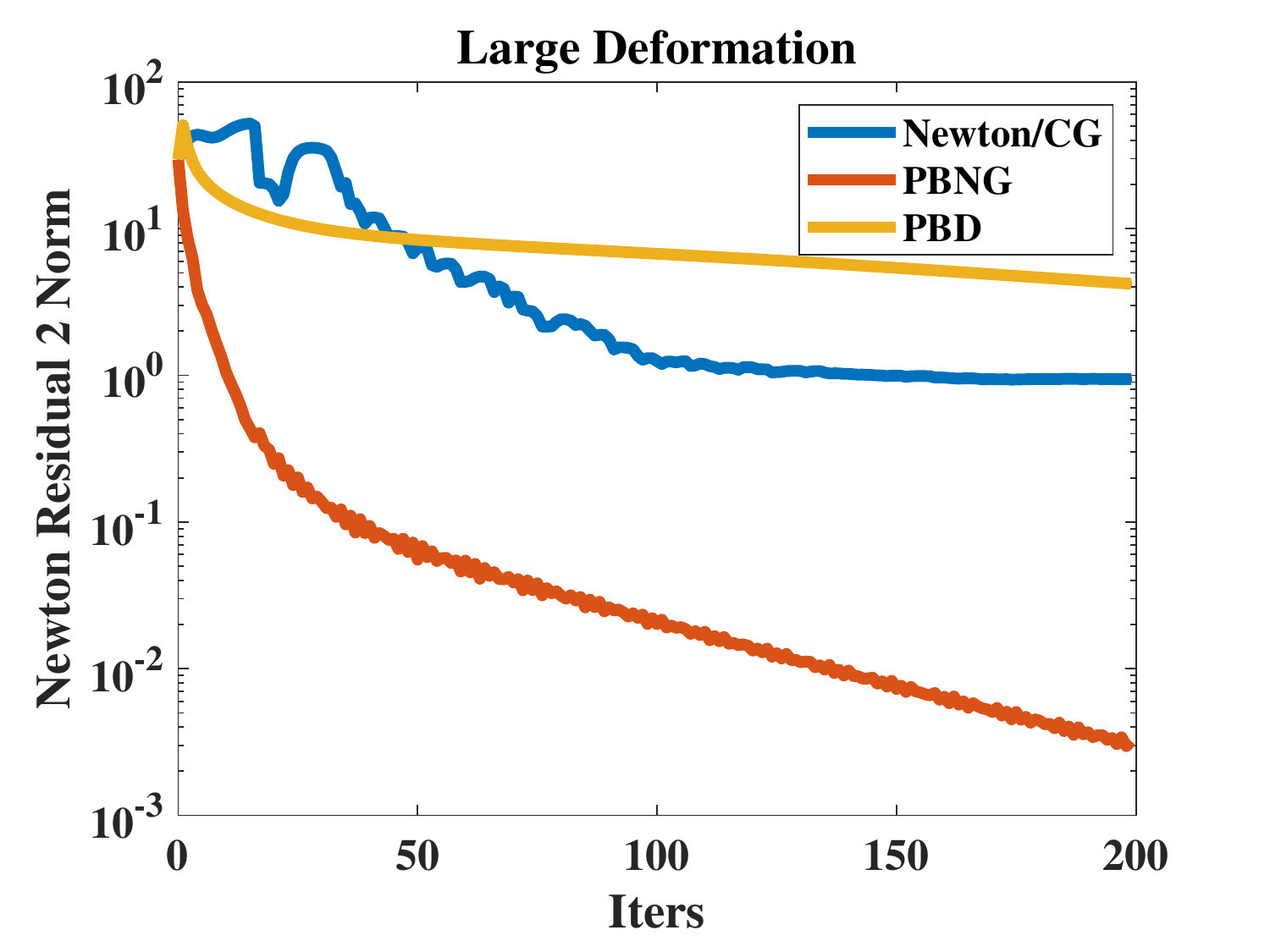}
	\includegraphics[draft=\mydraft,width=0.6\columnwidth,trim={0px 0px 0px 50px},clip]{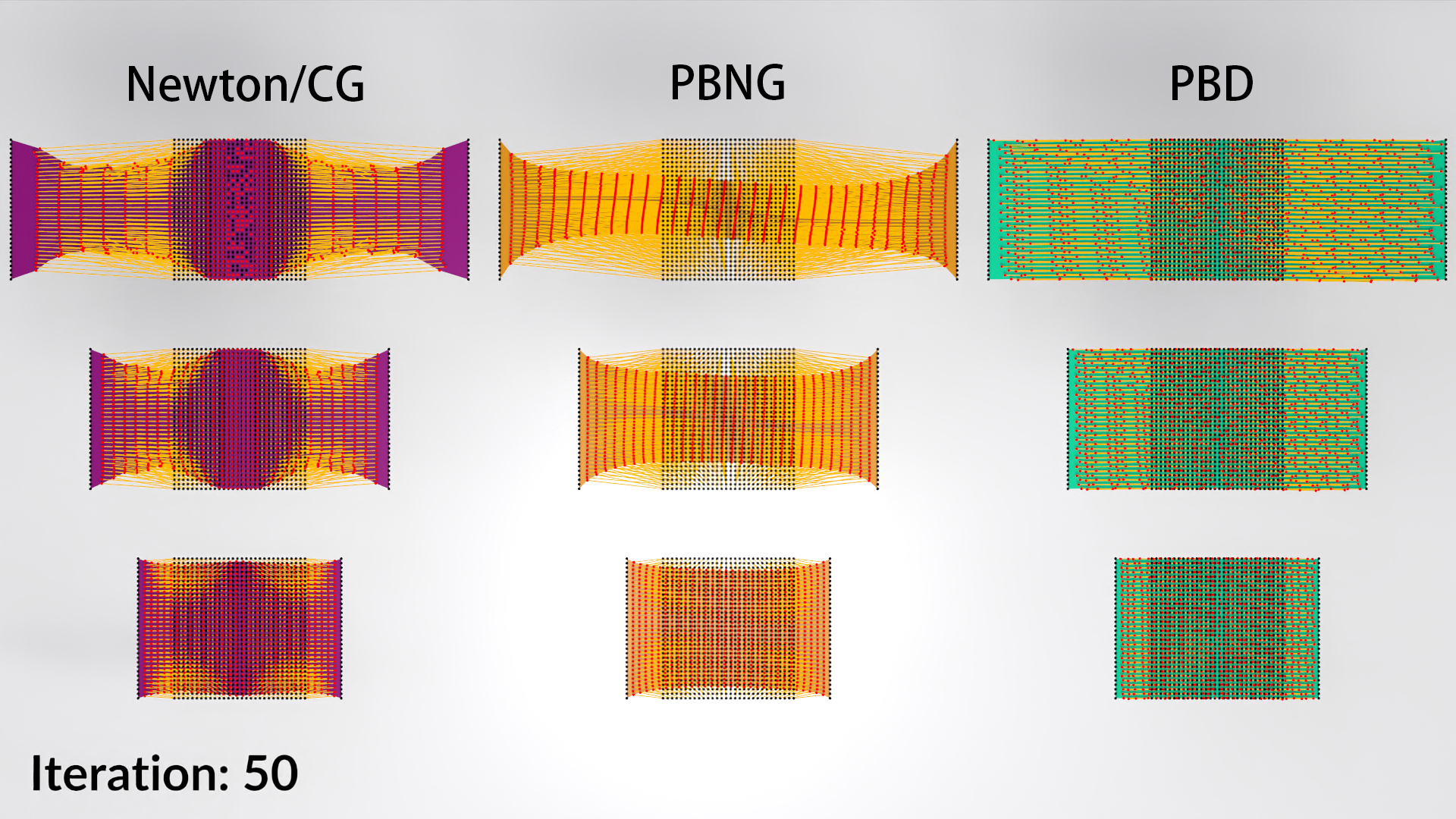}
	\includegraphics[draft=\mydraft,width=0.39\columnwidth,trim={5px 0px 35px 0px},clip]{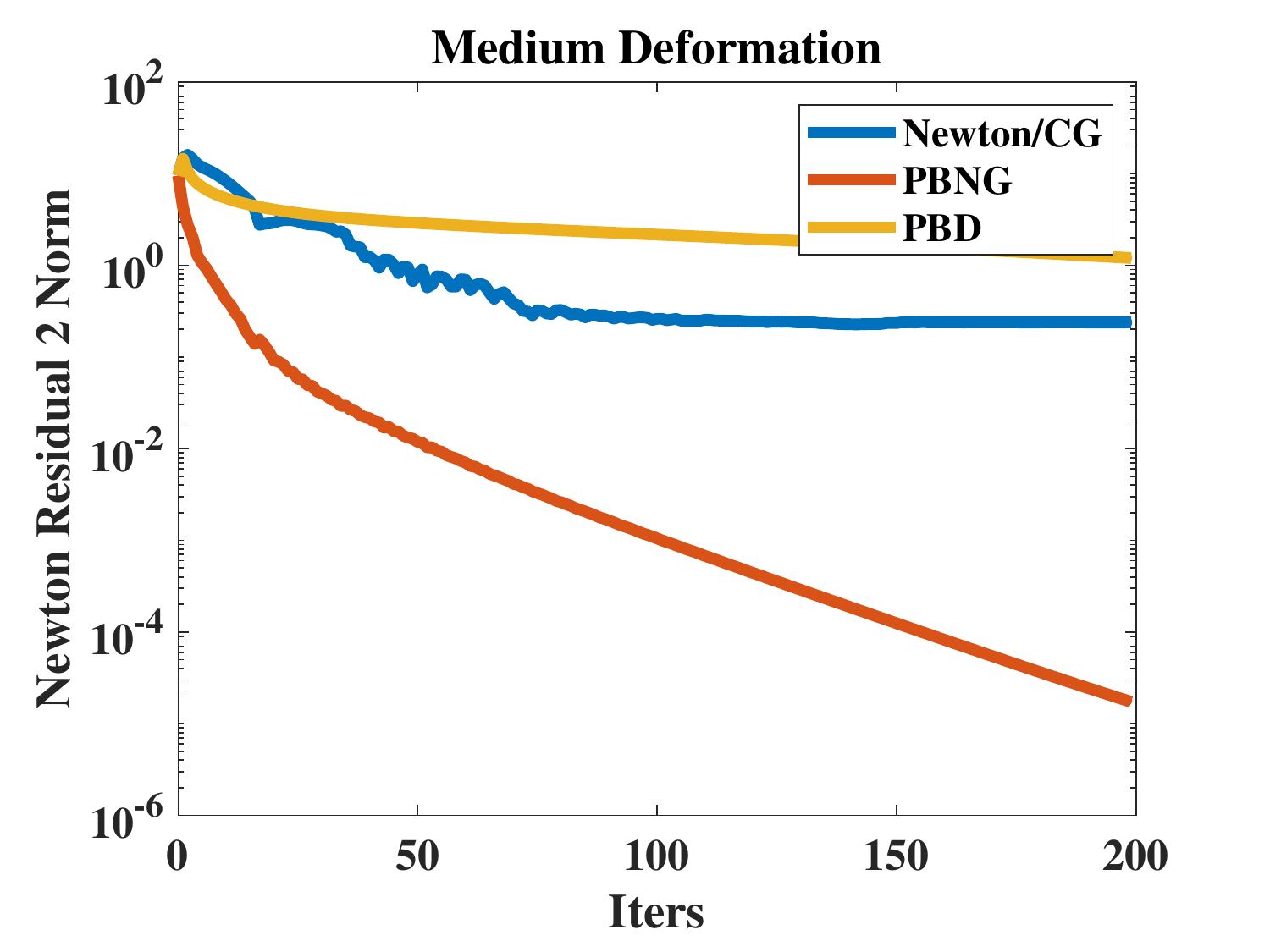}
	\includegraphics[draft=\mydraft,width=0.6\columnwidth,trim={0px 0px 0px 50px},clip]{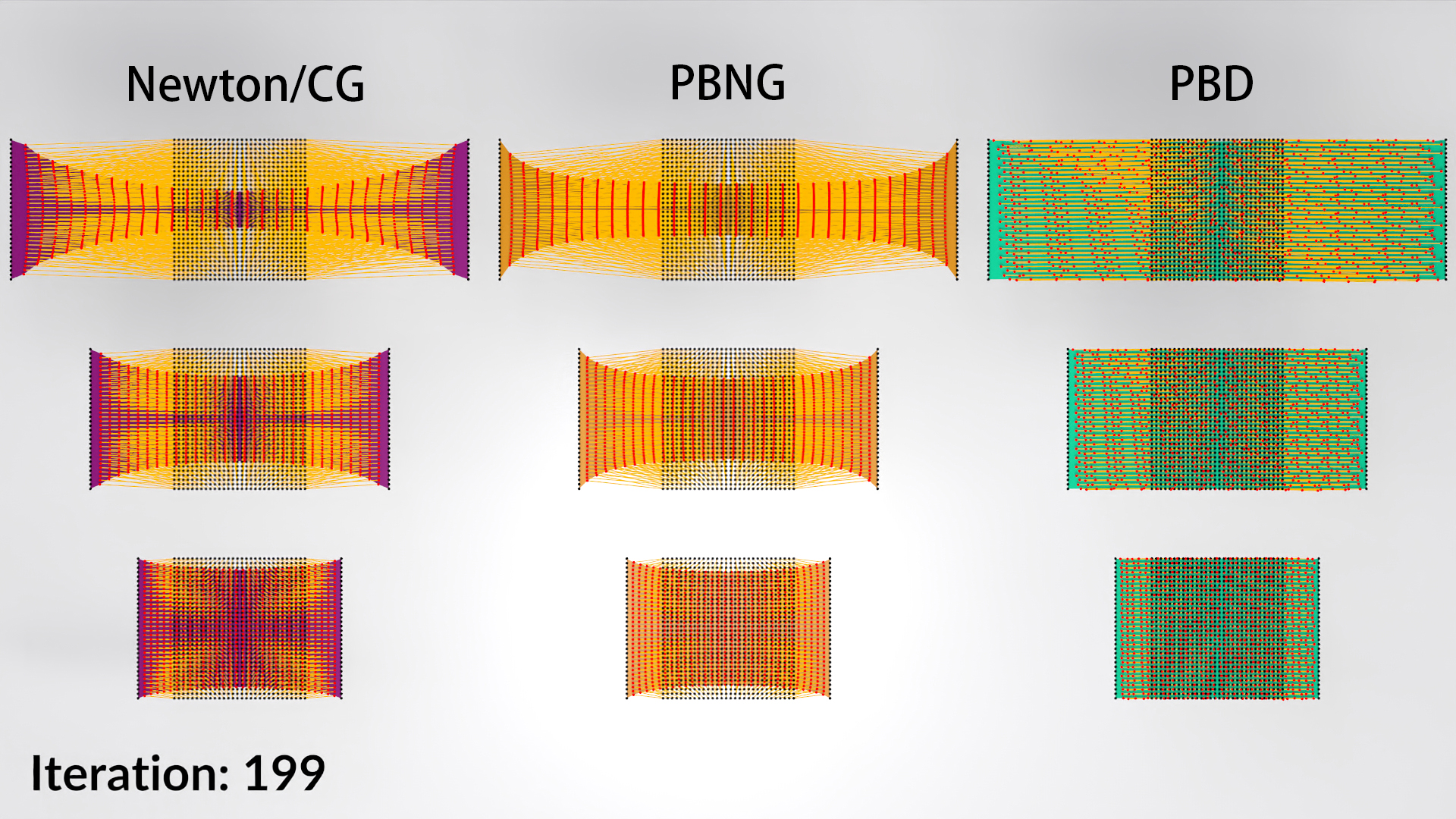}
	\includegraphics[draft=\mydraft,width=0.39\columnwidth,trim={5px 0px 35px 0px},clip]{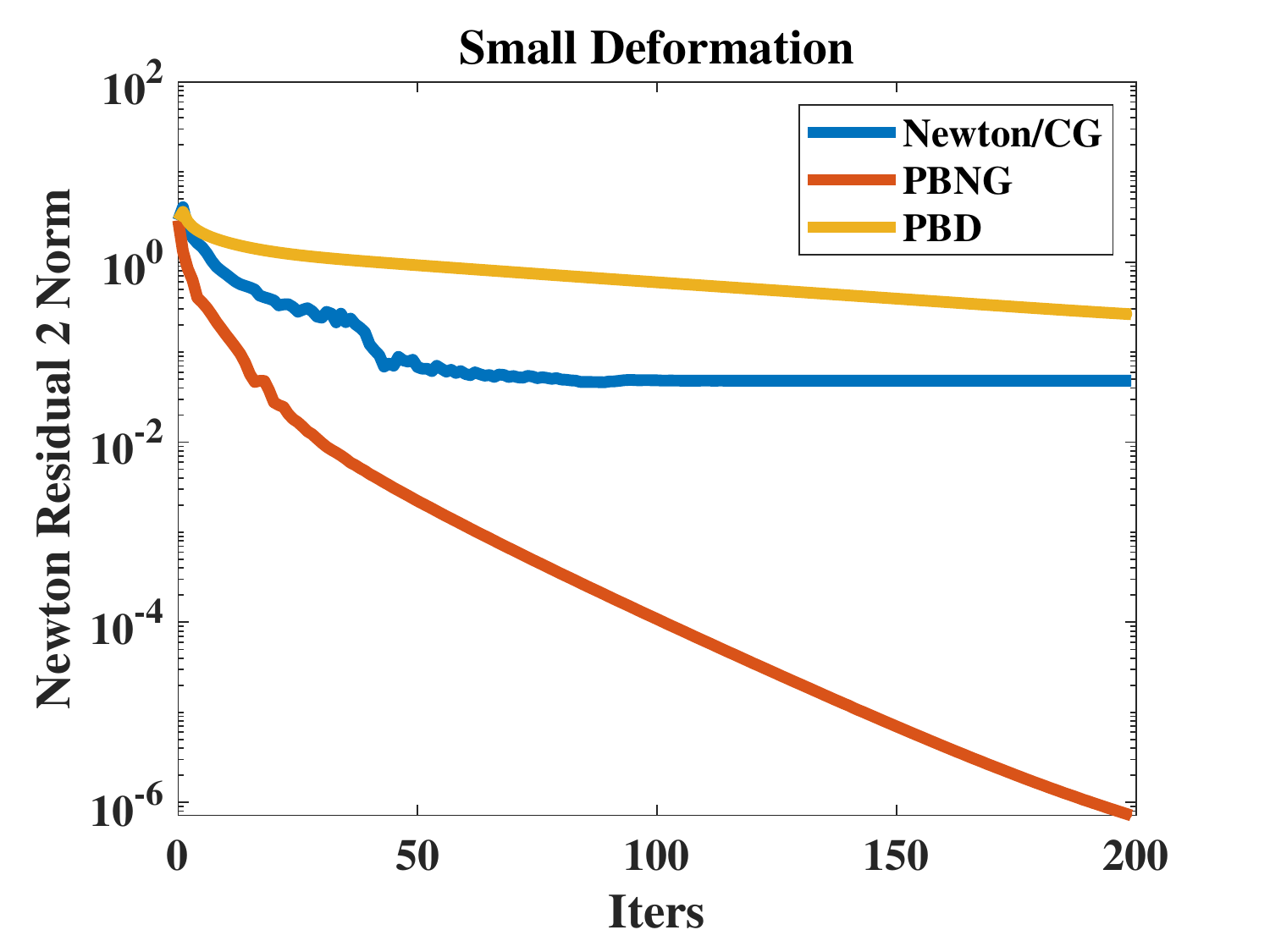}
	\caption{\textbf{Deformation Propagation Visualization}. A square block is initially stretched on its sides. 
	\textbf{Left column}: visual results of the blocks after certain iterations. 
	Black points are the initial positions. Red points are positions at the current iteration. Yellow line segments indicate the displacement of each node.
	Each method is color coded - purple is Newton, orange is PBNG, and green is PBD.
	Each row shows the results of large, medium, and small deformations respectively.
	PBNG converges to a visually plausible result in fewer iterations than one Newton step with increasing CG iterations.
	PBD fails to shrink in the transverse direction.
	\textbf{Right column}: $2$-norm of the Newton residual vector. PBNG outperforms Newton's method and PBD.}
	\label{fig:2d_visualization}
\end{figure}

\begin{table*}[t]
	
	\resizebox{\textwidth}{!}{%
		\begin{tabular}{@{}lllllll@{}}
			\toprule
			Example & \# Vertices & \# Elements. & PBGN Runtime / Frame & PBNG \# Iter/Frame & \# Substeps & Model \\ \midrule
			Box Stretching (low budget) & 32K & 150K & 170ms & 6 & 1  & Corotated\\
			Box Stretching (big budget) & 32K & 150K & 1300ms & 40 & 1& Corotated\\
			Muscle with collisions & 284k & 1097K & 67000ms & 510 & 17& Corotated \\
			Res 64 Box Stretching & 260K & 1250K & 1300ms & 20 & 1 & Corotated\\
			Res 128 Box Stretching & 2097K & 10242K & 61000ms & 40 & 1 & Corotated\\
			Dropping Simple Shapes Into Box & 256K & 1069K & 49800ms & 136 & 17 & Corotated\\
			Two moving blocks colliding & 8.2K & 33K & 1630ms & 136 & 17 & Corotated\\
			Box Stretching & 32K & 150K & 1300ms & 40 & 1& Stable Neo-Hookean\\
			Box Stretching & 32K & 150K & 825ms & 40 & 1& Neo-Hookean\\

			\bottomrule
		\end{tabular}
	}
	\caption{Performance Table of PBNG: runtime is measured for each frame (averaged over the course of the simulation). Each frame is written after advancing time .033. }
	\label{tbl:perf}
\end{table*}

\subsection{PBD}\label{sec:ex_pbd}
In this example, we show how PBD eliminates the effects of external forcing as the number of iterations increases. 
We clamp the left side of a simple bar mesh. 
We run a quasistatic simulation with gravity (acceleration $-9.8$ in the $y-$direction). 
The bar has $R^0=10$ and Young's modulus $E=1000$. 
As shown in Figure ~\ref{fig:bar_under_gravity}, PBD converges to a rigid bar configuration. 
PBNG converges to a plausible solution. 
XPBD-QS appears to resolve the issues with PBD and quasistatics. However, XPBD-QS with 10 iterations per pseudo-time step appears more converged than XPBD-QS with 1 iteration per pseudo-time step. 

\subsection{XPBD}\label{sec:ex_xpbd}
We run a simple dynamics example to show that XPBD does not converge numerically, as discussed in Section~\ref{sec:xpbd_conv}. 
We take a simple block with the left side clamped. 
It falls under gravity and oscillates.
The simulation scene is shown on the top of Figure ~\ref{fig:xpbd_stagnation}. 
The block has 4.1K particles and 17K elements. 
In this simple simulation, we compare the convergence behavior between PBNG and XPBD. 
As shown in Figure~\ref{fig:xpbd_stagnation}(b), XPBD stagnates, while PBNG converges to the tolerance. 
We demonstrate the reason for XPBD's stagnation in Figure ~\ref{fig:xpbd_stagnation}(a). 
XPBD omits the primary residual terms, which results in the stagnation of residual reduction. 
Though XPBD reduces the secondary residual, the true residual stagnates. 

\subsection{PBNG vs. PBD and Limited Newton}
We run a simple quasistatic example to illustrate the convergence propagation behavior of PBNG compared to each conjugate gradient (CG) iteration in Newton's method as well as PBD. 
In Figure ~\ref{fig:2d_visualization}, a block has its two sides stretched and then clamped. 
We compute the quasistatic equilibrium using Newton's method with 1 Newton iteration, 
PBD and PBNG. 
PBD does not converge to the right solution. 
After 50 iterations, PBNG looks visually plausible, but Newton's method is visually not converged. 
The residual plots are presented in Figure ~\ref{fig:2d_visualization}. 
PBNG iterations are comparable to CG iterations in Newton's method, but they have more favorable deformation propagation behavior. 

\section{Discussion and Limitations}
We show that a node-based Gauss-Seidel approach for the nonlinear equations of quasistatic and backward Euler time stepping has remarkably stable behavior.
While we generate visually plausible behaviors with restricted computational budgets in a manner that surpasses the PBD and XPBD state-of-the-art for quasistatic problems, our approach (even with Chebyshev and SOR acceleration) will still lose (in terms of numerical residual reduction) to a standard Newton's method when the computational budget is expanded.
A multigrid or domain decomposition approach could be combined with our approach to address this in future work.

% Bibliography
\bibliographystyle{ACM-Reference-Format}
\bibliography{./paper.bib}


\begin{thebibliography}{52}

%%% the Web address.

\ifx \showCODEN    \undefined \def \showCODEN     #1{\unskip}     \fi
\ifx \showDOI      \undefined \def \showDOI       #1{#1}\fi
\ifx \showISBNx    \undefined \def \showISBNx     #1{\unskip}     \fi
\ifx \showISBNxiii \undefined \def \showISBNxiii  #1{\unskip}     \fi
\ifx \showISSN     \undefined \def \showISSN      #1{\unskip}     \fi
\ifx \showLCCN     \undefined \def \showLCCN      #1{\unskip}     \fi
\ifx \shownote     \undefined \def \shownote      #1{#1}          \fi
\ifx \showarticletitle \undefined \def \showarticletitle #1{#1}   \fi
\ifx \showURL      \undefined \def \showURL       {\relax}        \fi
% invisible to TeX
\providecommand\bibfield[2]{#2}
\providecommand\bibinfo[2]{#2}
\providecommand\natexlab[1]{#1}
\providecommand\showeprint[2][]{arXiv:#2}

\bibitem[Anonymous(2023)]%
        {pbng:techdoc}
\bibfield{author}{\bibinfo{person}{Anonymous}.}
  \bibinfo{year}{2023}\natexlab{}.
\newblock \bibinfo{journal}{\emph{Supplementary Technical Document}}
  (\bibinfo{year}{2023}).
\newblock


\bibitem[Bailey et~al\mbox{.}(2018)]%
        {bailey:2018:fdd}
\bibfield{author}{\bibinfo{person}{S. Bailey}, \bibinfo{person}{D. Otte},
  \bibinfo{person}{P. Dilorenzo}, {and} \bibinfo{person}{J. O'Brien}.}
  \bibinfo{year}{2018}\natexlab{}.
\newblock \showarticletitle{Fast and Deep Deformation Approximations}.
\newblock \bibinfo{journal}{\emph{ACM Trans Graph}} \bibinfo{volume}{37},
  \bibinfo{number}{4} (\bibinfo{date}{Aug.} \bibinfo{year}{2018}),
  \bibinfo{pages}{119:1--12}.
\newblock
\urldef\tempurl%
\url{https://doi.org/10.1145/3197517.3201300}
\showDOI{\tempurl}


\bibitem[Baraff and Witkin(1998)]%
        {baraff:1998:cloth}
\bibfield{author}{\bibinfo{person}{D. Baraff} {and} \bibinfo{person}{A.
  Witkin}.} \bibinfo{year}{1998}\natexlab{}.
\newblock \showarticletitle{Large Steps in Cloth Simulation}. In
  \bibinfo{booktitle}{\emph{Proc ACM SIGGRAPH}}
  \emph{(\bibinfo{series}{SIGGRAPH '98})}. \bibinfo{pages}{43--54}.
\newblock


\bibitem[Bertiche et~al\mbox{.}(2021)]%
        {bertiche:2021:pbns}
\bibfield{author}{\bibinfo{person}{H. Bertiche}, \bibinfo{person}{M. Madadi},
  {and} \bibinfo{person}{S. Escalera}.} \bibinfo{year}{2021}\natexlab{}.
\newblock \bibinfo{title}{PBNS: Physically Based Neural Simulator for
  Unsupervised Garment Pose Space Deformation}.
\newblock
\newblock
\showeprint[arxiv]{2012.11310}~[cs.CV]


\bibitem[Bertsekas(1997)]%
        {bertsekas:1997:nonlinear}
\bibfield{author}{\bibinfo{person}{D. Bertsekas}.}
  \bibinfo{year}{1997}\natexlab{}.
\newblock \showarticletitle{Nonlinear programming}.
\newblock \bibinfo{journal}{\emph{J Op Res Soc}} \bibinfo{volume}{48},
  \bibinfo{number}{3} (\bibinfo{year}{1997}), \bibinfo{pages}{334--334}.
\newblock


\bibitem[Bonet and Wood(2008)]%
        {bonet:2008:continuum}
\bibfield{author}{\bibinfo{person}{J. Bonet} {and} \bibinfo{person}{R. Wood}.}
  \bibinfo{year}{2008}\natexlab{}.
\newblock \bibinfo{booktitle}{\emph{Nonlinear continuum mechanics for finite
  element analysis}}.
\newblock \bibinfo{publisher}{Cambridge University Press}.
\newblock


\bibitem[Bouaziz et~al\mbox{.}(2014)]%
        {bouaziz:2014:pdf}
\bibfield{author}{\bibinfo{person}{S. Bouaziz}, \bibinfo{person}{S. Martin},
  \bibinfo{person}{T. Liu}, \bibinfo{person}{L. Kavan}, {and}
  \bibinfo{person}{M. Pauly}.} \bibinfo{year}{2014}\natexlab{}.
\newblock \showarticletitle{Projective Dynamics: Fusing Constraint Projections
  for Fast Simulation}.
\newblock \bibinfo{journal}{\emph{ACM Trans Graph}} \bibinfo{volume}{33},
  \bibinfo{number}{4} (\bibinfo{year}{2014}), \bibinfo{pages}{154:1--154:11}.
\newblock


\bibitem[Boyd et~al\mbox{.}(2011)]%
        {boyd:2011:distributed}
\bibfield{author}{\bibinfo{person}{S. Boyd}, \bibinfo{person}{N. Parikh},
  \bibinfo{person}{E. Chu}, \bibinfo{person}{B. Peleato}, {and}
  \bibinfo{person}{J. Eckstein}.} \bibinfo{year}{2011}\natexlab{}.
\newblock \showarticletitle{Distributed optimization and statistical learning
  via the alternating direction method of multipliers}.
\newblock \bibinfo{journal}{\emph{Foundations and Trends in Machine Learning}}
  \bibinfo{volume}{3}, \bibinfo{number}{1} (\bibinfo{year}{2011}),
  \bibinfo{pages}{1--122}.
\newblock


\bibitem[Chao et~al\mbox{.}(2010)]%
        {chao:2010:geom}
\bibfield{author}{\bibinfo{person}{I. Chao}, \bibinfo{person}{U. Pinkall},
  \bibinfo{person}{P. Sanan}, {and} \bibinfo{person}{P. Schr\"{o}der}.}
  \bibinfo{year}{2010}\natexlab{}.
\newblock \showarticletitle{A Simple Geometric Model for Elastic Deformations}.
\newblock \bibinfo{journal}{\emph{ACM Trans Graph}} \bibinfo{volume}{29},
  \bibinfo{number}{4}, Article \bibinfo{articleno}{38} (\bibinfo{year}{2010}),
  \bibinfo{numpages}{6}~pages.
\newblock


\bibitem[eng et~al\mbox{.}(2020)]%
        {geng:2020:ml}
\bibfield{author}{\bibinfo{person}{Z. eng}, \bibinfo{person}{D. Johnson}, {and}
  \bibinfo{person}{R. Fedkiw}.} \bibinfo{year}{2020}\natexlab{}.
\newblock \showarticletitle{Coercing machine learning to output physically
  accurate results}.
\newblock \bibinfo{journal}{\emph{J Comp Phys}}  \bibinfo{volume}{406}
  (\bibinfo{year}{2020}), \bibinfo{pages}{109099}.
\newblock
\showISSN{0021-9991}
\urldef\tempurl%
\url{https://doi.org/10.1016/j.jcp.2019.109099}
\showDOI{\tempurl}


\bibitem[Etzmuss et~al\mbox{.}(2003)]%
        {etzmuss:2003:cont}
\bibfield{author}{\bibinfo{person}{O. Etzmuss}, \bibinfo{person}{J. Gross},
  {and} \bibinfo{person}{W. Strasser}.} \bibinfo{year}{2003}\natexlab{}.
\newblock \showarticletitle{Deriving a particle system from continuum mechanics
  for the animation of deformable objects}.
\newblock \bibinfo{journal}{\emph{IEEE Trans Vis Comp Graph}}
  \bibinfo{volume}{9}, \bibinfo{number}{4} (\bibinfo{date}{Oct.}
  \bibinfo{year}{2003}), \bibinfo{pages}{538--550}.
\newblock


\bibitem[Fan et~al\mbox{.}(2014)]%
        {fan:2014:avm}
\bibfield{author}{\bibinfo{person}{Y. Fan}, \bibinfo{person}{J. Litven}, {and}
  \bibinfo{person}{D. Pai}.} \bibinfo{year}{2014}\natexlab{}.
\newblock \showarticletitle{Active Volumetric Musculoskeletal Systems}.
\newblock \bibinfo{journal}{\emph{ACM Trans Graph}} \bibinfo{volume}{33},
  \bibinfo{number}{4} (\bibinfo{year}{2014}), \bibinfo{pages}{152:1--152:9}.
\newblock


\bibitem[Fratarcangeli et~al\mbox{.}(2016)]%
        {fratarcangeli:2016:vivace}
\bibfield{author}{\bibinfo{person}{M. Fratarcangeli}, \bibinfo{person}{T.
  Valentina}, {and} \bibinfo{person}{F. Pellacini}.}
  \bibinfo{year}{2016}\natexlab{}.
\newblock \showarticletitle{Vivace: a practical gauss-seidel method for stable
  soft body dynamics}.
\newblock \bibinfo{journal}{\emph{ACM Trans Graph}} \bibinfo{volume}{35},
  \bibinfo{number}{6} (\bibinfo{date}{Nov} \bibinfo{year}{2016}),
  \bibinfo{pages}{1–9}.
\newblock
\showISSN{0730-0301, 1557-7368}
\urldef\tempurl%
\url{https://doi.org/10.1145/2980179.2982437}
\showDOI{\tempurl}


\bibitem[Gast et~al\mbox{.}(2016)]%
        {gast:2016:svd}
\bibfield{author}{\bibinfo{person}{T. Gast}, \bibinfo{person}{C. Fu},
  \bibinfo{person}{C. Jiang}, {and} \bibinfo{person}{J. Teran}.}
  \bibinfo{year}{2016}\natexlab{}.
\newblock \bibinfo{booktitle}{\emph{Implicit-shifted Symmetric QR Singular
  Value Decomposition of 3x3 Matrices}}.
\newblock \bibinfo{type}{{T}echnical {R}eport}.
  \bibinfo{institution}{University of California Los Angeles}.
\newblock


\bibitem[Gast et~al\mbox{.}(2015)]%
        {gast:2015:tvcg}
\bibfield{author}{\bibinfo{person}{T. Gast}, \bibinfo{person}{C. Schroeder},
  \bibinfo{person}{A. Stomakhin}, \bibinfo{person}{C. Jiang}, {and}
  \bibinfo{person}{J. Teran}.} \bibinfo{year}{2015}\natexlab{}.
\newblock \showarticletitle{Optimization Integrator for Large Time Steps}.
\newblock \bibinfo{journal}{\emph{IEEE Trans Vis Comp Graph}}
  \bibinfo{volume}{21}, \bibinfo{number}{10} (\bibinfo{year}{2015}),
  \bibinfo{pages}{1103--1115}.
\newblock


\bibitem[Gonzalez and Stuart(2008)]%
        {gonzalez:2008:continuum}
\bibfield{author}{\bibinfo{person}{O. Gonzalez} {and} \bibinfo{person}{A.
  Stuart}.} \bibinfo{year}{2008}\natexlab{}.
\newblock \bibinfo{booktitle}{\emph{A first course in continuum mechanics}}.
\newblock \bibinfo{publisher}{Cambridge University Press}.
\newblock


\bibitem[Hecht et~al\mbox{.}(2012)]%
        {hecht:2012:chol}
\bibfield{author}{\bibinfo{person}{F. Hecht}, \bibinfo{person}{Y. Lee},
  \bibinfo{person}{J. Shewchuk}, {and} \bibinfo{person}{J. O'Brien}.}
  \bibinfo{year}{2012}\natexlab{}.
\newblock \showarticletitle{Updated Sparse Cholesky Factors for Corotational
  Elastodynamics}.
\newblock \bibinfo{journal}{\emph{ACM Trans Graph}} \bibinfo{volume}{31},
  \bibinfo{number}{5}, Article \bibinfo{articleno}{123} (\bibinfo{year}{2012}),
  \bibinfo{numpages}{13}~pages.
\newblock


\bibitem[Jin et~al\mbox{.}(2020)]%
        {jin:2020:ml}
\bibfield{author}{\bibinfo{person}{N. Jin}, \bibinfo{person}{Y. Zhu},
  \bibinfo{person}{Z. Geng}, {and} \bibinfo{person}{R. Fedkiw}.}
  \bibinfo{year}{2020}\natexlab{}.
\newblock \showarticletitle{A Pixel-Based Framework for Data-Driven Clothing}.
  In \bibinfo{booktitle}{\emph{Proc ACM SIGGRAPH/Eurographics Symp Comp Anim}}
  (Virtual Event, Canada) \emph{(\bibinfo{series}{SCA '20})}.
  \bibinfo{publisher}{Eurographics Association}, Article
  \bibinfo{articleno}{13}, \bibinfo{numpages}{10}~pages.
\newblock
\urldef\tempurl%
\url{https://doi.org/10.1111/cgf.14108}
\showDOI{\tempurl}


\bibitem[Jin et~al\mbox{.}(2022)]%
        {jin:2022:mls}
\bibfield{author}{\bibinfo{person}{Y. Jin}, \bibinfo{person}{Y. Han},
  \bibinfo{person}{Z. Geng}, \bibinfo{person}{J. Teran}, {and}
  \bibinfo{person}{R. Fedkiw}.} \bibinfo{year}{2022}\natexlab{}.
\newblock \showarticletitle{Analytically Integratable Zero-Restlength Springs
  for Capturing Dynamic Modes Unrepresented by Quasistatic Neural Networks}. In
  \bibinfo{booktitle}{\emph{ACM SIGGRAPH 2022 Conf Proc}} (Vancouver, BC,
  Canada) \emph{(\bibinfo{series}{SIGGRAPH '22})}. \bibinfo{publisher}{ACM},
  \bibinfo{address}{New York, NY, USA}, Article \bibinfo{articleno}{37},
  \bibinfo{numpages}{9}~pages.
\newblock
\showISBNx{9781450393379}
\urldef\tempurl%
\url{https://doi.org/10.1145/3528233.3530705}
\showDOI{\tempurl}


\bibitem[Kovalsky et~al\mbox{.}(2016)]%
        {kovalsky:2016:aqp}
\bibfield{author}{\bibinfo{person}{S. Kovalsky}, \bibinfo{person}{M. Galun},
  {and} \bibinfo{person}{Y. Lipman}.} \bibinfo{year}{2016}\natexlab{}.
\newblock \showarticletitle{Accelerated Quadratic Proxy for Geometric
  Optimization}.
\newblock \bibinfo{journal}{\emph{ACM Trans Graph}} \bibinfo{volume}{35},
  \bibinfo{number}{4}, Article \bibinfo{articleno}{134} (\bibinfo{year}{2016}),
  \bibinfo{numpages}{11}~pages.
\newblock


\bibitem[Li et~al\mbox{.}(2019)]%
        {li:2019:dot}
\bibfield{author}{\bibinfo{person}{M. Li}, \bibinfo{person}{M. Gao},
  \bibinfo{person}{T. Langlois}, \bibinfo{person}{C. Jiang}, {and}
  \bibinfo{person}{D. Kaufman}.} \bibinfo{year}{2019}\natexlab{}.
\newblock \showarticletitle{Decomposed Optimization Time Integrator for
  Large-Step Elastodynamics}.
\newblock \bibinfo{journal}{\emph{ACM Trans Graph}} \bibinfo{volume}{38},
  \bibinfo{number}{4} (\bibinfo{date}{jul} \bibinfo{year}{2019}),
  \bibinfo{numpages}{10}~pages.
\newblock
\urldef\tempurl%
\url{https://doi.org/10.1145/3306346.3322951}
\showDOI{\tempurl}


\bibitem[Liu et~al\mbox{.}(2008)]%
        {liu:2008:arap}
\bibfield{author}{\bibinfo{person}{L. Liu}, \bibinfo{person}{L. Zhang},
  \bibinfo{person}{Y. Xu}, \bibinfo{person}{C. Gotsman}, {and}
  \bibinfo{person}{S. Gortler}.} \bibinfo{year}{2008}\natexlab{}.
\newblock \showarticletitle{A Local/Global Approach to Mesh Parameterization}.
  In \bibinfo{booktitle}{\emph{Proc Symp Geom Proc}}
  \emph{(\bibinfo{series}{SGP '08})}. \bibinfo{publisher}{Eurograph Assoc},
  \bibinfo{pages}{1495?1504}.
\newblock


\bibitem[Liu et~al\mbox{.}(2013)]%
        {liu:2013:fsm}
\bibfield{author}{\bibinfo{person}{T. Liu}, \bibinfo{person}{A. Bargteil},
  \bibinfo{person}{J. O'Brien}, {and} \bibinfo{person}{L. Kavan}.}
  \bibinfo{year}{2013}\natexlab{}.
\newblock \showarticletitle{Fast Simulation of Mass-Spring Systems}.
\newblock \bibinfo{journal}{\emph{ACM Trans Graph}} \bibinfo{volume}{32},
  \bibinfo{number}{6} (\bibinfo{year}{2013}), \bibinfo{pages}{209:1--7}.
\newblock


\bibitem[Liu et~al\mbox{.}(2017)]%
        {liu:2017:lbfgs}
\bibfield{author}{\bibinfo{person}{T. Liu}, \bibinfo{person}{S. Bouaziz}, {and}
  \bibinfo{person}{L. Kavan}.} \bibinfo{year}{2017}\natexlab{}.
\newblock \showarticletitle{Quasi-Newton Methods for Real-Time Simulation of
  Hyperelastic Materials}.
\newblock \bibinfo{journal}{\emph{ACM Trans Graph}} \bibinfo{volume}{36},
  \bibinfo{number}{4}, Article \bibinfo{articleno}{116a}
  (\bibinfo{year}{2017}), \bibinfo{numpages}{16}~pages.
\newblock


\bibitem[Luo et~al\mbox{.}(2020)]%
        {luo:2020:ml}
\bibfield{author}{\bibinfo{person}{R. Luo}, \bibinfo{person}{T. Shao},
  \bibinfo{person}{H. Wang}, \bibinfo{person}{W. Xu}, \bibinfo{person}{X.
  Chen}, \bibinfo{person}{K. Zhou}, {and} \bibinfo{person}{Y. Yang}.}
  \bibinfo{year}{2020}\natexlab{}.
\newblock \bibinfo{title}{NNWarp: Neural Network-Based Nonlinear Deformation}.
\newblock , \bibinfo{numpages}{1745-1759}~pages.
\newblock
\urldef\tempurl%
\url{https://doi.org/10.1109/TVCG.2018.2881451}
\showDOI{\tempurl}


\bibitem[Macklin and Muller(2021)]%
        {Macklin:2021:neohookean_xpbd}
\bibfield{author}{\bibinfo{person}{M. Macklin} {and} \bibinfo{person}{M.
  Muller}.} \bibinfo{year}{2021}\natexlab{}.
\newblock \showarticletitle{A Constraint-based Formulation of Stable
  Neo-Hookean Materials}. In \bibinfo{booktitle}{\emph{Motion, Interaction and
  Games}}. \bibinfo{publisher}{ACM}, \bibinfo{pages}{1–7}.
\newblock
\showISBNx{9781450391313}
\urldef\tempurl%
\url{https://doi.org/10.1145/3487983.3488289}
\showDOI{\tempurl}


\bibitem[Macklin et~al\mbox{.}(2016)]%
        {macklin:2016:xpbd}
\bibfield{author}{\bibinfo{person}{M. Macklin}, \bibinfo{person}{M.
  M\"{u}ller}, {and} \bibinfo{person}{N. Chentanez}.}
  \bibinfo{year}{2016}\natexlab{}.
\newblock \showarticletitle{XPBD: Position-Based Simulation of Compliant
  Constrained Dynamics}. In \bibinfo{booktitle}{\emph{Proc 9th Int Conf Motion
  Games}} (Burlingame, California) \emph{(\bibinfo{series}{MIG '16})}.
  \bibinfo{publisher}{ACM}, \bibinfo{pages}{49?54}.
\newblock


\bibitem[Martin et~al\mbox{.}(2011)]%
        {martin:2011:ebem}
\bibfield{author}{\bibinfo{person}{S. Martin}, \bibinfo{person}{B.
  Thomaszewski}, \bibinfo{person}{E. Grinspun}, {and}
  \bibinfo{person}{M.Gross}.} \bibinfo{year}{2011}\natexlab{}.
\newblock \showarticletitle{Example-Based Elastic Materials}. In
  \bibinfo{booktitle}{\emph{ACM SIGGRAPH 2011}} (Vancouver, British Columbia,
  Canada) \emph{(\bibinfo{series}{SIGGRAPH '11})}. \bibinfo{publisher}{ACM},
  Article \bibinfo{articleno}{72}, \bibinfo{numpages}{8}~pages.
\newblock


\bibitem[McAdams et~al\mbox{.}(2011)]%
        {mcadams:2011:mge}
\bibfield{author}{\bibinfo{person}{A. McAdams}, \bibinfo{person}{Y. Zhu},
  \bibinfo{person}{A. Selle}, \bibinfo{person}{M. Empey}, \bibinfo{person}{R.
  Tamstorf}, \bibinfo{person}{J. Teran}, {and} \bibinfo{person}{E. Sifakis}.}
  \bibinfo{year}{2011}\natexlab{}.
\newblock \showarticletitle{Efficient Elasticity for Character Skinning with
  Contact and Collisions}.
\newblock \bibinfo{journal}{\emph{ACM Trans Graph}} \bibinfo{volume}{30},
  \bibinfo{number}{4} (\bibinfo{year}{2011}), \bibinfo{pages}{37:1--37:12}.
\newblock


\bibitem[Modi et~al\mbox{.}(2021)]%
        {modi:2021:emu}
\bibfield{author}{\bibinfo{person}{V. Modi}, \bibinfo{person}{L. Fulton},
  \bibinfo{person}{A. Jacobson}, \bibinfo{person}{S. Sueda}, {and}
  \bibinfo{person}{D. Levin}.} \bibinfo{year}{2021}\natexlab{}.
\newblock \showarticletitle{Emu: Efficient muscle simulation in deformation
  space}. In \bibinfo{booktitle}{\emph{Comp Graph Forum}},
  Vol.~\bibinfo{volume}{40}. Wiley Online Library, \bibinfo{pages}{234--248}.
\newblock


\bibitem[M.Tournier et~al\mbox{.}(2015)]%
        {tournier:2015:stable_dynamics}
\bibfield{author}{\bibinfo{person}{M.Tournier}, \bibinfo{person}{M. Nesme},
  \bibinfo{person}{B. Gilles}, {and} \bibinfo{person}{F. Faure}.}
  \bibinfo{year}{2015}\natexlab{}.
\newblock \showarticletitle{Stable Constrained Dynamics}.
\newblock \bibinfo{journal}{\emph{ACM Trans Graph}} (\bibinfo{year}{2015}),
  \bibinfo{pages}{1–10}.
\newblock


\bibitem[M{\"u}ller et~al\mbox{.}(2002)]%
        {muller:2002:stable}
\bibfield{author}{\bibinfo{person}{M. M{\"u}ller}, \bibinfo{person}{J. Dorsey},
  \bibinfo{person}{L. McMillan}, \bibinfo{person}{R. Jagnow}, {and}
  \bibinfo{person}{B. Cutler}.} \bibinfo{year}{2002}\natexlab{}.
\newblock \showarticletitle{Stable real-time deformations}. In
  \bibinfo{booktitle}{\emph{Proc 2002 ACM SIGGRAPH/Eurograph Symp Comp Anim}}.
  \bibinfo{pages}{49--54}.
\newblock


\bibitem[M\"{u}ller and Gross(2004)]%
        {muller:2004:ivm}
\bibfield{author}{\bibinfo{person}{M. M\"{u}ller} {and} \bibinfo{person}{M.
  Gross}.} \bibinfo{year}{2004}\natexlab{}.
\newblock \showarticletitle{Interactive virtual materials}. In
  \bibinfo{booktitle}{\emph{Proc Graph Int}}. \bibinfo{publisher}{Canadian
  Human-Computer Communications Society}, \bibinfo{pages}{239--246}.
\newblock


\bibitem[M{\"u}ller et~al\mbox{.}(2007)]%
        {muller:2007:pbd}
\bibfield{author}{\bibinfo{person}{M. M{\"u}ller}, \bibinfo{person}{B.
  Heidelberger}, \bibinfo{person}{M. Hennix}, {and} \bibinfo{person}{J.
  Ratcliff}.} \bibinfo{year}{2007}\natexlab{}.
\newblock \showarticletitle{Position based dynamics}.
\newblock \bibinfo{journal}{\emph{J Vis Comm Im Rep}} \bibinfo{volume}{18},
  \bibinfo{number}{2} (\bibinfo{year}{2007}), \bibinfo{pages}{109--118}.
\newblock


\bibitem[Narain et~al\mbox{.}(2016)]%
        {narain:2016:admm}
\bibfield{author}{\bibinfo{person}{R. Narain}, \bibinfo{person}{M. Overby},
  {and} \bibinfo{person}{G. Brown}.} \bibinfo{year}{2016}\natexlab{}.
\newblock \showarticletitle{ADMM Projective Dynamics: Fast Simulation of
  General Constitutive Models}. In \bibinfo{booktitle}{\emph{Proc ACM
  SIGGRAPH/Eurograph Symp Comp Anim}} (Zurich, Switzerland)
  \emph{(\bibinfo{series}{SCA '16})}. \bibinfo{publisher}{Eurograph Assoc},
  \bibinfo{pages}{21?28}.
\newblock
\showISBNx{9783905674613}


\bibitem[Neuberger(1985)]%
        {neuberger:1985:sgd}
\bibfield{author}{\bibinfo{person}{J. Neuberger}.}
  \bibinfo{year}{1985}\natexlab{}.
\newblock \showarticletitle{Steepest descent and differential equations}.
\newblock \bibinfo{journal}{\emph{J Math Soc Japan}} \bibinfo{volume}{37},
  \bibinfo{number}{2} (\bibinfo{year}{1985}), \bibinfo{pages}{187--195}.
\newblock


\bibitem[Nocedal and Wright(2006)]%
        {nocedal:2006:cg}
\bibfield{author}{\bibinfo{person}{J. Nocedal} {and} \bibinfo{person}{S.
  Wright}.} \bibinfo{year}{2006}\natexlab{}.
\newblock \showarticletitle{Conjugate gradient methods}.
\newblock \bibinfo{journal}{\emph{Num Opt}} (\bibinfo{year}{2006}),
  \bibinfo{pages}{101--134}.
\newblock


\bibitem[Rabinovich et~al\mbox{.}(2017)]%
        {rabinovich:2017:injective}
\bibfield{author}{\bibinfo{person}{M. Rabinovich}, \bibinfo{person}{R.
  Poranne}, \bibinfo{person}{D. Panozzo}, {and} \bibinfo{person}{O.
  Sorkine-Hornung}.} \bibinfo{year}{2017}\natexlab{}.
\newblock \showarticletitle{Scalable Locally Injective Mappings}.
\newblock \bibinfo{journal}{\emph{ACM Trans Graph}} \bibinfo{volume}{36},
  \bibinfo{number}{2}, Article \bibinfo{articleno}{16} (\bibinfo{year}{2017}),
  \bibinfo{numpages}{16}~pages.
\newblock


\bibitem[Schmedding and Teschner(2008)]%
        {schmedding:2008:inversion}
\bibfield{author}{\bibinfo{person}{R. Schmedding} {and} \bibinfo{person}{M.
  Teschner}.} \bibinfo{year}{2008}\natexlab{}.
\newblock \showarticletitle{Inversion handling for stable deformable modeling}.
\newblock \bibinfo{journal}{\emph{Vis Comp}} \bibinfo{volume}{24},
  \bibinfo{number}{7-9} (\bibinfo{year}{2008}), \bibinfo{pages}{625--633}.
\newblock


\bibitem[Sifakis and Barbic(2012)]%
        {sifakis:2012:course}
\bibfield{author}{\bibinfo{person}{E. Sifakis} {and} \bibinfo{person}{J.
  Barbic}.} \bibinfo{year}{2012}\natexlab{}.
\newblock \showarticletitle{FEM simulation of 3D deformable solids: a
  practitioner's guide to theory, discretization and model reduction}. In
  \bibinfo{booktitle}{\emph{ACM SIGGRAPH 2012 Courses}}
  \emph{(\bibinfo{series}{SIGGRAPH '12})}. \bibinfo{publisher}{ACM},
  \bibinfo{pages}{20:1--20:50}.
\newblock


\bibitem[Smith et~al\mbox{.}(2019)]%
        {smith:2019:analytic}
\bibfield{author}{\bibinfo{person}{B. Smith}, \bibinfo{person}{F. Goes}, {and}
  \bibinfo{person}{T. Kim}.} \bibinfo{year}{2019}\natexlab{}.
\newblock \showarticletitle{Analytic eigensystems for isotropic distortion
  energies}.
\newblock \bibinfo{journal}{\emph{ACM Trans Graph (TOG)}} \bibinfo{volume}{38},
  \bibinfo{number}{1} (\bibinfo{year}{2019}), \bibinfo{pages}{1--15}.
\newblock


\bibitem[Smith et~al\mbox{.}(2018)]%
        {smith:2018:stable}
\bibfield{author}{\bibinfo{person}{B. Smith}, \bibinfo{person}{F.~De Goes},
  {and} \bibinfo{person}{T. Kim}.} \bibinfo{year}{2018}\natexlab{}.
\newblock \showarticletitle{Stable neo-hookean flesh simulation}.
\newblock \bibinfo{journal}{\emph{ACM Trans Grap (TOG)}} \bibinfo{volume}{37},
  \bibinfo{number}{2} (\bibinfo{year}{2018}), \bibinfo{pages}{1--15}.
\newblock


\bibitem[Sorkine and Alexa(2007)]%
        {sorkine:2007:as-rigid-as-possible}
\bibfield{author}{\bibinfo{person}{O. Sorkine} {and} \bibinfo{person}{M.
  Alexa}.} \bibinfo{year}{2007}\natexlab{}.
\newblock \showarticletitle{As-Rigid-As-Possible Surface Modeling}. In
  \bibinfo{booktitle}{\emph{EUROGRAPHICS SYMPOSIUM ON GEOMETRY PROCESSING}}.
\newblock


\bibitem[Stern and Desbrun(2006)]%
        {stern:2006:int}
\bibfield{author}{\bibinfo{person}{A. Stern} {and} \bibinfo{person}{M.
  Desbrun}.} \bibinfo{year}{2006}\natexlab{}.
\newblock \showarticletitle{Discrete Geometric Mechanics for Variational Time
  Integrators}. In \bibinfo{booktitle}{\emph{ACM SIGGRAPH 2006 Courses}}
  (Boston, Massachusetts) \emph{(\bibinfo{series}{SIGGRAPH '06})}.
  \bibinfo{publisher}{ACM}, \bibinfo{pages}{75?80}.
\newblock


\bibitem[Stomakhin et~al\mbox{.}(2012)]%
        {stomakhin:2012:invertible}
\bibfield{author}{\bibinfo{person}{A. Stomakhin}, \bibinfo{person}{R. Howes},
  \bibinfo{person}{C. Schroeder}, {and} \bibinfo{person}{J. Teran}.}
  \bibinfo{year}{2012}\natexlab{}.
\newblock \showarticletitle{Energetically consistent invertible elasticity}. In
  \bibinfo{booktitle}{\emph{Proc Symp Comp Anim}}. \bibinfo{pages}{25--32}.
\newblock


\bibitem[Teran et~al\mbox{.}(2005a)]%
        {teran:2005:muscle}
\bibfield{author}{\bibinfo{person}{J. Teran}, \bibinfo{person}{E. Sifakis},
  \bibinfo{person}{S. Blemker}, \bibinfo{person}{V. Ng-Thow-Hing},
  \bibinfo{person}{C. Lau}, {and} \bibinfo{person}{R. Fedkiw}.}
  \bibinfo{year}{2005}\natexlab{a}.
\newblock \showarticletitle{Creating and simulating skeletal muscle from the
  visible human data set}.
\newblock \bibinfo{journal}{\emph{IEEE Trans Vis Comp Graph}}
  \bibinfo{volume}{11}, \bibinfo{number}{3} (\bibinfo{year}{2005}),
  \bibinfo{pages}{317--328}.
\newblock


\bibitem[Teran et~al\mbox{.}(2005b)]%
        {teran:2005:robust}
\bibfield{author}{\bibinfo{person}{J. Teran}, \bibinfo{person}{E. Sifakis},
  \bibinfo{person}{G. Irving}, {and} \bibinfo{person}{R. Fedkiw}.}
  \bibinfo{year}{2005}\natexlab{b}.
\newblock \showarticletitle{Robust quasistatic finite elements and flesh
  simulation}. In \bibinfo{booktitle}{\emph{Proc 2005 ACM SIGGRAPH/Eurograph
  Symp Comp Anim}}. \bibinfo{pages}{181--190}.
\newblock


\bibitem[Wang et~al\mbox{.}(2020)]%
        {wang:2020:acs}
\bibfield{author}{\bibinfo{person}{B. Wang}, \bibinfo{person}{M. Zheng}, {and}
  \bibinfo{person}{J. Barbi\v{c}}.} \bibinfo{year}{2020}\natexlab{}.
\newblock \showarticletitle{Adjustable Constrained Soft-Tissue Dynamics}.
\newblock \bibinfo{journal}{\emph{Pac Graph 2020 and Comp Graph Forum}}
  \bibinfo{volume}{39}, \bibinfo{number}{7} (\bibinfo{year}{2020}).
\newblock


\bibitem[Wang(2015)]%
        {wang:2015:cheby}
\bibfield{author}{\bibinfo{person}{H. Wang}.} \bibinfo{year}{2015}\natexlab{}.
\newblock \showarticletitle{A Chebyshev Semi-Iterative Approach for
  Accelerating Projective and Position-Based Dynamics}.
\newblock \bibinfo{journal}{\emph{ACM Trans Graph}} \bibinfo{volume}{34},
  \bibinfo{number}{6}, Article \bibinfo{articleno}{246} (\bibinfo{date}{nov}
  \bibinfo{year}{2015}), \bibinfo{numpages}{9}~pages.
\newblock


\bibitem[Wang and Yang(2016)]%
        {Wang:2016:des_gpu}
\bibfield{author}{\bibinfo{person}{H. Wang} {and} \bibinfo{person}{Y. Yang}.}
  \bibinfo{year}{2016}\natexlab{}.
\newblock \showarticletitle{Descent methods for elastic body simulation on the
  GPU}.
\newblock \bibinfo{journal}{\emph{ACM Trans Graph}} \bibinfo{volume}{35},
  \bibinfo{number}{6} (\bibinfo{date}{Nov} \bibinfo{year}{2016}),
  \bibinfo{pages}{1–10}.
\newblock
\showISSN{0730-0301, 1557-7368}
\urldef\tempurl%
\url{https://doi.org/10.1145/2980179.2980236}
\showDOI{\tempurl}


\bibitem[Witemeyer et~al\mbox{.}(2021)]%
        {witemeyer:2021:qlb}
\bibfield{author}{\bibinfo{person}{B. Witemeyer}, \bibinfo{person}{N. Weidner},
  \bibinfo{person}{T. Davis}, \bibinfo{person}{T. Kim}, {and}
  \bibinfo{person}{S. Sueda}.} \bibinfo{year}{2021}\natexlab{}.
\newblock \showarticletitle{QLB: Collision-Aware Quasi-Newton Solver with
  Cholesky and L-BFGS for Nonlinear Time Integration}. In
  \bibinfo{booktitle}{\emph{Proc 14th ACM SIGGRAPH Conf Mot Int Games}}
  \emph{(\bibinfo{series}{MIG '21})}. \bibinfo{publisher}{ACM}, Article
  \bibinfo{articleno}{14}, \bibinfo{numpages}{7}~pages.
\newblock


\bibitem[Zhu et~al\mbox{.}(2018)]%
        {zhu:2018:qn}
\bibfield{author}{\bibinfo{person}{Y. Zhu}, \bibinfo{person}{R. Bridson}, {and}
  \bibinfo{person}{D. Kaufman}.} \bibinfo{year}{2018}\natexlab{}.
\newblock \showarticletitle{Blended Cured Quasi-Newton for Distortion
  Optimization}.
\newblock \bibinfo{journal}{\emph{ACM Trans Graph}} \bibinfo{volume}{37},
  \bibinfo{number}{4}, Article \bibinfo{articleno}{40} (\bibinfo{date}{jul}
  \bibinfo{year}{2018}), \bibinfo{numpages}{14}~pages.
\newblock


\end{thebibliography}

\end{document}